\RequirePackage{amsmath}

\documentclass[epj]{svjour}
\usepackage{graphicx}

\usepackage{latexsym}
\usepackage{graphics}
\usepackage[usenames, dvipsnames]{color}
\usepackage{hyperref}
\usepackage{multirow}
\usepackage{booktabs}
\usepackage{amssymb}
\usepackage{cite}
\usepackage[dvipsnames]{xcolor}

\hypersetup{linkcolor=cyan, colorlinks = true, citecolor=magenta}

\DeclareMathOperator{\Tr}{Tr}
\DeclareMathOperator{\Diag}{Diag}
\begin{document}
\title{Lattice Boltzmann Methods and Active Fluids}
\author{Livio Nicola Carenza\inst{1} \and Giuseppe Gonnella\inst{1}  \and Antonio Lamura\inst{2} \and Giuseppe Negro\inst{1}
\and Adriano Tiribocchi\inst{3}}
%
%
\institute{Dipartimento di Fisica, Universit\`a degli Studi di Bari, and INFN, Sezione di Bari, Via Amendola 173, Bari 70126, Italy \and Istituto Applicazioni Calcolo, CNR, Via Amendola 122/D, 70126 Bari, Italy \and Center for Life Nano Science@La Sapienza, Istituto Italiano di Tecnologia, 00161 Roma, Italy}
\date{}
%
\abstract{We review the state of the art of active fluids with particular attention to hydrodynamic continuous models and to the use of Lattice Boltzmann Methods (LBM) in this field. We present the thermodynamics of active fluids, in terms of liquid crystals modelling adapted to describe large-scale organization of active systems, as well as other effective phenomenological models. We discuss how LBM can be implemented to solve the hydrodynamics of active matter, starting from the case of a simple fluid, for which we explicitly recover the continuous equations by means of Chapman-Enskog expansion. Going beyond this simple case, we summarize how LBM can be used to treat complex and active fluids.
We then review recent developments concerning some relevant topics in active matter that have been studied by means of LBM: spontaneous flow, self-propelled droplets, active emulsions, rheology, active turbulence, and active colloids.} 
\maketitle
\section{Introduction}
\label{intro}

The goal of this article is to describe the use of Lattice Boltzmann Methods in the study of large scale properties of active fluids~\cite{toner2005,ramaswamy2010,marchetti2013,elgeti2015,
gonnella2015,bechinger2016,doostmohammadi2018}, 
also showing  recent progress in few relevant topics.
Active fluids are living matter or biologically inspired systems with the common
characteristic of being  composed by self-propelled (or active) units that burn stored or ambient energy and turn it into work giving rise, eventually, to systematic movement. An example in nature is given by the cell cytoskeleton or, in laboratory,  by synthetic suspensions of cell extracts with molecular motors (e.g. myosin or kinesin)~\cite{surrey2001,bendix2008}. 
Molecular motors exert forces on cytoskeletal filaments (actin filaments and microtubules)~\cite{howard2003} and trigger their motion in the surrounding fluid. These forces, exchanged through transient and motile contact points between filaments and motor proteins, result from the conversion of chemical energy, typically coming from ATP hydrolysis, into mechanical work. 

Active systems show many interesting physical properties, of general character, related to their collective behavior, remarkable especially when compared with their analogue  in passive or equilibrium systems. Pattern formation is an example.
A disordered array of microtubules may arrange into spiral or aster configurations when the concentration of motor proteins like kinesin is sufficiently high~\cite{surrey2001}.
Suspensions of bacteria, despite their low Reynolds numbers, can exhibit turbulent flow patterns~\cite{dombrowski2004,dunkel2013}, characterized  by traveling \textit{jets} of high collective velocities and surrounding \textit{vortices}.
Active fluids can be classified according to their swimming mechanism as extensile or contractile, if they respectively push or pull the surrounding fluid. This difference marks all the phenomenology of active fluids and, in particular, has important effects on
the rheological properties. Activity is either capable to develop
shear-thickening properties in contractile systems~\cite{hatwalne2004,marenduzzo2007,foffano2012,foffanoPRL,sokolov2009}, or to induce a \textit{superfluidic} regime under suitable conditions in extensile suspensions~\cite{cates2008,loisy2018,Guo201722505}.  Simulations of extensile active emulsions under constant shear have shown the occurrence of velocity profiles (for the component of velocity in the direction of the applied flow) with inverted gradient 
(\textit{negative viscosity}) and also jumps in the sign of apparent viscosity~\cite{cates2008,loisy2018,Guo201722505}.  

Other striking properties have emerged in the study of fluctuation statistics~\cite{simha2002,chate2006,narayan2007,deseigne2010,kumar2011,cagnetta2017} and of  order-disorder phase transitions
~\cite{bechinger2016,cates,paoluzzi2016}.
Fluctuations and phase transitions  have been  mainly  analyzed in the context   of agent-based models.
The flocking transition~\cite{cavagna2014}, for instance, was the first one to be  studied  in  a model of point-like particles moving at  fixed speed and with
aligning interaction~\cite{vicsek1995}.
Activity alone  actually favors aggregation and can  induce a phase transition, often called Motility Induced Phase Separation (MIPS)~\cite{cates2015}.  This has been numerically studied by using simple 
models of active colloids with  excluded volume interactions and various shapes~\cite{peruani2006,baskaran2008,yang2010,fily2012,lowen2013, redner2013,cugliandolo2017,digregorio2018}.
The particle description has been  also largely used in other contexts, to simulate, for example, the self-organization of cytoskeleton filaments described as semiflexible filaments~\cite{nedelec1997}.

By a different approach,
 large scale behavior and macroscopic material properties of active fluids have been 
largely studied using 
 coarse-grained descriptions based on general symmetry arguments  and conservation laws.
The first continuum description in terms of density and  polarization field,
with interactions favoring  alignment with   polar order,
was  proposed in~\cite{toner1995}.    In this model, as  in others where nematic interactions
were considered~\cite{gruler1999,ramaswamy2003,baskaran2010}, the medium in which particles are supposed to move
does not  contribute with its own dynamics to the evolution of the system.  Hence the environment of the active system can be considered as a momentum-absorbing substrate so that momentum is not conserved.  On the other hand, there are systems in which the dynamics of the solvent can  be relevant in a certain  interval of length scales~\cite{schaller2010} and must be
incorporated in the description. 
The action of the active components on  the solvent  is taken into account by introducing an active stress into a generalized form of the Navier-Stokes equation. Suitable advection terms depending on the self-propulsion velocity of active units also appear in the dynamical equations for the order parameters
describing the orientation of the active material (nematic or polar) or its concentration.  
A  useful form  for the active stress was first proposed in~\cite{pedley1990}  and later developed 
in the context of a coarse-grained model  in~\cite{simha2002} and, for active filaments or orientable particles, 
in~\cite{kruse2004,julicher2007}. 
The total  stress 
also includes elastic contributions, depending on the polar or nematic character of the system, stemming from an appropriate free-energy expression, as in the passive or equilibrium counterpart of the systems in exam usually called active gels.
The resulting  dynamical  description consists of
 non-linear coupled partial differential equations that require   numerical methods to be solved. 
A suitable approach, largely used to study multicomponents and complex fluids whose dynamics obey such equations, is the Lattice Boltzmann Method (in the following we will refer to it as LBM or LB)~\cite{higuera1989,succi1991,succi2001}, a computational fluid dynamics scheme for  solving the Navier-Stokes equation, eventually  coupled to advection-relaxation equations~\cite{yeomans2009,swift1996,gonnella1997,lamura1999,tiribocchi2009,denniston2001}. Among its features, this method is found to correctly capture the coupling between hydrodynamics and orientational order of liquid crystals (often known as backflow~\cite{leslie1968,martin1972,beris1994}), a crucial requirement to simulate the dynamics of active gels~\cite{orlandini2008,marenduzzo2007}.




In this article we will review the way LBM can be used to describe collective properties of active fluids, describing also recent developments concerning  issues where hydrodynamics plays a relevant role.
We will initially review the thermodynamics of active fluids whose internal constituents are orientable objects, such as active liquid crystals. After shortly introducing the order parameters and the free-energy usually adopted to describe their properties, we will show how the active behavior enters the model and how hydrodynamic equations can be written to correctly capture the physics. This will be done in Section~\ref{sec:2}.

Afterwards we will discuss different LB strategies used to study simple and structured fluids, convenient for active fluids generalization. For a simple fluid, LBM solves a minimal Boltzmann kinetic equation governing the evolution of a single set of variables (the distribution functions), in terms of which hydrodynamic quantities can be written~\cite{mcNamara1988,higuera1989}. A detailed description of the LB methods for a single fluid can be found in~\cite{succi2001,Chen1992,succi2018}. For structured fluids, a full LBM approach can be followed by introducing a further set of distribution functions for the  order parameter that follow the dynamics of appropriate  lattice Boltzmann equations to be  added to those describing the dynamics of the density and velocity of the fluid~\cite{swift1996}. Then, interactions can be implemented by specific collision rules introduced on a phenomenologically ground or by making reference to a specific free-energy model that sets the thermodynamics of the system~\cite{swift1995,swift1996,li2012,li2012_2}. The first approach, in numerous variants, has been largely used in the context of binary mixtures, due to its practical convenience, with the collision step designed in order to favor separation of the A and B components of the mixture~\cite{shan1993}. When the fluid  structure becomes more complex, the second approach becomes almost mandatory. The characteristics of a specific system will enter the lattice dynamic equations through a chemical potential and a pressure tensor that can be obtained by a given free-energy functional. Liquid crystals~\cite{denniston2001}, but also ternary mixtures with surfactant~\cite{lamura1999} or other kinds of complex fluids~\cite{Cates2099,tiribocchi2011}, have been largely studied in this way. Finite difference methods, with possible  numerical advantages, can be also applied to simulate the order parameter dynamical equations~\cite{tiribocchi2009} and have been implemented in hybrid approaches coupled to  LBM used as a solver for Navier-Stokes equations. These different options  will be reviewed in Section~\ref{sec:LBM}, in relation with the modeling of active fluids proposed in Section \ref{sec:2}, and  with details on possible  algorithms and numerical implementations.

The following Sections will be dedicated to discuss some relevant topics in active fluids in which LBM has played an essential role. In Section~\ref{sec:spont_fl} the main numerical results concerning the hydrodynamic instabilities generated by spontaneous flows~\cite{voituriez2005,marenduzzo2010} will be reviewed. Understanding how this occurs is fundamental, for instance, to assess the dynamics of topological defects as well as the physics of self-propelled droplets, objects which can capture some relevant features of motile cells~\cite{tjhung2012,tjhung2015}.
Section~\ref{sec:self_prop} will be devoted to review  relevant  results 
on the modeling of self-propelled droplets and of systems with many droplets such as active emulsions.
The latter is  a new subject of research with new fascinating perspectives. Active emulsions can be potentially realized by dispersing sticky bacteria~\cite{linek2012} or self attractive cytoskeleton gels~\cite{sanchez2012,guillamat2016} in water, or encapsulating an active nematic gel within a water-in-oil emulsion~\cite{sanchez2012,guillamat2016}.
Another stimulating field of research
concerns the study of the rheological response of an active fluid to  externally imposed flows. In Section~\ref{sec:rheology} we will review the most recent and pioneering achievements in this 
field, in which, for example, an active gel has been predicted to have either a shear-tickening or superfluid-like behavior depending on the nature, extensile or contractile, of the flow~\cite{PhysRevE.83.041910}. 
As a further topic we will illustrate the results obtained via LBM simuations to investigate ``active'' turbulence~\cite{dombrowski2004, wensink2012, creppy2015, putzig2016}, a \emph{turbulent}-like behavior observed in active fluids at low Reynolds numbers (Section~\ref{sec:turb}).

The versatility of LBM as a solver for hydrodynamics  has been also used to study the flow generated by different kinds of swimmers, treated as discrete particles coupled to the surrounding fluid by proper boundary conditions~\cite{pooley2008}. In this case the solvent is described as a simple fluid whose dynamics can be solved by LBM. Although the study of the collective properties of these systems can be difficult within this approach due to the complicated structure of the flows induced by the swimmers, in a few cases this shortcoming has been overcome by using a mixed particle-continuum description~\cite{ramachandran2006,Llopis2006,alarcon2013,earl2007,smith2007,degraaf2016,degraaf2016_2}. Section~\ref{sec:coll} will be  dedicated to describe how LBM has been extended to include active particles.


\section{Active fluid models}
\label{sec:2}

In this section we will focus on fluids whose internal units have an orientable character, a feature that crucially affects their reciprocal interactions, especially when a high density sample of active units is considered. In such cases the emerging orientational order on  macroscopic scales can be captured by proper order parameters, such as the polarization vector  $\vec P(\vec r,t)$ and the tensor ${\underline{\underline{\vec Q}}}(\vec r,t)$, often used to describe ordering in liquid crystals. These quantities will be introduced in~\ref{sec:2a}.

The thermodynamics of these systems is usually described via a Landau-like free energy functional, depending upon powers of the order parameter and its gradients, respecting the symmetries of the disordered phase. The different free-energy terms describing bulk and elastic properties of the active fluid will be discussed in~\ref{sec:2b}, while~\ref{sec:2c} will be dedicated to describe how activity is introduced in continuum models. In~\ref{sec:2d} we will briefly discuss the thermodynamics of a fluid mixture with an active component, with and without alignment interaction. The latter case has been recently considered for the study of the motility induced phase separation in active fluids~\cite{tiribocchi2015}.

Finally the hydrodynamic equations describing both the evolution of the order parameter and of the velocity field will be shown in~\ref{sec:2e}.
\subsection{Order parameters}
\label{sec:2a}

\begin{figure*}
\center
\includegraphics[width=1\textwidth]{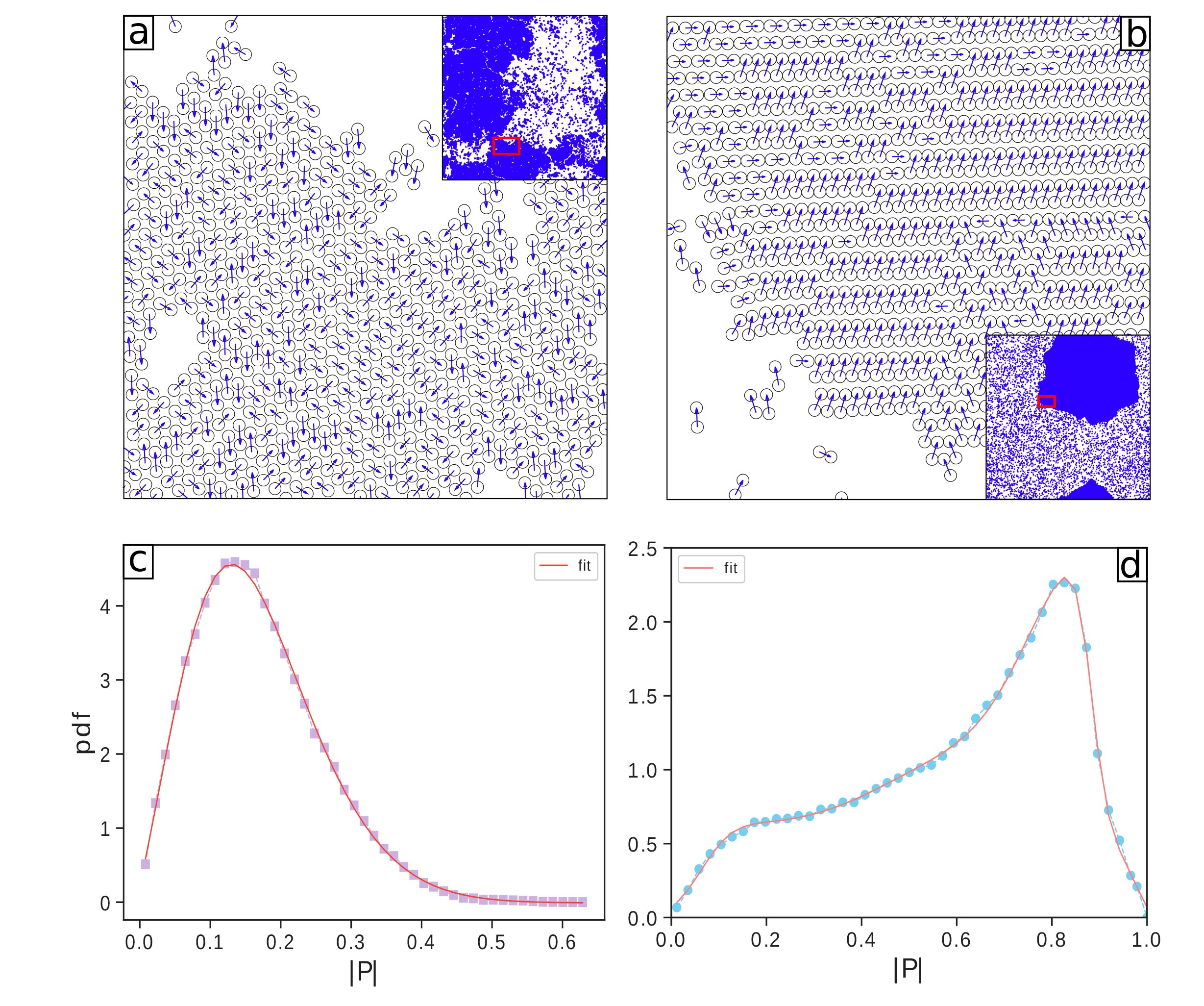}
\caption{System of self-propelled Brownian dumbbells for total covered  fraction area $\phi= 0.5$ and  different values of the self-propulsion force corresponding to the P\'eclet number $Pe=10$ and $Pe=40$,  in  panels (a) and (b), respectively. 
For the definition of the model and  detailed meaning of parameters see~\cite{petrelli2018,cugliandolo2015}. Dumbbells have a tail and a head; the blue vectors represent the directions of self-propulsion  of each  dumbbell, related to the tail-head axis. The snapshots represent small portions (red boxes) of the larger systems shown in the insets.
Both cases correspond to points in the phase diagram where a dilute and a  more dense aggregated phase coexist. Note that for 
small P\'eclet number  polar order is not present in the aggregated phase that only shows  hexatic order, while for higher P\'eclet   the hexatic phase is  polarized.  The probability distributions (\emph{pdf}) of the local coarse-grained polarization field confirm 
this behavior. At small P\'eclet the \emph{pdf} (panel (c)) shows a maximum at a polarization  magnitude  $|P| \approx 0.15$ while at $Pe=40$ the \emph{pdf} (panel (d)) can be interpreted as taking contributions from two distributions with maxima at $|P| \approx 0.18$ and $|P| \approx 0.8$,  
respectively~\cite{petrelli2018}.
} \label{fig:peclet_10_40_dumbbells}   
\end{figure*}

Active fluids whose internal constituents have an anisotropic shape (such as an elongated structure) encompass diverse systems ranging from bacterial colonies and algae suspensions~\cite{marchetti2013} to the cytoskeleton of eukaryotic cells~\cite{bray2000}. Depending upon the symmetries of such microscopic agents and upon their reciprocal interactions, these active fluids generally fall into two wide categories. 
The first one is the active polar fluid composed of elongated self-propelled particles, characterized by a head and a tail, whose interactions have polar symmetry. Such systems may order
either in polar states, when all the particles are on average aligned along the same direction,
as in the case of bacteria self-propelled along the direction of their head~\cite{dombrowski2004}. Nevertheless systems of intrinsic polar particles, such as actin filaments cross-linked with myosin~\cite{bray2000,tjhung2011,tjhung2012,nature2015} or
microtubule bundles coupled with kinesin motors~\cite{surrey2001,Decamp2015,sanchez2012}, may still arrange in a nematic
fashion, restoring head-tail symmetry, when interactions favour alignment regardless of the
polarity of the individual particles.
Fig.~\ref{fig:peclet_10_40_dumbbells} shows, for example, the aggregated phase of a system of self-propelled Brownian polar dumbbells~\cite{suma2014,suma2014_2,gonnella2014,cugliandolo2015} which, depending on the strength of the self-propulsion force, may arrange in a polar state (right) or in an isotropic state (left), a behavior also found in bacterial colonies~\cite{dellarciprete2018}. The second class includes head-tail symmetric, or {\it apolar}, particles that may move back and forth with no net motion, and order in nematic states. Examples of realizations in nature include \textit{melanocytes}~\cite{Kemkemer2000}, i.e. melanin producing cells in human body, and \textit{fibroblasts}~\cite{Duclos2014}, cells playing a central role in wound healing, both spindle-shaped with no head-tail distinction. 

The continuum fields describing polar and nematic order are the vector field $P_\alpha(\vec r,t)$ and the tensor field $Q_{\alpha \beta}(\vec r,t )$ respectively (Greek subscripts denote the Cartesian components). They emerge either from a coarse grained description of a microscopical model~\cite{degennes1993} or from a theory based on general symmetry arguments~\cite{martin1972,beris1994}. Following, for instance, the former approach, for a system of rod-like particles the polarization field can be defined as
\begin{equation}
\vec{P}(\vec{r},t) = \langle \vec{\nu} (\vec{r},t) \rangle  = \int \mathrm{d \Omega} f_P(\vec{\nu}, \vec{r},t) \vec{\nu},
\end{equation}
where $f_P(\vec{\nu}, \vec{r},t)$ is the probability density, encoding  all the information coming from the microscopical model, of finding a particle at position $\vec{r}$ and at time $t$ oriented along the direction $\vec{\nu}$, and the integration is carried out over the solid angle $\mathrm{\Omega}$. The polarization can be also written as
\begin{equation}
\vec{P}(\vec{r},t) = P(\vec{r},t) \vec{n}(\vec{r},t),
\end{equation}
where $\vec{n}(\vec{r},t)$ is a unit vector defining the local mean orientation of particles in the neighborhood of $\vec{r}$, and  $P(\vec{r},t)$ is a measure of the local degree of alignment, ranging from $0$ (in an isotropic state) to $1$ (in a perfectly polarized state).

Differently, the  nematic phase cannot be described by a vector field, as both orientations $\vec{\nu}$ and $-\vec{\nu}$ equally contribute to the same ordered state, due to the head-tail symmetry of the constituents. For a system of rod-like particles, the order is described by a nematic tensor which, in the uniaxial approximation (i.e. when a liquid crystal is rotationally symmetric around a single preferred axis), can be defined as
\begin{equation}
\label{def:particle_nematic_tensor}
Q_{\alpha \beta}(\vec{r},t) = \langle \nu_\alpha \nu_\beta - \dfrac{1}{d} \delta_{\alpha \beta} \rangle  = \int \mathrm{d \Omega} f_Q(\vec{\nu}, \vec{r},t) (\nu_\alpha \nu_\beta-\dfrac{1}{d}\delta_{\alpha \beta}).
\end{equation}
Again $f_Q(\vec{\nu}, \vec{r},t)$ is the probability density to find a nematic particle oriented along $\vec{\nu}$ at position $\vec{r}$ and time $t$, while $d$ is the dimensionality of the system. As for the polarization field, the nematic tensor can be also written in terms of the versor $\vec{n}$ (usually called \emph{director} field) defining the local mean orientation of the particles
\begin{equation}
\label{def:nematic_tensor}
Q_{\alpha \beta}(\vec{r},t) = S(\vec{r},t)\left[n_\alpha (\vec{r},t) n_\beta(\vec{r},t) - \frac{1}{d} \delta_{\alpha \beta}\right].
\end{equation}
Note that, by defining the nematic tensor in this way, one can separate local anisotropic features out of isotropic ones. Indeed, the only scalar quantity that can be derived from a tensorial object, \emph{i.e.} its trace, is identically null. In Eq.~\eqref{def:nematic_tensor} $S(\vec{r},t)$ plays the same role of $P(\vec{r},t)$ in defining the degree of alignment of the molecules in the nematic phase.  In fact, by multiplying Eq.~\eqref{def:particle_nematic_tensor} and~\eqref{def:nematic_tensor} by $n_\alpha n_\beta$, summing over spatial components and comparing them, one gets (in three dimensions)
\begin{equation}
\label{def:S}
S(\vec{r},t) = \dfrac{1}{2} \langle 3 \cos^2 \theta - 1 \rangle,
\end{equation}
where $\cos \theta= \vec{n} \cdot \vec{\nu}$ is a measure of the local alignment of particles. The scalar order parameter $S$ achieves its maximum in the perfectly aligned state, where $\langle  \cos^2 \theta \rangle=1$, while it falls to zero in the isotropic phase where the probability density $f_Q$ is uniform over the solid angle and $\langle  \cos^2 \theta \rangle=1/3$. Assuming $\vec{n}$ to be parallel to a Cartesian axis, one can soon verify from Eq.~\eqref{def:nematic_tensor} that $Q_{\alpha\beta}$ has two degenerate eigenvalues $\lambda_2=\lambda_3=-S/3$ (whose associated eigenvectors lie in the plane normal to the particle axes) and a third non-degenerate one $\lambda_1=2S/3$, greater in module than $\lambda_2$ and $\lambda_3$ and related to the director itself. Such formalism can be also extended to treat the case of biaxial nematics, \emph{i.e.} liquid crystals with three distinct optical axis. Unlike an uniaxial liquid crystal which has an axis of rotational symmetry (such as the director ${\vec n}$), a biaxial liquid crystal has no axis of complete rotational symmetry. As such theory is out of the scope of this review, we briefly mention it in Appendix A, focusing, in particular, on how biaxiality is included in the tensor order parameter and on the role it plays in the localization of topological defects. 
\begin{figure}
\center
\resizebox{1.0\columnwidth}{!}{\includegraphics{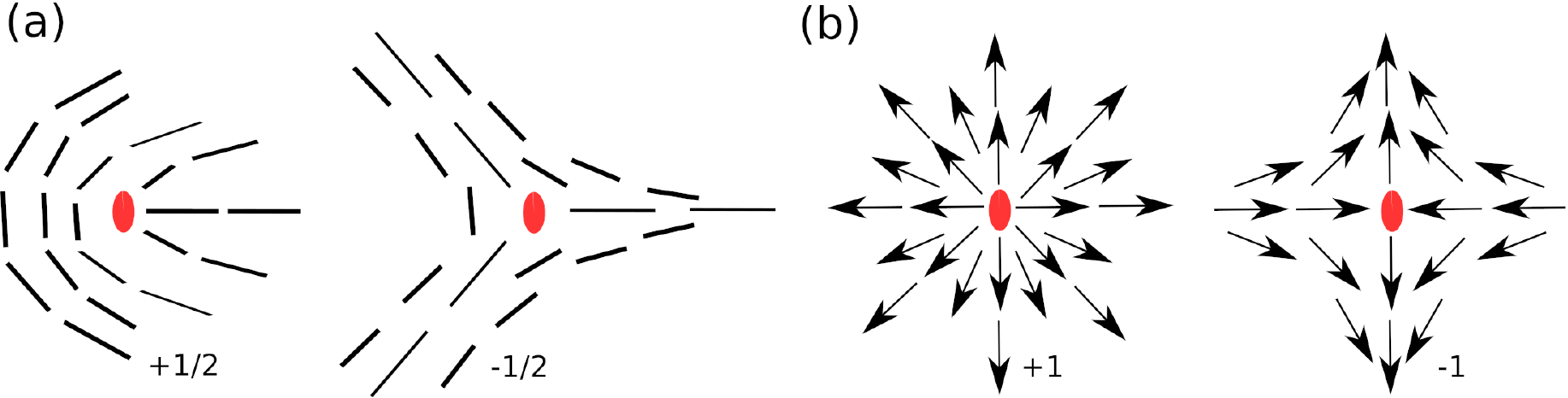}}
\caption{Sketch of (a)  half-integer topological defects in 2D nematic liquid crystals, and (b) integer topological defects in polar liquid crystals. These can only host defects with integer winding number (see main text).}
\label{fig:1} 
\end{figure}

\begin{figure*}
\center
\resizebox{1.0\textwidth}{!}{\includegraphics{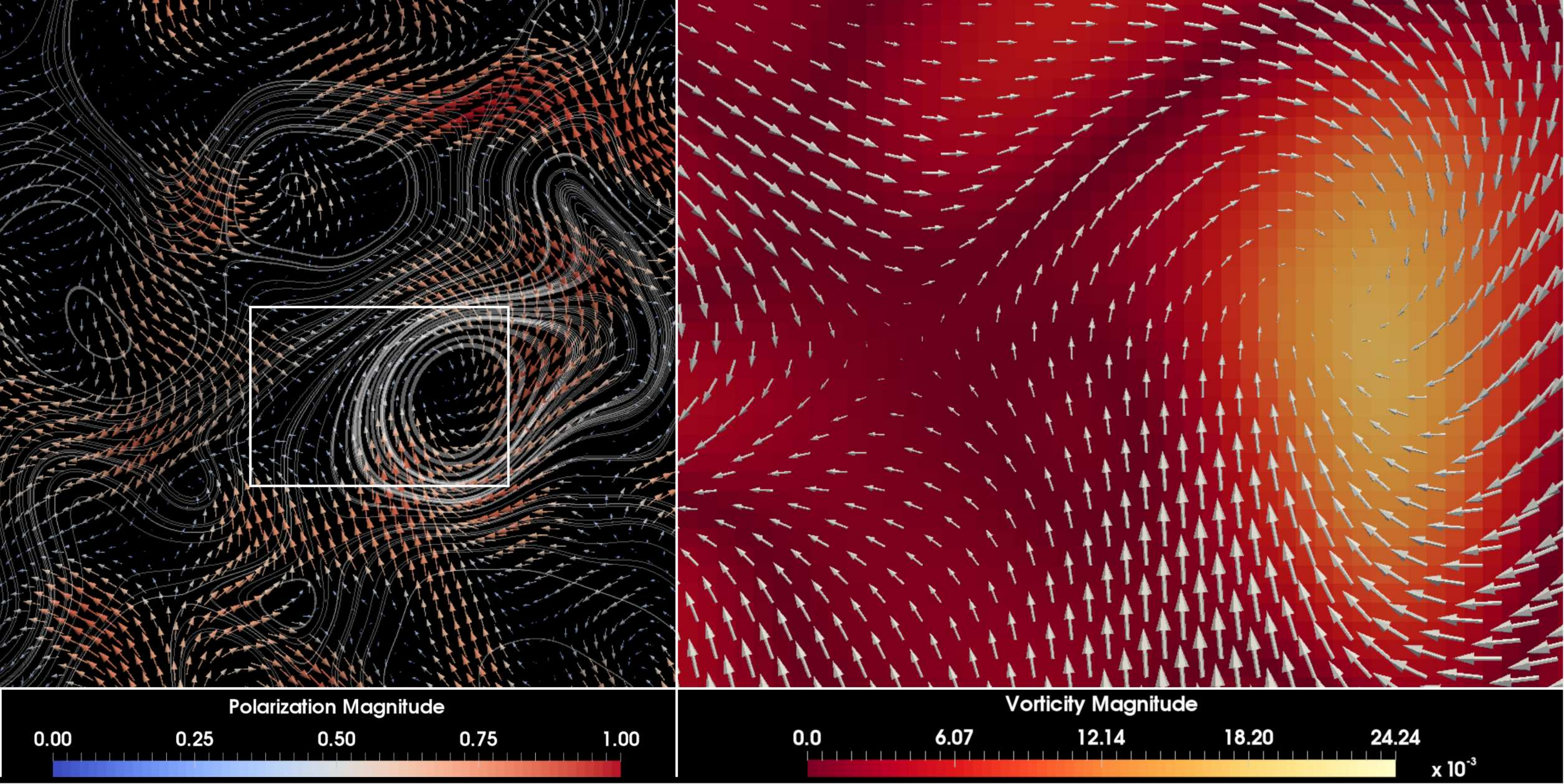}}
\caption{Defect dynamics in active polar systems. The left panel shows the polarization field, represented by arrows,  with the superposition of some velocity streamlines; red/long arrows correspond to ordered regions, while blue/short arrows are associated with the presence of topological defects, surrounded by regions with strong deformations of the polarization. Note that $ + 1$  defects act as a source of vorticity: indeed, most of the  closed streamlines wrap the core of a defect. This is also  shown in the right panel with  the polarization field superimposed to the vorticity contour plot in the region highlighted by the white box in the left panel. 
Here two defects of charge $ \pm 1$ are close. In proximity of the defect cores the polarization magnitude is approximately null 
and order is locally lost. These simulations have been performed by the authors of this paper using a lattice  Boltzmann approach applied to  the model described by the free energy in Eq.~\eqref{free-energy-mixture}, initializing the system uniformly in the active phase ($\phi(\vec{r})=\phi_0$).}
\label{fig:defects_simulation}  
\end{figure*}

\subsubsection{Topological defects}

Topological defects (\emph{disclinations} in nematic/polar and cholesteric liquid crystals) are regions where the order parameter cannot be defined~\cite{degennes1993,chaikin1995}. A crucial difference between allowed polar and nematic systems really lies on the nature of the topological defects. As they play a relevant role in the dynamics of the velocity field in active fluids, we provide here a brief introduction about the theory of topological defects and address the reader to more specialized books (such as~\cite{chaikin1995}) for further details. 

A topological defect can be characterized by looking at the configuration of the order parameter far from its core. This can be done by computing the \emph{winding} number (or topological charge), which is a measure of the strength of the topological defect and is defined as the number of times that the order parameter turns of an angle of $2\pi$ while moving along a close contour surrounding the defect core. Hence possible values of defect strengths critically depend upon the nature of the order parameter: indeed polar systems only admit topological defects with integer winding numbers (Fig.~\ref{fig:1}b), while nematic systems offer a wider scenario; in fact by virtue of the  head-tail symmetry, the headless nematic director can give rise to disclination patterns that also allows for half-integer winding numbers (Fig.~\ref{fig:1}a).  Fig.~\ref{fig:defects_simulation} shows, for example, two defects of charge $\pm 1$ in an active  contractile polar system: their mutual attracting interaction, due to elastic deformations,  couples to the hydrodynamics generating a backflow~\cite{Thot2002,giomi2013} that moves the two defects closer and leads to their annihilation. Fig.~\ref{fig:defects_simulation} also shows how  defects act as a source of vorticity with the velocity field tilted with respect to polarization.
On the contrary, if the system is extensile, activity drives defects of opposite topological charge apart and suppresses pair annihilation~\cite{giomi2013,sanchez2012}. In simulations the correct position of a topological defect can be tracked either by looking at the polarization (or director for nematics) field profile or, only for nematics, by locating the regions where the scalar order parameter of the tensor field drops down. In the latter case, a further method, based on computing the degree of biaxiality around the defect core is briefly discussed in Appendix A. Indeed, regions close to the defect core display biaxiality~\cite{PhysRevLett.59.2582}. 
Note that, although defects appearing in the active fluid of Fig.~\ref{fig:defects_simulation} are points, other structures are possible. 

Defects are said to be topologically stable if a non uniform configuration of the order parameter cannot be reduced to a uniform state by a continuous transformation.
A general criterion to establish whether a defect is topologically stable or not, is to look at the dimension $n$ of the order parameter. In a $d$-dimensional space, the condition that all the $n$ components of the order parameter must vanish at the defect core defines a ``surface'' of dimension $d-n$. Hence defects exist if $n\leq d$. In Fig.~\ref{fig:defects_simulation}, for example, we have a two-dimensional system ($d=2$) with  an order parameter (the polarization $\vec P$) having  two components ($n=2$), and the defects allowed are points (or vortices). However,  point defects can be unstable in quasi-$2d$ systems, i.e. when the order parameter fully lives in the three-dimensional space, as in such case one would have $n>d$: indeed the vector field in proximity of a vortex is always capable to escape out of the plane aligning with itself, thus removing the defect. In three-dimensional systems ($d=3$) one may have either point defects (if $n=3$) or lines (if $n=2$).

\subsection{Free energy}
\label{sec:2b}

\begin{table*}[h]
\centering
\caption{The table summarizes bulk and elastic contributions to free energy for polar and nematic, both uniaxial and biaxial (see Appendix A), systems. Splay, twist and bending contributions have been written explicitly in terms of different elastic constants $\kappa_i$ ($i=1,2,3$) for both polar and uniaxial nematic gels, while in the most general case of a biaxial nematic we did not distinguish between different contributions. The last line in the Table shows how the elastic contribution looks like assuming that the medium is elastically isotropic, \textit{i.e.}, $\kappa_1=\kappa_2=\kappa_3=\kappa$.}
\label{tab:1}
\begin{tabular}{lllrc}
\toprule
Free energy contributions  &  & \multicolumn{1}{l|}{Polar Gel} & \multicolumn{2}{c}{Nematic Gel} \\ \cmidrule{4-5}
                                     &  & \multicolumn{1}{l|}{} & Uniaxial & Biaxial         \\ \cmidrule{1-5}
\multicolumn{2}{l}{Bulk}             & $a \vec{P}^2+b \vec{P}^4 $  & $r S^2 -wS^3 + u S^4 $ & $\tilde{r} Q_{ij}Q_{ji} -\tilde{w} Q_{ij}Q_{jk}Q_{ki} + \tilde{u} (Q_{ij}Q_{ji})^2 $                 \\ \hline

\multicolumn{1}{l|}{\multirow{3}{*}{Elastic}} & Splay & $\dfrac{\kappa_1}{2} (\nabla \cdot \vec{P})^2 $  & \multicolumn{1}{l|}{$\dfrac{\kappa_1}{2} (\nabla \cdot \vec{n})^2$}  & \multirow{3}{*}{} \\ \cmidrule{2-4}
\multicolumn{1}{l|}{}                  & Twist & $\dfrac{\kappa_2}{2} (\vec{P} \cdot \nabla \times \vec{P})^2 $ &  \multicolumn{1}{l|}{$\dfrac{\kappa_2}{2} (\vec{n} \cdot \nabla \times \vec{n})^2 $} & $ \dfrac{L_1}{2}(\partial_k Q_{ij})^2+ \dfrac{L_2}{2}(\partial_j Q_{ij})^2  + \dfrac{L_3}{2} Q_{ij} (\partial_i Q_{kl})(\partial_j Q_{kl}) $       \\ \cmidrule{2-4}
\multicolumn{1}{l|}{}                  & Bend & $\dfrac{\kappa_3}{2} (\vec{P} \times \nabla \times \vec{P})^2 $ & \multicolumn{1}{l|}{$\dfrac{\kappa_3}{2} (\vec{n} \times \nabla \times \vec{n})^2 $} &                   \\ \cmidrule{1-5}
\begin{tabular}[c]{@{}l@{}}Single constant\\ approximation\end{tabular} &  & $\kappa(\nabla \vec{P})^2$ & $\kappa(\nabla \vec{n})^2$ & $L_1 (\partial_k Q_{ij})^2$   \\ \bottomrule
\end{tabular}
\end{table*}

In this Section we will shortly review the free-energy expressions generally used to describe polar and nematic suspensions and often employed in studying  active fluids, built from the order parameters previously discussed.

Bulk properties and order-disorder phase transitions can be derived by a free energy functional with terms respecting the symmetries of the disordered phase, in the spirit of Landau approach. Free-energy $F$ will only contain scalar terms invariant under space rotations, proportional to the order parameters and their powers. For a vectorial order parameter, scalar objects of the form $\vec{P}^{2 m}$ can be considered, with  $m$ positive integer, usually  arresting the expansion to the fourth order. For the nematic order parameter scalar quantities are  
of the form $\Tr (\vec {\underline{\underline{Q}}}^m)$; note that there is no impediment here to odd power terms, by virtue of the invariance of $\vec {\underline{\underline{Q}}}$ under inversions, but no linear term will appear in the expansion since $\Tr \vec {\underline{\underline{Q}}}$ is identically null by definition. The presence of a third order term will lead to a first order nematic-isotropic transition through the establishment of metastable regions in the phase diagram~\cite{chaikin1995}. Table~\ref{tab:1} summarizes the bulk contributions to free energy for both polar and nematic systems; note that the uniaxial free energy can be derived from the biaxial case by writing the $\vec {\underline{\underline{Q}}}$ tensor through Eq.~\eqref{def:nematic_tensor}.

In order to take into account the energetic cost due to continuous deformations of the order parameters,  elastic terms are also introduced in the free energy functional. In both polar and nematic systems three different kinds of deformations can be identified: splay, twist and bending, gauged to the theory through (in general) different elastic constants $\kappa_1,\kappa_2,\kappa_3$. While splay is related to the formation of fan-out patterns of the director and polarization field, bending generates rounded circular patterns. Instabilities associated to such deformations underlie the establishment of defects of different strength. Twist is forbidden in pure bidimensional systems, since this kind of deformation implies the director to coil around an axis, normal to the director itself. Table~\ref{tab:1} also provides a picture of the energetic cost due to different kinds of deformations in terms of $\vec{P}$ and $\vec{n}$, respectively for polar systems and uniaxial nematics, under the assumption of uniform ordering ($S=cost$). The most general case is provided by the elastic contributions in biaxial nematics and still applies to the uniaxial case with $S=S(\vec{r})$.
In order to exploit which terms are related to which deformations, one should expand the $\underline{\underline{\vec Q}}$ tensor into the elastic biaxial free energy in terms of the director through Eq.~\eqref{def:nematic_tensor}; doing so and grouping splay, twist and bend contributions one finds, after some algebric effort, that
\begin{equation*}
\begin{array}{l}
\displaystyle L_1 = \dfrac{\kappa_3+2\kappa_2-\kappa_1}{9S^2},\\
\\
\displaystyle L_2 = \dfrac{4(\kappa_1-\kappa_2)}{9S^2},\\
\\
\displaystyle L_3 = \dfrac{2(\kappa_3-\kappa_1)}{9S^3},
\end{array}
\label{eq:xdef}
\end{equation*}
given that the Frank constants $\kappa_i$ fullfill the condition $\kappa_3 \geqslant \kappa_1 \geqslant \kappa_2$ to guarantee the positivity of $L_i$~\cite{schiele1983}.
In many practical situations it is convenient to adopt the single constant approximation, consisting in setting all elastic constants equal to the same value, leading to a much simpler form for the elastic free energy~\cite{chaikin1995}.

\subsection{Active Forces}
\label{sec:2c}

\begin{figure}[b]
\center
\resizebox{.8\columnwidth}{!}{%
  \includegraphics{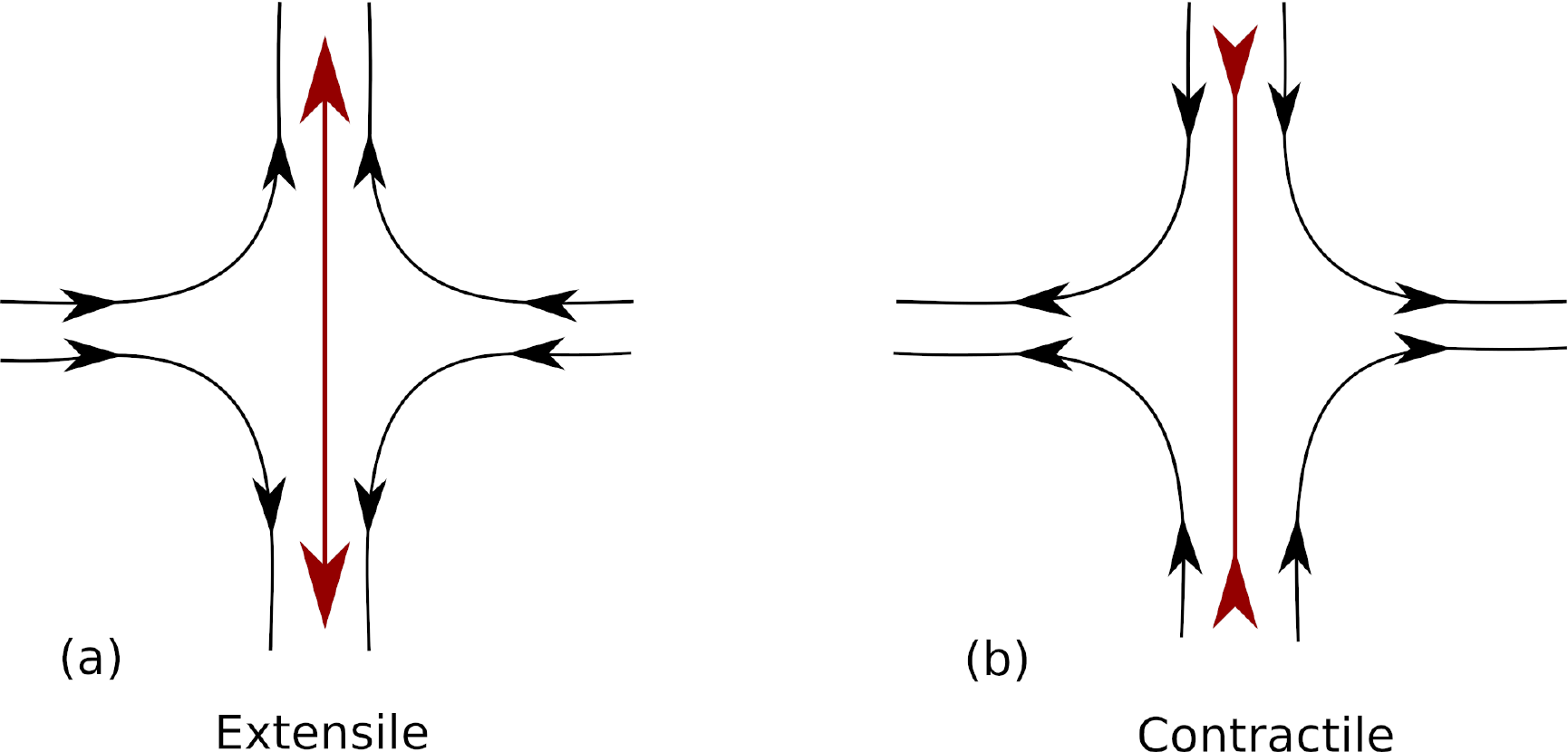}
}
\caption{ Cartoon of  \textbf{(a)} extensile  and \textbf{(b)} contractile flow (black lines), and force dipoles (red arrows).}
\label{fig:2}       
\end{figure}

So far we reviewed the well known theoretical description for liquid crystals and  fluids with anisotropic order parameter. 
We  will see now how the active behavior of the constituents of the fluid can be expressed  into the theoretical framework.
The most direct way to develop the equations of motion for active systems at  continuum level 
is by explicitly coarse-graining more detailed particle-based models~\cite{ramaswamy2003,marchetti2013}. Therefore, before starting the theoretical description, we spend few words in describing the swimming mechanism of some microorganisms. 

In general, the propulsive motion of  active agents dispersed in a fluid creates a circulating flow pattern around each swimmer. The specific swimming mechanism of bacteria, for example, causes fluid to be expelled both forwards and backwards along the fore-aft axis, and drawn inwards radially towards this axis, creating an \textit{extensile} flow pattern (Fig.~\ref{fig:2}). In some cytoskeleton extracts (such as the actomyosin protein complex), motor proteins can pull the filaments amongst themselves, causing them to contract lengthwise and giving rise to a \textit{contractile} flow opposite to that of the previous example (Fig.~\ref{fig:2})\footnote{A more detailed description of the hydrodynamics of swimmers is given in~\cite{marchetti2013,ramaswamy2010,yeomans2017}.}. Typically, activity creates a flow pattern that can be complicated in the near field, but whose far field is generically equivalent, at the lowest order, to the action of a force dipole~\cite{pedley1992} and can be represented as such. By summing the contributions from each force dipole and coarse-graining~\cite{simha2002}, it is possible to show that the stress exerted by the active particles on the fluid has the form
\begin{equation} \label{stressnematic}
\sigma^{\textrm{active}}_{\alpha \beta} =- \zeta \phi Q_{\alpha \beta},
\end{equation}
where $\zeta$ is a phenomenological parameter that measures the activity strength, being negative for contractile systems and positive for extensile ones, while $\phi$ represents the concentration of the active material. Usually only terms linearly proportional to $\zeta$ are considered. In the case of polar active liquid crystals, the description can be carried out considering only the polarization field, re-expressing $\underline{\underline{\vec{Q}}}$  as a function of $\vec{P}$. The active stress in terms of the dynamical variable $\vec{P}(\vec{r},t)$ takes the form
\begin{equation}
\mathbf{\sigma}^{\textrm{active}}_{\alpha \beta}=- \zeta \phi \left( P_\alpha P_\beta - \frac{1}{d} |\vec{P}|^2  \delta_{\alpha \beta}\right)\ \ .
\label{stresspolar}
\end{equation}
The expressions Eq.~\eqref{stressnematic})  and Eq.~\eqref{stresspolar}, as we will see later, have been largely applied in the study of active fluids.

Many biological systems also  display a local chirality~\cite{Zhou2014,Naganathan2014}. Actin filaments, for example, are twisted in a right-handed direction~\cite{Depue1965} so that myosin motors tend also to rotate them while pulling, creating a torque dipole. A concentrated solution of DNA has long been known to exhibit a cholesteric or blue-phase in different salt conditions~\cite{livolant1986,livolant1991}. Such system can be made motile if interacting with DNA- or RNA-polymerases or with motor proteins. 
The effect of chirality, more than being taken into account by a suitable cholesteric term in the free energy\footnote{Chirality can be modelled~\cite{wright1989} introducing a suitable term $(\nabla \times \underline{\underline{\mathbf{Q}}} + 2 q_0 \underline{\underline{\mathbf{Q}}})^2$ in the free energy.
This contribution favors the formation of an helix in the director pattern, whose pitch $p_0=2 \pi/q_0$.}, can be  incorporated in the description adding to the active stress extra terms, providing a source of angular momentum. For instance, if the active particles act on the surrounding fluid with a net torque monopole, a coarse-graining procedure~\cite{furthauer2012} shows that a suitable choice for the nematic chiral stress tensor is given by $\zeta_2 \epsilon_{\alpha \mu}  Q_{\mu \beta}$~\cite{maitra2019}, where $\epsilon_{\alpha \mu}$ is the second order Levi-Civita tensor. Analogously, if the net torque is null but torque dipoles do not vanish, the corresponing stress tensor is given by $\zeta'_2 \epsilon_{\alpha\beta\mu}\partial_{\nu}\phi(P_{\mu}P_{\nu})$~\cite{Tjhung2017}, with $\epsilon_{\alpha \beta \mu}$ the third order Levi-Civita tensor. The sign of the second activity parameter $\zeta_2$ or $\zeta'_2$ determines whether the stress generates a flux parallel ($\zeta_2,\zeta'_2>0$) or antiparallel 	($\zeta_2,\zeta'_2<0$) with respect to the helicity of the twisting deformation.
These terms drive the system out of equilibrium by injecting energy into it, and, as those of Eq.~\eqref{stressnematic} and Eq.~\eqref{stresspolar}, cannot be derived from a free energy functional.
In this approach the active stress tensor enters the hydrodynamic equations governing the motion of the  self-propelled particles suspension, as discussed in Section~\ref{sec:2e}. These are constructed from general principles, by assuming that an active gel is described by (a) ``conserved'' variables, which are the fluctuations of the local concentration of suspended particles and the total (solute plus solvent) momentum density, and (b) ``broken-symmetry'' variables, which, in the nematic phase, is the deviation of the director field from the ground state.

A more  general way to construct the equations of motion at a coarse-grained level, is to generalize the forces-and-fluxes approach~\cite{Degroot2013} to active systems~\cite{kruse2004}.
Considering for example an active gel characterized by  polarization $\vec{P}$ and velocity $\vec{v}$, or equivalently by the strain rate tensor $u_{\alpha\beta}=(\partial_{\alpha}v_{\beta}+\partial_{\beta}v_{\alpha})/2$, the generalized hydrodynamic equations can be  derived using Onsager relations,  thus expanding fluxes $\partial_t \vec{P}$ and the stress tensor in terms of their conjugate forces $-\delta{F}/\delta\vec{P}$ and $u_{\alpha\beta}$ respectively, with $F$ polarization free energy.  Active dynamics is obtained holding the system out of equilibrium by introducing a further pair of conjugate variables, namely the chemical potential difference between ATP and hydrolysis products and the rate of ATP consumption~\cite{kruse2004}.
This approach can be further generalized~\cite{1742-5468-2017-5-054002} including thermal fluctuations, recasting the forces-and-fluxes approach  in the language of coupled generalized Langevin equations~\cite{mazenko2006nonequilibrium}.

Finally, we mention a more phenomenological model used to show self organization and scale selection for the flow pattern
in active matter. This approach is inspired by the use of the Brazovskii model~\cite{braz1975,Gonnella1998} for 
describing system with periodic order parameter and by 
dynamical approaches in  studies regarding the role of hydrodynamics~\cite{Swift1977}
in the onset of convection Rayleigh-B\'enard instability~\cite{chandrasekhar1981hydrodynamic}.
Higher order derivatives of the velocity gradients are considered in the stress tensor in addition to the usual dissipative terms:
\begin{equation}
\underline{\underline{\sigma}}=(\Gamma_2\nabla^2 + \Gamma_4 \nabla^4)\left[\nabla {\vec v}+(\nabla {\vec{v}})^T\right].
\label{eqn:swifthonenbergh_st}
\end{equation}
If $\Gamma_2$ is chosen to be negative, this corresponds to the injection of energy in a definite range of wavelengths, while $\Gamma_4>0$ corresponds to hyperviscosity flow damping. This is obtained by truncating a long-wavelength expansion of the stress tensor~\cite{bellout2013incompressible}.
The resulting generalized Navier-Stokes equations have been proven to capture experimentally observed bulk vortex dynamics of bacterial suspensions and some rheological properties of active matter~\cite{dunkel2013_2,dunkel2013,wensink2012}.  

\subsection{Fluid mixtures with an active component}
\label{sec:2d}

The active stress expressions of Eqs.~\eqref{stressnematic} and~\eqref{stresspolar} depend on the concentration of the active material.  This quantity  in turn can be a dynamical field if one would like to take into account a inhomogeneous presence of the active material in the solution. At level of particle description, different kinds of models for mixtures of self propelled and passive units have been considered. For example, Brownian-like simulations~\cite{mccandlish2012,grosberg2015,stenhammar2015}  focused on the role of activity in separating the two components of the mixtures. In a continuum description, binary fluids with an active component have been studied in~\cite{tjhung2011,tjhung2012,nature2015,blow2014} showing that the active part may cause instabilities on an active-passive interface.  Here we only introduce,  as an example among the different models that can be used to describe fluid mixtures with an active component, the free-energy for a binary mixture where the active component is a polar gel~\cite{tjhung2011}.
It is given by
\begin{eqnarray}\label{free-energy-mixture}
&F&[\phi,\vec{P}]
=\int d\vec{r}\,\{\frac{a}{4\phi_{cr}^4}\phi^{2}(\phi-\phi_0)^2+\frac{k}{2}\left|\nabla \phi\right|^{2}
\\
&-&\frac{\alpha}{2} \frac{(\phi-\phi_{cr})}{\phi_{cr}}\left|\vec{P}\right|^2+ \frac{\alpha}{4}\left|\vec{P}\right|^{4}+\frac{\kappa}{2}(\nabla\vec{P})^{2}
+\beta\vec{P}\cdot\nabla\phi\} \nonumber\ \ .
\end{eqnarray}

The first term, multiplied by the phenomenological constant $a > 0$, describes the bulk properties of the fluid; it is chosen in order to create two free-energy minima, one ($\phi = 0$) corresponding to the passive material and the other one ($\phi=\phi_0$) corresponding to the active phase. The second one determines the interfacial tension between the passive and active phase, with $k$ positive constant. The third and the fourth terms control the bulk properties of the polar liquid crystal. Here $\alpha$ is a positive constant and $\phi_{\textrm{cr}} = \phi_0/2$  is the critical concentration for the transition from isotropic $(|\vec{P}| = 0)$ to polar $(|\vec{P}| > 0)$ states. The choice of $\phi_{\textrm{cr}}$ is made to break the symmetry between the two phases and to confine the polarization field in the active phase $\phi > \phi_{\textrm{cr}}$. The  term proportional to $(\nabla{\vec{P}})^2$ describes the energetic cost due to elastic deformations in the liquid crystalline phase (see Table~\ref{tab:1}) in the single elastic constant approximation. Finally, the last term is a dynamic anchorage energy and takes into account the orientation of the polarization at the interface between the two phases. If $\beta\ne 0$, $\vec{P}$ preferentially points perpendicularly to the interface (normal anchoring): towards the passive (active) phase if $\beta>0$ ($\beta<0$).
This  choice for the anchoring is suggested by experimental observations. For instance,   bacterial orientation at  water-oil interfaces results from a relatively hydrophobic portion of each cell being rejected from the aqueous phase of the system~\cite{Marshall1973}.

Such model can be also extended to study active nematic gels, by using the nematic tensor in place of the polarization field~\cite{blow2014,giomi2013,giomi2014,doostmohammadi2018}. In this case the coefficients of the expansion of $Tr(\vec{\underline{\underline{Q}}}^n)$ in bulk free energy (see Table 1) would depend on the scalar field $\phi$ and the elasticity, again written in the single elastic constant approximation, would include a term of the form $L\partial_{\alpha}\phi Q_{\alpha\beta}\partial_{\beta}\phi$ (with $L$ constant) to guarantee a perpendicular anchoring of the liquid crystal at the interface.

We finally mention a recent generalization of such models where emulsification of the active component is favored by the presence of surfactant added to the mixture~\cite{bone2017}. This is done by allowing negative values of the binary fluid elastic constant $k$ and by including a term of the form $\frac{c}{2}(\nabla^2\phi)^2$ (with $c$ positive constant) to guarantee the stability of the free-energy.


A different continuum model, specifically introduced to study the motility induced phase separation (MIPS) without direct appeal to orientational order parameters $\vec{P}$ or $\underline{\underline{\vec{Q}}}$, but only to the scalar concentration field $\phi$, is the so called Active-model H~\cite{tiribocchi2015}.  In the old classification by Hohenberg and Halperin~\cite{hohenberg1977}, the  passive model H considers a diffusing, conserved, phase separating order parameter $\phi$ coupled to an isothermal and incompressible fluid flow through the advection-diffusion equation that will be introduced in Section~\ref{sec:2e}. The chemical potential that enters the dynamic equation of the passive model H is given by
\begin{equation}
 \mu=\frac{\delta F}{\delta \phi}= a\phi + b\phi^3 - k \nabla ^2 \phi\ ,
\end{equation} 
with $a$, $b$, $k$ constants appearing in the Landau free energy for binary mixtures~\cite{bray1994} (with $a$ negative in order to have phase separation between the two fluid components and $b$ and $k$ positive for stability). The same terms appear in Eq.~\eqref{free-energy-mixture} without the polarization contributions. The active model is then constructed  by adding  a leading order time-reversal breaking active  term of the form $\mu_a=\lambda_a \left(\nabla \phi \right)^2$ (with $\lambda_a$ constant), not stemming from the free energy functional~\cite{tiribocchi2015}. The  deviatoric stress $\underline{\underline{\sigma}}$, that enters in the NS equations for the fluid flow, is, in $d$ dimensions,
\begin{equation}\label{dev_stress}
\sigma_{\alpha\beta}^a=-\hat\zeta \left(\partial_{\alpha} \phi \partial_{\beta} \phi - \frac{1}{d}\left(\nabla \phi\right)^2\delta_{\alpha\beta}\right),
\end{equation}
and can be obtained from the free energy, according to the formula reported in the second row of Table~\ref{terms}, only if $\hat\zeta=k$. If $\hat\zeta\neq k$ this is not true anymore and Eq.~\eqref{dev_stress} is the sole leading-order contribution to the deviatoric stress for scalar active matter. Again here, $\zeta<0$ describes contractile systems while $\zeta>0$ the extensile ones. While $\mu_a$ has been found to create a jump in the thermodynamic pressure across interfaces and to alter the static phase diagram~\cite{wittkowski2014}, the active stress $\underline{\underline{\sigma}}^a$ creates a negative interfacial tension in contractile systems that arrests the coarsening~\cite{tiribocchi2015}. 

\begin{table*}
\centering
\caption{Explicit expressions of the elastic (first row) and the interface (second row) stress, and of
the term $\vec{S}$ in the Beris-Edwards equation \eqref{eq:adv_diff} (fourth row) for polar and nematic gels.
The molecular field $\mathbf{\Xi}$ is a vector, with components $h_{\alpha}$,  
for polar gels and a tensor $H_{\alpha \beta}$, for nematic gels, as shown in the third row. $\kappa$ is the elastic constant of the liquid crystal; the flow-alignment parameters $\xi$ and $\xi'$ are respectively related to the polarization field $\vec{P}$ and to the nematic tensor $\underline{\underline{\vec Q}}$ and depend on the geometry of the microscopic constituents (for instance $\xi>0$, $\xi<0$ and $\xi=0$ for rod-like, disk-like and spherical particles, respectively). In addition, these parameters establish whether the fluid is flow aligning ($|\xi|>1$) or flow tumbling ($|\xi|<1$) under shear.
 $\underline{\underline{D}}=(\underline{\underline{W}}+\underline{\underline{W}}^T)/2$ and $\underline{\underline{\Omega}}=(\underline{\underline{W}}-\underline{\underline{W}}^T)/2$ represent the symmetric and the antisymmetric parts of the velocity gradient tensor $W_{\alpha\beta}=\partial_{\beta}v_{\alpha}$. }
\label{terms}
\resizebox{\textwidth}{!}{
\begin{tabular}{l|l|l}
\toprule
                                   & Polar Gel                                                                                                                                                                 & Nematic Gel                                                                                                                                                                                                                                                                                                                                                                                                                                  \\
\midrule
\rule{0pt}{6ex}
$\sigma_{\alpha\beta}^{elastic}$   & $\frac{1}{2}(P_{\alpha}h_{\beta}-P_{\beta}h_{\alpha})-\frac{\xi}{2}(P_{\alpha}h_{\beta}+P_{\beta}h_{\alpha})-\kappa\partial_{\alpha}P_{\gamma}\partial_{\beta}P_{\gamma}$ & \begin{tabular}[c]{@{}l@{}} $2\xi'\left(Q_{\alpha \beta}-\frac{\delta_{\alpha \beta}}{3}\right)Q_{\gamma \nu}H_{\gamma \nu}-\xi' H_{\alpha \gamma}\left(Q_{\gamma \beta}+\frac{\delta_{\gamma \beta}}{3}\right)$\\$-\xi'\left(Q_{\alpha \gamma}+\frac{\delta_{\alpha\gamma}}{3}\right)H_{\gamma\beta}-\partial_{\alpha}Q_{\gamma\nu}\frac{\delta\mathcal{F}}{\delta\partial_{\beta}Q_{\nu\gamma}}+Q_{\alpha\gamma}H_{\gamma\beta}-H_{\alpha\gamma}Q_{\gamma\beta}$\end{tabular} \\
\rule{0pt}{5ex}

$\sigma_{\alpha\beta}^{interface}$ & $\left( f-\phi\frac{\delta F}{\delta\phi} \right)\delta_{\alpha\beta} - \frac{\partial f}{\partial\left(\partial_{\beta}\phi\right)} \partial_{\alpha}\phi$               &  $\left( f-\phi\frac{\delta F}{\delta\phi} \right)\delta_{\alpha\beta} - \frac{\partial f}{\partial\left(\partial_{\beta}\phi\right)} \partial_{\alpha}\phi$                                                                                                                                                                                                                                                                                                                                                                                                                                             \\
\rule{0pt}{5ex}

$\mathbf{\Xi}$ & $  h_{\alpha} = \dfrac{\delta F}{\delta P_\alpha} $ & $  H_{\alpha \beta} = \dfrac{\delta F}{\delta Q_{\alpha \beta}} - \left( \dfrac{\delta F}{\delta Q_{\gamma \gamma}} \right) \delta_{\alpha \beta}$                                                                                                                                                                                                                                                                                                                                                                                                                                             \\
\rule{0pt}{5ex}

$\vec{S}$ & $ -\Omega_{\alpha \beta} P_{\beta} +\xi D_{\alpha \beta} P_{\beta}$                                                                              & \begin{tabular}[c]{@{}l@{}} $ [\xi' D_{\alpha \gamma} +\Omega_{\alpha \gamma}]\left(Q_{\gamma \beta}+\frac{\delta_{\gamma \beta}}{3}\right)+\left(Q_{\alpha \gamma}+\frac{\delta_{\alpha \gamma}}{3}\right)[\xi' D_{\gamma \beta} -\Omega_{\gamma \beta}]$\\$-2\xi'\left(Q_{\alpha \beta}+\frac{\delta_{\alpha \beta}}{3}\right)\left(Q_{\gamma \delta} \ \partial_{\gamma} v_{\delta} \right)$ \end{tabular}    \\

\bottomrule
\end{tabular}
}
\end{table*}

\subsection{Hydrodynamic equations}
\label{sec:2e}
We can now introduce the hydrodynamic equations for active liquid crystals. Evolution equations for mass density $\rho (\vec{r},t)$ and velocity $\vec{v}(\vec{r},t)$ are given by
\begin{eqnarray}
\partial_t \rho + \nabla \cdot ( \rho \vec{v}) & = & 0, \label{eqn:cont_eq}\\
\rho\left(\partial_t+\vec{v}\cdot\nabla\right)\vec{v} & = & -\nabla p + \nabla\cdot\underline{\underline{\sigma}}, \label{nav}
\end{eqnarray}
with the energy balance equation generally neglected in this context.
Eq.~\eqref{eqn:cont_eq} is the continuity equation for mass density. 
In most of active matter systems  Mach numbers $Ma$, defined as the ratio of the stream velocity and the speed of sound, is small; in such limit,
this equation reduces to the solenoidal condition for the velocity field
\begin{equation}
\nabla \cdot \vec{v}=0 + \mathcal{O}(Ma^2),
\label{cont}
\end{equation}
so that the fluid in this regime can be assumed at all practical effects as incompressible.
Eq.~\eqref{nav} is the Navier-Stokes equation, where $p$ is the ideal fluid pressure
 and $\underline{\underline{\sigma}}$ is the stress tensor~\cite{beris1994} that can be split into the equilibrium/passive and non-equilibrium/active contributions:
\begin{equation}
\underline{\underline{\sigma}}=\underline{\underline{\sigma}}^{\textit{passive}}+\underline{\underline{\sigma}}^{\textit{active}}.
\end{equation}
The passive part is, in turn, the sum of three terms:
\begin{equation}
\underline{\underline{\sigma}}^{\textit{passive}}=\underline{\underline{\sigma}}^{\textit{viscous}}+\underline{\underline{\sigma}}^{\textit{elastic}}+\underline{\underline{\sigma}}^{\textit{interface}} \mbox{.}
\end{equation}
The first term is the viscous stress, written as  $\sigma_{\alpha\beta}^{viscous}=\eta(\partial_{\alpha}v_{\beta}+\partial_{\beta}v_{\alpha})$, where $\eta$ is the shear viscosity\footnote{In the compressible case, the viscous stress tensor also includes a term proportional to the divergence of the velocity, such that:
\begin{equation}
\sigma_{\alpha\beta}^{viscous}=\eta(\partial_{\alpha}v_{\beta}+\partial_{\beta}v_{\alpha}) + \left( \tilde{\zeta} - \frac{2 \eta}{d} \right) \partial_\gamma v_\gamma \delta_{\alpha \beta},
\label{eqn:viscous_stress_tensor_incompressible}
\end{equation}
where we denoted the bulk viscosity with $\tilde{\zeta}$.
}. An explicit form for the elastic and interface stress is reported for the polar and nematic cases in Table~\ref{terms}.

The order parameter $\mathbf{\Psi}$ of the active liquid crystal  (that is $\underline{\underline{\vec{Q}}}$ for nematics and $\vec{P}$ for polar systems) evolves  according to
\begin{equation}\label{eq:adv_diff}
 \left(\partial_t+\vec{v}\cdot\nabla\right) \mathbf{\Psi} - \vec{S}= -\Gamma \mathbf{\Xi}\ ,
\end{equation}
known as \emph{Beris-Edwards equation}, within the theory of liquid crystal hydrodynamics described through the $\vec{\underline{\underline{Q}}}$-tensor. The term $\vec{S}$ accounts for the response of the orientational order to the extensional and rotational components of the velocity gradient and is reported for the polar~\cite{stark2003,marchetti2006} and nematic~\cite{degennes1993} case in the fourth row of Table~\ref{terms}.
The molecular field $\mathbf{\Xi}$ governs the relaxation of the orientational order to equilibrium, and is multiplied by a collective rotational-diffusion constant $\Gamma$. Its expressions are given in the third row of Table~\ref{terms}. The left-hand side of Eq.~\eqref{eq:adv_diff} is commonly addressed as material derivative of the order parameter $\mathbf{\Psi}$, and can be formally derived making use of Liouville equations. In fact one can write $D_t\mathbf{\Psi}=\partial_t{\mathbf{\Psi}}+ \lbrace \mathbf{\Psi}, \mathcal{H}\rbrace$, where $\lbrace...\rbrace$ are the Poisson brackets and the Hamiltonian is $\mathcal{H} = F + \frac{1}{2} \int \rho \vec{v}^2 $.

A more phenomenological procedure to derive the material derivative explicitly is based on the fact that order parameters can be advected by the fluid. Here we outline the procedure referring only to the polarisation field. We first note that the relative position $\vec{\tilde{r}}$ of two close points in the fluid evolves according to the following equation:
\begin{equation}
D_t \vec{\tilde{r}} = \partial_t \vec{\tilde{r}} + (\vec{v} \cdot \nabla) \vec{\tilde{r}} + \underline{\underline{D}} \cdot \vec{\tilde{r}} + \underline{\underline{\Omega}}\cdot \vec{\tilde{r}},
\label{eqn:material_derivative_reconstruction}
\end{equation}
where $\underline{\underline{D}}$ and $\underline{\underline{\Omega}}$ have been defined in Table~\ref{terms}.
The first two contributions are the usual lagrangian derivative terms, while the third and fourth ones account respectively for rigid rotations and deformations of the fluid element. Thus the material derivative for the polarisation field will include the first three terms since a vector advected by the flow is capable to follow any rigid motion; for what concerns the last term in Eq.~\eqref{eqn:material_derivative_reconstruction}, this cannot enter directly into the material derivative of a vector field, but it must be weighted through an alignment parameter $\xi$, ruling the dynamical behavior of the vector field under enlargement and/or tightening of flow tubes. This allows us to obtain the material derivative for the polarization field simply substituting $\vec{P}$ in place of $\vec{\tilde{r}}$.	

Finally the time evolution of the concentration field $\phi(\vec{r},t)$ of the active material is governed by an advection-diffusion equation
\begin{equation}
\partial_t \phi+\nabla\cdot\left(\phi\vec{v}\right)=\nabla \cdot \left( M\nabla\frac{\delta F}{\delta \phi}\right),\label{conc_eq}
\end{equation}
where $M$ is the mobility and $\delta F/\delta\phi$ is the chemical potential. A more generalized form of the material derivative has been used to model self advective phenomena, for example, actin polymerization in motile eukaryotic cells~\cite{tjhung2012}, by substituting $\nabla\cdot(\phi\vec{v})\rightarrow\nabla\cdot(\phi\vec{v}+w\vec{P})$, where $w$ is a constant related to the velocity of actin polymerization.


\section{Lattice Boltzmann Method}
	\label{sec:LBM}
A certain number of approaches are feasible when dealing with the description
of fluid systems; each of them can be classified according to the level
of spatial approximation.
A molecular approach would hardly access the time and space scales relevant for
a complete hydrodynamic description of the systems here considered.
At a mesoscopic level, kinetic theory furnishes a description of irreversible
and non-equilibrium thermodynamic phenomena in terms of a set of distribution functions encoding
all necessary informations related to space positions and velocities of particles.
Continuum equations give a description of irreversible phenomena by using macroscopic
variables slowly varying in time and space. This last approach has the not-negligible advantage that
one has to deal with a few fields. 
On the other hand, when considering continuous equations, one has to face some technical
issues arising from the stability of numerical implementation and discretization schemes~\cite{LeVeque2007}. 
Moreover, many numerical methods aimed at solving the continuous equations, exhibit
criticalities in the amount of computational resources, mostly in terms of processing times and memory requirement, or in the implementation of boundary conditions in complex geometries.
To avoid these issues lattice-gas-automaton (LGA) models were
first developed starting from the pioneering work of Frisch 
\emph{et al.}~\cite{Frisch1986}. This kinematic approach to
hydrodynamics is based on the description of the dynamics of a 
number of particles moving on a suitable lattice. 
An exclusion principle is imposed 
to restrict the number of particles with a given velocity
at a certain lattice point to be $0$ or $1$. This latter feature
allows for a description of the local particle equilibrium
through the Fermi-Dirac statistics~\cite{alexander1992}. Despite LGA proved to
be very efficient  in simulating the Navier-Stokes equation
from a computational point of view and
in managing boundary conditions, LGA simulations are intrinsically noisy due to large fluctuations of local density. Moreover, they 
suffer from non-Galilean invariance, due to density dependence of the convection 
coefficient and from an unphysical velocity dependence of the
pressure, arising directly from the discretization procedure~\cite{Chen1992}.

Lattice Boltzmann methods were then developed to overcome these 
difficulties~\cite{higuera1989}. Particles in the LGA approach are formally substituted
by a discretized set of distribution functions,
so that hydrodynamic variables are indeed expressed at each lattice point 
in terms of such distribution functions.
Despite the fact that lattice Boltzmann is a mesoscopic numerical method, it has a number of advantages that
resulted in
a broad usage 
in many branches of hydrodynamics. 
Firstly LB algorithms are appreciably stable and they are
characterized by their simplicity in the treatment of boundary 
conditions. Not to be neglected is the fact that LB algorithms
are particularly suitable to parallel approach.

In the following of this Section we will first provide a simple  overview of the method, 
without getting too technical,
in order to convey to the reader the purpose of this approach.
In Section~\ref{sec:general_features_LBM} we will first
introduce LBM for a simple fluid, while Section~\ref{sec:LBM_sf} will be devoted to
recover the continuum hydrodynamic equations, already presented in Section~\ref{sec:2e}.
In Section~\ref{sec:forcing_in_LBM} we will describe some routes
to adapt LBM to the case of complex fluids 
by introducing either a forcing term or properly fixing the second moment of the equilibrium distribution functions. In Section~\ref{sec:lbm4advection} different approaches to deal with  the advection-relaxation equations for  order parameters coupled to the momentum  equations will be examined.
In Section~\ref{sec:LBM4activefluids}
we will focus on some algorithms that have  been recently
used in the numerical investigation of active matter.
Finally, in Section~\ref{sec:performance} we will focus on the computational performance and stability of the method.

\subsection{General features of lattice Boltzmann method}
\label{sec:general_features_LBM}

\begin{figure}
\center
\resizebox{0.95\columnwidth}{!}{\includegraphics{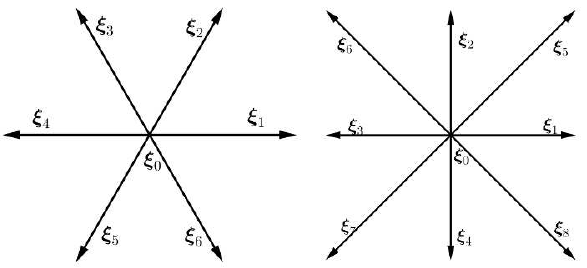}}
\caption{Graphical representation of lattice velocities for the triangular $d2Q7$ and face centered squared $d2Q9$ lattices, respectively shown in the left and right panels. Cartesian components of lattice vectors $\vec{\xi}_i$ are found in Table~\ref{tab:Lattice2D3D}.}
\label{img:lattices}
\end{figure}

The lattice Boltzmann approach to hydrodynamics is based on a phase-space discretized
form of the Boltzmann equation~\cite{Benzi1992, Rothman1997, Chopard1998, Wolf-Gladrow2000, succi2001}
for the distribution function $f(\vec{r}, \vec{\xi}, t)$, describing the
fraction of fluid mass at position $\vec{r}$ moving with velocity $\vec{\xi}$ at  time $t$.
Since space and velocities are discretized, the algorithm is expressed in terms of a set of discretized
distribution functions $\lbrace f_i (\vec{r}_\alpha,t)\rbrace$, defined on each lattice site $\vec{r}_\alpha$ and related to a discrete set of
$N$ lattice speeds $\lbrace \vec{\xi}_i \rbrace$, labelled with an index $i$ that varies from $1$ to $N$ (see Fig.~\ref{img:lattices}).
In the case of the \emph{collide and stream} version of the algorithm,
the evolution equation for the distribution functions has the form
\begin{equation}
\label{eqn:evolution_equation_LBM}
f_i(\vec{r}+\vec{\xi}_i\Delta t, t+\Delta t) - f_i(\vec{r}, t)  = \mathcal{C}(\lbrace f_i \rbrace, t),
\end{equation}
where $\mathcal{C}(\lbrace f_i \rbrace, t)$ is the collisional operator that drives the system towards equilibrium, represented 
by  a set of equilibrium distribution functions,  and depends on
the distribution functions; its explicit form will depend upon the particular implementation
of the method. 
Eq.~\eqref{eqn:evolution_equation_LBM} describes how
fluid particles collide in the lattice nodes and 
move afterward  along the lattice links in the time step $\Delta t$
towards neighboring sites at distance $\Delta x = \xi_i \Delta t$. This latter relationship
is no more considered in finite difference lattice Boltzmann models  (FDLBM)~\cite{Cao1997,Mei1998,  lee2001, Sofonea2003, Sofonea2004, Cristea2006, Cristea2010}.
In this kind of models the discrete velocity set can be chosen with more freedom,
making possible to use non uniform grids, selecting lattice velocities independently from the lattice 
structure\footnote{When dealing with FDLBM it is useful to introduce more than only one set of distribution functions $\lbrace f_{ki} \rbrace$, where the extra index $k$ labels different sets of discrete velocities $\lbrace \vec{\xi}_{ki} \rbrace$, with index $i$ still denoting the streaming direction.
The evolution equation for distribution functions for the FDLBM reads:
\begin{equation}
\partial_t f_{ki} + (\vec{\xi}_{ki} \cdot \nabla) f_{ki} = \mathcal{C}(\lbrace f_{ki} \rbrace, t).
\end{equation}
Here differential operators must be discretized: Runge-Kutta or midpoint schemes can be used to compute the time derivative while there are several possibilities to compute the advective term on the left-hand side of the previous equation. 
For the reader interested in details of the implementation we suggest to refer to~\cite{Sofonea2003,watari2003}.}.
This result is found to be extremely useful when it is necessary to release the constraint of having
a constant temperature in the system~\cite{watari2003, gonnella2007}. Moreover it might
be also helpful in the case of LB models for multicomponent systems where
the components have different masses and this would result in having different lattice speeds, one for each fluid species.
Beside the wider range of applicability of the FDLBM with respect to the LBM, the latter furnishes a simple and efficient way to solve hydrodynamic equations; in addition we are not aware of any implementation of the FDLBM algorithm developed
to study active matter; for this reasons we will avoid any further discussion on this variant of LBM.

In the case of a simple fluid, in absence of any external force,
assuming the BGK approximation with a single relaxation time~\cite{bhatnagar1954}, one writes
\begin{equation}
\label{eqn:BGK_simple_fluid}
\mathcal{C}(\lbrace f_i \rbrace, t) = - \dfrac{1}{\tau} (f_i-f_i^{eq}),
\end{equation}
where $f_i^{eq}$ are the equilibrium distribution functions and $\tau$ is the relaxation time,
connected to the viscosity of the fluid, as it will be seen.
The mass and momentum density  are  defined as
\begin{align}
\rho(\vec{r}, t)  &= \sum_i f_i(\vec{r}, t), \label{eqn:mass_constraint} \\
\rho(\vec{r}, t)  \vec{v}(\vec{r}, t)  &= \sum_i f_i (\vec{r}, t) \vec{\xi}_i, \label{eqn:momentum_constraint}
\end{align}
where summations are performed over all discretized directions at each lattice point.
By assuming both mass and momentum density to be conserved in each collision,
it is found that conditions in Eq.~\eqref{eqn:mass_constraint},~\eqref{eqn:momentum_constraint} must hold also
for the equilibrium distribution functions:
\begin{align}
\rho(\vec{r}, t)  &= \sum_i f_i^{eq}(\vec{r}, t) \label{eqn:mass_constraint_eq}, \\
\rho(\vec{r}, t)  \vec{v}(\vec{r}, t)  &= \sum_i f_i ^{eq}(\vec{r}, t) \vec{\xi}_i. \label{eqn:momentum_constraint_eq}
\end{align}
Moreover, it is necessary to introduce further constraints on the second 
moment of the equilibrium distribution functions to
recover continuum equations,
as it will become  more evident in the following.
Further constraints on higher order moments may become necessary to simulate
more complex systems: for instance full compressible flows or supersonic adaptation of the algorithm may require the specification of moments up to the third, while for a complete hydrodynamic description
in which heat transfer is also taken into account, even the fourth moment
needs to be specified~\cite{watari2003}.
Active matter systems such as bacterial and microtubules suspensions reasonably fulfil the incompressible condition, so that in the following we will only impose constraints up to second order moments.

Another peculiar fact is that viscosity explicitly depends on the choice of a particular lattice.
Due to the fact that sufficient lattice symmetry is required
to recover the correct Navier-Stokes equation in the continuum limit~\cite{Frisch1986}, not all the possible lattice structures can be
adopted.
By denoting the space dimension by $d$ and
the number of lattice speeds by $Q$, 
Table~\ref{tab:Lattice2D3D} shows the velocities $\lbrace \vec{\xi}_i \rbrace$
and the corresponding weights in the equilibrium distribution functions 
(see next Section) for the most frequent choices. 
Here the quantity $c=\Delta x/ \Delta t$, 
connected to the speed of sound of the algorithm,
has been introduced as the ratio  between the lattice spacing $\Delta x$ 
and the time step $\Delta t$.
Figure~\ref{img:lattices} explicitly illustrates the lattice structures
in the two-dimensional case.

\begin{table}
\centering
\caption{Lattice speeds with their weights $\omega_i$ for spatial dimensions
$d=2$ and $d=3$ and number of neighboring nodes $Q$. }
\label{tab:Lattice2D3D}
\resizebox{\columnwidth}{!}{
\begin{tabular}{lcc}
\toprule
Lattice & $\vec{\xi}_i$  & $\omega_i$\\ \toprule
$d2Q7$        &   $(0,0) $         							 		         &   $1/2$      \\
                 &   $c (\cos(i\pi/3), \sin(i\pi/3)) $  	         &   $1/12$     \\ \midrule
$d2Q9$        &   $(0,0) $         							 		         &   $4/9$      \\
                 &   $(\pm c, 0) \, \ (0,\pm c)  $  			     &   $1/9$     \\
                 &   $(\pm c, \pm c)$                 			         &   $1/36$     \\  \bottomrule \toprule
$d3Q15$       &   $(0,0,0) $         							 			  &   $2/9$      \\
                 &   $(\pm c, 0,0) \, \ (0,\pm c,0) \, \ (0,0,\pm c)  $   &   $1/9$       \\
                 &   $(\pm c,\pm c,\pm c)  $  		 	                          &   $1/72$      \\ \midrule
$d3Q19$       &   $(0,0,0) $         							 						    &   $1/3$      \\
                 &   $(\pm c, 0,0) \, \ (0,\pm c,0) \, \ (0,0,\pm c)  $   &   $1/18$       \\
                 &   $(\pm c, \pm c ,0) \, \ (\pm c,0, \pm c) \, \ (0,\pm c,\pm c)  $  	&   $1/36$      \\ \midrule
$d3Q27$      &   $(0,0,0) $         				                       &   $8/27$     \\
                 &   $(\pm c, 0,0) \, \ (0,\pm c,0) \, \ (0,0,\pm c)  $  &   $2/27$      \\
                 &   $(\pm c, \pm c ,0) \, \ (\pm c,0,\pm c) \, \ (0,\pm c,\pm c)  $   &   $1/54$      \\
                 &   $(\pm c, \pm c ,\pm c) $  &   $1/216$      \\ \bottomrule
\end{tabular}
}
\end{table}

\subsection{Lattice Boltzmann for a simple fluid}
\label{sec:LBM_sf}

In this Section we will present a basic lattice Boltzmann algorithm to solve the
hydrodynamic equations~\eqref{eqn:cont_eq} and~\eqref{nav} for a simple fluid; 
in this case the term on the right hand side of the Navier-Stokes equation~\eqref{nav} reduces to the pressure gradient plus the mere
viscous contribution $\partial_\beta \sigma_{\alpha \beta}^{viscous}$, if
no external force is acting on the fluid.

Conditions~\eqref{eqn:mass_constraint_eq} and~\eqref{eqn:momentum_constraint_eq} can be
satisfied by expanding the equilibrium distribution functions up to the second order in the fluid velocity $\vec{v}$~\cite{Chen1992}:
\begin{equation}
f_i^{eq}= A_s + B_{s} v_{\alpha} \xi_{i \alpha} + C_s v^2 + D_s v_{\alpha} v_{\beta} \xi_{i \alpha} \xi_{i \beta},
\label{eqn:expansion_eq_df_simple_fuid1}
\end{equation}
where index $s=|\vec{\xi}_i |^2/c^2$ relates the $i$-th distribution function to the square module
of the corresponding lattice velocity, and the greek index denotes the Cartesian component.
This expansion is valid
as far as the Mach number $Ma=v/c_s$ is kept small, $c_s$ being the speed of sound, whose explicit expression in turn depends upon the lattice discretization~\cite{shan2006}. The present assumption has the important consequence that LB models based on the previous expansion of the equilibrium distribution functions have great difficulty in simulating compressible Euler flows, that usually take place at high Mach numbers. This issue arises in standard LB approaches because of the appearence of third order nonlinear deviations from the Navier-Stokes equation~\cite{guangwu1998}. Qian and Orzsag demonstrated in~\cite{qian1993} that such nonlinear deviations grow together with $Ma^2$, so that they can be neglected in the low Mach number regime but become important in the compressible limit\footnote{In order to overcome the limit posed by the low Mach number regime, many variations of the standard LBM have been developed. Alexander \emph{et al.} proposed a model where the high Mach number regime could be achieved by decreasing the speed of sound~\cite{alexander1992}; discrete-velocity models~\cite{nadiga1995,sun1998} were later introduced allowing for simulation of the compressible Euler equation in a wider range of Mach numbers. Other implementations are based on a Taylor expansion of the equilibrium distributions up to higher orders together with suitable constraints on the third and fourth moments~\cite{shan2006,kataoka2004,kataoka2004_2}. }. 
For such reasons it is necessary to ensure that velocities never exceed a critical threshold that can be reasonably chosen such that $Ma \lesssim 0.3$~\cite{qian1993}.

Besides constraints expressed by Eq.~\eqref{eqn:mass_constraint_eq} and~\eqref{eqn:momentum_constraint_eq}, an additional condition
on the second moment of the equilibrium distribution
functions is imposed so that
\begin{equation}
\sum_i f_i^{eq} \xi_{i \alpha} \xi_{i \beta} = \dfrac{c^2}{3} \rho \delta_{\alpha \beta} + \rho v_{\alpha} v_{\beta} .
\label{eqn:II_moment_condition}
\end{equation}
This is a necessary condition to recover the Navier-Stokes equation in the continuum limit.
By substituting the expansion in Eq.~\eqref{eqn:expansion_eq_df_simple_fuid1} in constraints introduced in Eq.~\eqref{eqn:mass_constraint_eq}, \eqref{eqn:momentum_constraint_eq} and~\eqref{eqn:II_moment_condition}, a suitable choice for the expansion coefficients is found to be
\begin{eqnarray}
A_0= \rho-20A_2                  \quad  &A_1=4A_2  \quad  &A_2=\dfrac{\rho}{36}  \label{eqn:Icoeff'}\\
B_0=0                                    \quad  &B_1=4B_2   \quad  &B_2=\dfrac{\rho}{12 c^2} \label{eqn:Icoeff}\\
C_0= -\dfrac{2\rho}{3c^2}  \quad  &C_1=4C_2  \quad  &C_2=-\dfrac{\rho}{24 c^2} \\
D_0=0                                   \quad  &D_1=4D_2     \quad  &D_2=\dfrac{\rho}{8 c^4}, \label{eqn:Icoeff''}
\label{eqn:coefficient_simple_fluid}
\end{eqnarray}
where for the sake of clarity we have explicitly chosen a $d2Q9$ lattice geometry.
Requiring suitable isotropy conditions and Galilean invariance~\cite{Chen2008}, it is even possible to show analytically~\cite{Qian1992}, 
that the equilibrium distribution functions can be written in a more general way as
\begin{equation}
f_i^{eq}= \rho \omega_i \left[ 1 + 3 \frac{v_{\alpha} \xi_{i \alpha}}{c^2} - \frac{3}{2}\frac{v^2}{c^2} + 
\frac{9}{2} \frac{(v_{\alpha} \xi_{i \alpha})^2}{c^4} \right],
\label{eqn:expansion_eq_df_simple_fuid}
\end{equation}
where the weights $\omega_i$ are given in Table~\ref{tab:Lattice2D3D}. In Appendix~\ref{sec:appB} it will be shown that the algorithm here presented correctly reproduces Eqs.~\eqref{eqn:cont_eq} and \eqref{nav}.

We add for completeness that it is also possible 
to adopt a discretization in velocity space based on the quadrature of a 
Hermite polynomial
expansion of the Maxwell-Boltzmann distribution~\cite{shan2006}. One then gets a lattice Boltzmann equation that allows us to exactly recover a finite number of leading order moments of the equilibrium distribution functions. 
In this case the quantity $c$ is fixed and given by $c=2$ for the geometry $d2Q7$
and by $c=\sqrt{3}$ for the other geometries in Table~\ref{tab:Lattice2D3D}. 
For a detailed discussion the interested reader may refer to Ref.~\cite{shan2006}.
Finally, we mention that
it would be possible to  introduce small  thermal fluctuations 
into  the  algorithm, in a controlled way, by means of a
stochastic collision operator.
The fluctuation-dissipation theorem can then be satisfied by 
requiring consistency with fluctuating hydrodynamics \cite{Adhikari2005}. Since to the best of our knowledge 
there are no LB models for active systems including thermal noise, 
we do not give further details referring the interested reader to the Ref.~\cite{Dunweg2009}.

\subsection{LBM beyond simple fluids}
\label{sec:forcing_in_LBM}
So far we have implemented a lattice Boltzmann method for a simple fluid in  absence of any forcing term, with only viscous contribution to the stress tensor. On the other hand when dealing with more complex systems, such as multicomponents or multiphase fluids, the stress tensor may include further contributions (such as elastic and interfacial ones, see Table~\ref{terms}) which have a non-trivial dependence on order parameters and their derivatives. In this Section we will show two different strategies adopted to numerically implement such terms. Briefly, while in the first one they are included in an extra term, appearing in the second moment of the equilibrium distribution functions, in the second one they enter through an external forcing added to the collision operator in the lattice Boltzmann equation.

\subsubsection{First method}
\label{sec:additional_constraint}
To implement a general symmetric stress tensor contribution in the lattice Boltzmann scheme previously introduced, we again impose the constraints of Eq.~\eqref{eqn:mass_constraint_eq} and Eq. \eqref{eqn:momentum_constraint_eq} on the zeroth and on the first moment of the equilibrium distribution functions, while constraint on the second moment previously given in Eq.~\eqref{eqn:II_moment_condition} is modified according to the following relation
\begin{equation}
\sum_i f_i^{eq} \xi_{i \alpha} \xi_{i \beta} = - \sigma_{\alpha \beta} + \rho v_{\alpha} v_{\beta}.
\label{eqn:general_stress_II_moment}
\end{equation}
Here $\sigma_{\alpha \beta}$ stands for the total stress tensor including preassure contributions, but deprived of viscous ones. Note that, due to the symmetry of the left hand side of Eq.~\eqref{eqn:general_stress_II_moment}, this algorithm can be applied to models that involve only symmetric contributions to the stress tensor. For instance, this method is suitable to study binary mixtures, as the stress tensor associated to the concentration contribution is indeed symmetric, but not  liquid crystals, as the antisymmetric part of the relative stress tensor does not vanish (see Table~\ref{terms}). This latter case will be discussed in the following Sections. To satisfy Eq.~\eqref{eqn:general_stress_II_moment}~\cite{denniston2001, orlandini2008}, the equilibrium distribution functions can be expanded as follows
\begin{equation}
\begin{split}
f_i^{eq}= A_s + B_s v_{\alpha} \xi_{i \alpha} + C_s v^2 \qquad \qquad \qquad \\+ D_s v_{\alpha} v_{\beta} \xi_{i\alpha} \xi_{i\beta} + G^{\alpha\beta}_s\xi_{i \alpha} \xi_{i \beta},
\end{split}
\label{eqn:expansion_eq_df}
\end{equation}
where an extra term, quadratic in lattice velocities, has been added with respect to the case of a simple fluid (see Eq.~\eqref{eqn:expansion_eq_df_simple_fuid1}), to include a general stress tensor in the model. As for a simple fluid, the coefficients of the expansion can be calculated by imposing constraints of Eq.~\eqref{eqn:mass_constraint_eq}, Eq.~\eqref{eqn:momentum_constraint_eq} and Eq.~\eqref{eqn:general_stress_II_moment}. For a $d2Q9$ geometry a suitable choice is given by
\begin{equation}
\begin{split}
A_0= \rho-20A_2                  \quad  A_1&=4A_2  \quad  A_2=\dfrac{\Tr{\sigma}}{24c^2} \\
B_0=0                                    \quad  B_1&=4B_2   \quad  B_2=\dfrac{\rho}{12 c^2}\\
C_0= -\dfrac{2\rho}{3c^2}  \quad  C_1&=4C_2 \quad  C_2=-\dfrac{\rho}{24 c^2} \\
D_0=0                                    \quad  D_1&=4D_2    \quad  D_2=\dfrac{\rho}{8 c^4} \\
G^{\alpha\beta}_s=0           \quad  G^{\alpha\beta}_s&=4G^{\alpha\beta}_2    \quad  G^{\alpha\beta}_2=\dfrac{\sigma_{\alpha \beta}^{0}}{8c^2},
\end{split}
\label{eqn:coefficient}
\end{equation}
where we denoted by $\sigma_{\alpha \beta}^{0}$ the traceless part of $\sigma_{\alpha \beta}$.

One can now proceed to recover the Navier-Stokes equation by using a Chapman-Enskog expansion\footnote{The second moment constraint on the equilibrium  distribution functions is not necessary for the derivation of the continuity equation. Hence the procedure to recover this equation is not affected by the modifications introduced in the new version of the algorithm, with respect to the case of a simple fluid.}. Assuming that the fluid is flowing at small Mach numbers, so to ignore third-order terms in the fluid velocity, 
and taking the first moment of Eq.~\eqref{eqn:eps1}, one gets
\begin{equation}
\partial_{t_1}(\rho v_\alpha) +  \partial_{\beta_1} \left(\rho v_{\alpha} v_{\beta} \right) = \partial_{\beta_1} \sigma_{\alpha \beta} + \mathcal{O}(\epsilon),
\label{eqn:NS_eps}
\end{equation}
which is the Navier-Stokes equation at first order in Knudsen number. To recover the Navier-Stokes equation at second order, we start from Eq.~\eqref{eqn:eq50}, where we need to evaluate the second moment of $f_i^{(1)}$
\begin{equation}
\begin{split}
\sum_i f_i^{(1)} \xi_{i \alpha} \xi_{i \beta}= - \tau \Delta t (\partial_{t_1}+ \xi_{i \gamma} \partial_{\gamma_1} ) \left(\sum_i f_i^{eq} \xi_{i \alpha} \xi_{i \beta} \right) \\
=  - \tau \Delta t \left[ \partial_{t_1}\left(-\sigma_{\alpha \beta}+ \rho v_\alpha v_\beta\right) + \partial_{\gamma_1} \left(\sum_i f_i^{eq} \xi_{i \alpha} \xi_{i \beta} \xi_{i \gamma} \right) \right].
\end{split}
\label{eqn:eq66}
\end{equation}
The first time derivative in square brackets is negligible at the leading order, 
while
\begin{equation}
\partial_{t_1}(\rho v_\alpha v_\beta)= v_\alpha\partial_{t_1}(\rho  v_\beta)+v_\beta\partial_{t_1}(\rho  v_\alpha)
\end{equation}
that shows, together with  Eq.~\eqref{eqn:NS_eps}, that this term gives a null contribution.
Finally, using  Eq.~\eqref{eqn:eq52} we get
the same result of Eq.~\eqref{eqn:eq54} which allows one to restore the Navier-Stokes equation.

\subsubsection{Second method}
\label{sec:external_force}

An alternative route to the solution of the LB equation (\ref{eqn:evolution_equation_LBM}) relies on the use of a pure forcing method~\cite{li2007,tiribocchi2009}. In this case the total stress tensor enters the model via a forcing term $\mathcal{F}_i$, without any additional constraint on the second moment of the equilibrium distribution functions, with condition given in Eq.~\eqref{eqn:II_moment_condition}. The collision term $\mathcal{C}_{f_i}$ assumes the simple form of the BGK approximation supplemented by a forcing term
\begin{equation}
\mathcal{C}(\lbrace f_i \rbrace , t)=-\frac{1}{\tau}[f_i(\vec{r},t)-f_i^{eq}(\vec{r},t)] + \Delta t \mathcal{F}_i,
\label{eqn:collision_op_forcing}
\end{equation}
where the equilibrium distribution functions $f_i^{eq}$ are again expressed as a second-order expansion in the velocity $\vec{v}$ of the Maxwell-Boltzmann distribution~\cite{Qian1992}.
The fluid momentum is now given by the average between the pre- and post-collisional values of the velocity
$\vec{v}$, as usually done when using a forcing term~\cite{Shan1994, Buick2000}
\begin{equation}
\rho v_\alpha = \sum_i f_i \xi_{i \alpha} +\frac{1}{2}F_\alpha \Delta t,
\end{equation}
where $F_{\alpha}$ is the cartesian component of the force density acting on the fluid. The choice of the equilibrium distribution functions and their constraints is kept as in Section~\ref{sec:LBM_sf}, with coefficients given by Eqs.~\eqref{eqn:Icoeff'}-\eqref{eqn:Icoeff''} for a $d2Q9$ lattice.
The term $\mathcal{F}_i$ can be written as an expansion at the second order in the lattice velocity vectors~\cite{ladd2001}:
\begin{equation}
\mathcal{F}_i = \omega_i \left[ A + \dfrac{B_\alpha \xi_{i \alpha}}{c_s^2} + \dfrac{C_{\alpha \beta}(\xi_{i \alpha} \xi_{i \beta} - c_s^2 \delta_{\alpha \beta})}{2c_s^4} \right],
\end{equation}
where coefficients $A$, $B_{\alpha}$ and $C_{\alpha \beta}$ are functions of $F_{\alpha}$. In order to correctly reproduce hydrodynamic equations, the moments of the force term must fulfil the following relations
\begin{equation}
\begin{split}
\sum_i \mathcal{F}_i = A \qquad \qquad	 \sum_i \mathcal{F}_i \xi_{i \alpha} = B_\alpha \\ 
\sum_i \mathcal{F}_i \xi_{i \alpha}  \xi_{i \beta}  = c_s^2 A \delta_{\alpha \beta}  + \dfrac{1}{2} \left[ C_{\alpha \beta} + C_{\beta \alpha}  \right], 
\end{split}
\end{equation}
which lead to~\cite{guo2002}
\begin{equation}
\mathcal{F}_i = \left( 1-\frac{1}{2 \tau} \right) \omega_i
\left[\frac{\xi_{i \alpha} -v_\alpha}{c_s^2}
+\frac{ \xi_{i \beta} v_{\beta}}{c_s^4}{\bf \xi}_{i \alpha} \right] F_\alpha.
\label{eqn:force_exp}
\end{equation}
To recover the continuity~\eqref{eqn:cont_eq} and the Navier-Stokes~\eqref{nav} equations it suffices to require that
\begin{equation}
F_{\alpha}= \partial_\beta (\sigma^{total}_{\alpha \beta} - \sigma^{viscous}_{\alpha \beta}) .
\label{eqn:force4forcing}
\end{equation}
From the Chapman-Enskog expansion (see Appendix~\ref{sec:appC} for the details of the calculation) it results that the fluid viscosity in Eq.~\eqref{eqn:viscous_stress_tensor_incompressible} is $\eta=\frac{\rho \Delta t c_s^2}{3}(\tau-1/2)$. No extra contributions appear in the continuum equations~\eqref{eqn:cont_eq} and~\eqref{nav}, apart from a term of order $v^3$ which can be neglected if the Mach number is kept small. 

Other approaches to the numerical solution of the LB equation introduce spurious
terms which cannot always be kept under control.
For a complete discussion the interested reader may refer to Ref.~\cite{guo2002}.
The one presented here has proved to be effective for simple fluids~\cite{guo2002}, multicomponent~\cite{gonnella2010} and multiphase fluid systems~\cite{coclite2014, sofonea2018} even though, as far as we know, a full external forcing algorithm has not been applied to active systems yet. 
We add that boundary walls can be easily implemented as 
illustrated in the Appendix~\ref{sec:appD}.

\subsection{Coupling with advection-diffusion equation}
\label{sec:lbm4advection}
The aim of lattice Boltzmann methods goes far beyond the treatment of Navier-Stokes equation; indeed, it has proven to be a fundamental  tool to solve general conservation equations~\cite{ancona1994}. Moreover, beside many implementations devoted to hydrodynamics studies, such as the ones cited at the end of the previous Section, recently a LBM approach has also been used to solve Einstein equations for gravitational waves~\cite{Ilseven2016}.

We devote this Section to report on two characteristic ways to solve the dynamics of order parameters coupled to hydrodynamics in a fluid system. Because of its relevance in the study of complex fluids we will focus on the treatment of the advection-diffusion equation~\eqref{conc_eq}  for a concentration field. The first possibility is to develop a full LBM approach in which the advection-diffusion equation is solved by introducing a new set of distribution functions $\lbrace g_i (\vec{r}, t) \rbrace $ connected to the concentration field, beside the distribution functions $\lbrace f_i (\vec{r}, t) \rbrace $ needed to solve the Navier-Stokes equation. Another route is to follow a hybrid approach where the advection-diffusion equation is solved via a standard finite difference algorithm while hydrodynamics is still solved through a LB algorithm.

\paragraph{Full LBM approach}
To solve the hydrodynamic equations for a binary system through a full LB approach the introduction of a new set of distribution functions $\lbrace g_i (\vec{r}, t) \rbrace $ is needed~\cite{xu2003,xu2004}. The index $i$ again assigns each distribution function to a particular lattice direction indicated by the velocity vector $\vec{\xi}_i$. The concentration field $\phi(\vec{r}, t)$ is thus defined as
\begin{equation}
\phi(\vec{r}, t) = \sum_i g_i (\vec{r}, t).
\label{eqn:LBMconcentration}
\end{equation}
As in Eq.~\eqref{eqn:evolution_equation_LBM}, distribution functions $g_i$ evolve according to the following equation
\begin{equation}
g_i(\vec{r}+\vec{\xi}_i\Delta t, t+\Delta t) - g_i(\vec{r}, t)  = -\dfrac{1}{\tau_\phi} ( g_i - g_i^{eq} ),
\label{eqn:concentration_distribution_evolution}
\end{equation}
where the BGK approximation for the collisional operator has been used. A new relaxation time $\tau_\phi$ has been introduced since the relaxation  dynamics of the concentration field may consistently differ from that of the underlying fluid. In Eq.~\eqref{eqn:concentration_distribution_evolution} we have also introduced the set of equilibrium distribution functions $\lbrace g_i^{eq} (\vec{r}, t) \rbrace $ that fulfill the following relation
\begin{equation}
\sum_i g_i^{eq} (\vec{r}, t)=\phi(\vec{r}, t).
\label{eqn:zeroth_moment_concentration}
\end{equation}
This ensures that the concentration field is conserved during the evolution.

To recover the advection-diffusion equation in the continuum limit, it is necessary  to impose the following constraints on the first and second moments of the equilibrium distribution functions
\begin{align}
&\sum_ig_i^{eq}\xi_{i \alpha} = \phi v_\alpha,  \label{eqn:first_moment_concentration}\\
&\sum_i g_i^{eq} \xi_{i \alpha} \xi_{i \beta} =\phi v_\alpha v_\beta + c^2 \chi \mu  \delta_{\alpha \beta}. \label{eqn:second_moment_concentration}
\end{align}
Here the mobility parameter $\chi$ tunes the diffusion constant $M$ that appears on the right-hand side of the advection-diffusion equation, while $\mu$ is the chemical potential.
A suitable choice of the distribution function which fulfills Eq.~\eqref{eqn:zeroth_moment_concentration}, Eq.~\eqref{eqn:first_moment_concentration} and  Eq.~\eqref{eqn:second_moment_concentration} can be written as a power expansion up to the second order in the velocity
\begin{equation}
g_i^{eq}= H_s + J_s v_{\alpha} \xi_{i \alpha} + K_s v^2 + M_s v_{\alpha} v_{\beta} \xi_{\alpha} \xi_{\beta},
\label{eqn:expansion_eq_df_concentration}
\end{equation}
where 
the coefficients of the expansion can be computed from Eqs.~\eqref{eqn:coefficient} 
through the formal substitution
\begin{equation}
\rho \rightarrow \phi \qquad \sigma_{\alpha \beta} \rightarrow - c^2 \chi \mu \delta_{\alpha \beta}.
\label{eqn:formal_substitution}
\end{equation}
The continuum limit of the advection-diffusion equation can be performed through a Taylor expansion of the left-hand side of Eq.~\eqref{eqn:concentration_distribution_evolution} and by using Eqs.~\eqref{eqn:zeroth_moment_concentration}-\eqref{eqn:second_moment_concentration}~\cite{yeomans2009}. This leads to the following expression of the  diffusion constant 
\begin{equation}
M = \chi c^2  \Delta t \left( \tau_\phi -\dfrac{1}{2}	 \right).
\label{eqn:mobility-diffusion}
\end{equation}

This algorithm can be generalized to describe the evolution of more complex order parameters, such as the nematic tensor $Q_{\alpha\beta}$, whose dynamics is governed by the Beris-Edwards equation of motion (Eq.~\eqref{eq:adv_diff}). Since $Q_{\alpha\beta}$ is a traceless symmetric tensor, in $d$ dimensions, at least $d(d+1)/2-1$ extra distribution functions  $\lbrace G_{i,\alpha\beta}(\vec{r}, t) \rbrace $ are needed, which are related to $Q_{\alpha\beta}$  through
\begin{equation}
Q_{\alpha \beta} = \sum_i  G_{i,\alpha \beta}.
\label{eqn:constraintQ}
\end{equation}
The rest of the algorithm can be thus developed as the one presented for the concentration field. In Section~\ref{sec:LBM4activefluids} we will go back to LBM for liquid crystal dynamics and we will present another algorithm that employs a predictor-corrector numerical scheme.

\paragraph{Hybrid LBM approach}
An alternative approach to solve the Navier-Stokes equation and an advection-diffusion equation for an order parameter is based on a hybrid method, in which a standard LBM solves the former while a finite-difference scheme integrates the latter equation. 

Let us consider, for instance, the evolution Eq.~\eqref{conc_eq} of the concentration field $\phi(\vec{r},t)$. Space $\vec{r}$ and time $t$ can be discretized by defining a lattice step $\Delta x_{FD}$ and a time step $\Delta t_{FD}$ for which $\Delta x_{FD}=\Delta x_{LB}$ (namely the scalar field is defined on the nodes of the same lattice used for the LB scheme) and $\Delta t_{LB}=m\Delta t_{FD}$, with $m$ positive integer. At each time step the field $\phi$ evolves according to Eq.~\eqref{conc_eq} and is updated in two partial steps.
\begin{enumerate}
\item Update of the convective term by means of an explicit Euler algorithm
\begin{equation}
\phi^* ( \vec{r}_\alpha ) = \phi  - \Delta t_{FD} ( \phi \partial_\alpha v_\alpha +   v_\alpha \partial_\alpha \phi),
\end{equation}
where all variables appearing at the right-hand side are computed at position $\vec{r}_\alpha$ and time $t$. Note that the velocity field $\vec{v}$ is obtained from the lattice Boltzmann equation.
\item Update of the diffusive part 
\begin{equation}
\phi ( \vec{r}_\alpha, t + \Delta t_{FD} ) = \phi^* + \Delta t_{FD} \left(\nabla^2 M  \dfrac{\delta \mathcal{F}}{\delta \phi}\right)_{\phi=\phi^*} .
\end{equation}
\end{enumerate}
Note that one could use more elaborate methods to solve convection-diffusion equations. For instance, one can combine predictor-corrector schemes for the treatment of the advective term with a wealth of finite-difference schemes for the numerical solution of parabolic equations~\cite{Evans2000}. Nevertheless one has to always  keep in mind consistency between the order of accuracy of combined different numerical schemes used.
However, the method here described, besides being relatively simple to implement, combines a good numerical stability with a reduced 
memory requirement with respect to the full LBM approach~\cite{tiribocchi2009}, as it will be discussed in Section~\ref{sec:performance}. 



\subsection{LBM for Active Fluids}
\label{sec:LBM4activefluids}

As outlined in Section~\ref{sec:2}, many properties of active matter are captured by liquid crystal hydrodynamics. Here we describe a LB method that solves both the Navier-Stokes equation and the Beris-Edwards equation through a full LB approach, a method often employed to numerically investigate active matter~\cite{denniston2001,Cates2099}.


As the liquid crystal stress tensor entering the Navier-Stokes equation is generally not symmetric, one could either (i) build an algorithm in which it is fully included through an external forcing term (as described in Section~\ref{sec:external_force}) or (ii) separate the symmetric part from the antysimmetric one, by including the former in the second moment of the equilibrium distribution functions and treating the latter as an external forcing term. Although the two procedures are equivalent, only the second approach, first introduced by Denniston \emph{et al.}~\cite{denniston2001}, has been developed so far.

In this method two sets of distribution functions, $\lbrace f_{i}\rbrace $ and $\lbrace G_{i,\alpha \beta} \rbrace $, are defined and are connected to the hydrodynamic variables (i.e. density, momentum and order parameter) through Eqs.~\eqref{eqn:mass_constraint},~\eqref{eqn:momentum_constraint} and~\eqref{eqn:constraintQ}.
Their evolution equations are solved by using a predictor-corrector-like scheme
\begin{equation}
\begin{split}
&f_i(\vec{r}+\vec{\xi}_i\Delta t, t+\Delta t) - f_i(\vec{r}, t)  \\&  \qquad \qquad =\dfrac{\Delta t}{2} \left[\mathcal{C}(\lbrace f_i \rbrace, \vec{r}, t)  +\mathcal{C}(\lbrace f_i^{*} \rbrace, \vec{r}+\vec{\xi}_i \Delta t ,t+\Delta t) \right] , 
\end{split}  \label{eqn:evolution_pc_f} 
\end{equation}
\begin{equation}
\begin{split}
G_{i,\alpha \beta}(\vec{r}+\vec{\xi}_i &\Delta t, t+\Delta t) - G_{i,\alpha \beta}(\vec{r}, t)  \\&= \dfrac{\Delta t}{2} \left[\mathcal{C}(\lbrace G_{i,\alpha \beta} \rbrace , \vec{r}, t) \right. \\ & \left. \qquad \qquad +\mathcal{C}(\lbrace G^{*}_{i,\alpha \beta} \rbrace \vec{r}+\vec{\xi}_i \Delta t, t) \right] ,
\end{split}  \label{eqn:evolution_pc_G}
\end{equation}
where $f_{i}^{*}$ and  $G_{i,\alpha \beta}^{*}$ are respectively first order approximation to $f_{i}^{*}(\vec{r}+\vec{\xi}_i \Delta t ,t+\Delta t)$ and  $G_{i,\alpha \beta}^{*}( \vec{r}+\vec{\xi}_i \Delta t ,t+\Delta t)$ obtained by setting $f_i^* \equiv f_i$ and $G_{i,\alpha \beta}^{*} \equiv G_{i,\alpha \beta}$ in Eq.~\eqref{eqn:evolution_pc_f} and~\eqref{eqn:evolution_pc_G}. The collisional terms are given by a combination of the usual collision operator in the BGK approximation plus a forcing term
\begin{eqnarray}
\mathcal{C}(\lbrace f_i \rbrace, \vec{r}, t) &=& -\dfrac{1}{\tau_f} (f_i-f_i^{eq}) + p_i, \label{eqn:collision_bis}\\
\mathcal{C}(\lbrace G_{i,\alpha \beta} \rbrace, \vec{r}, t) &=& -\dfrac{1}{\tau_G} (G_{i,\alpha \beta}-G_{i,\alpha \beta}^{eq}) + M_{i,\alpha \beta},
\end{eqnarray}
where $\tau_f$ and $\tau_G$ are two distinct relaxation times, and $p_i$ and $M_{i,\alpha\beta}$ are the two additional forcing terms.

In order to recover continuum equations one must impose constraints on the zeroth, first and second moments of the equilibrium distribution functions and on the forcing terms. The local conservation of mass and momentum is ensured by~\eqref{eqn:mass_constraint_eq} and~\eqref{eqn:momentum_constraint_eq}, while the second moment is given by Eq.~\eqref{eqn:general_stress_II_moment}, in which the stress tensor on the right hand side includes the sole symmetric part. The antisymmetric contribution $\sigma^{anti}_{\alpha\beta}$ is introduced through the forcing term $p_i$, which fulfills the following relations
\begin{equation}
\sum_i p_i= 0, \qquad \sum_i p_i \xi_{i \alpha}= \partial_\beta \sigma^{anti}_{\alpha \beta}, \qquad \sum_i p_i \xi_{i \alpha} \xi_{i \beta}=0.
\label{eqn:constraint_forcing_term}
\end{equation}
The remaining distribution functions $G_{i,\alpha \beta}$ obey the following equations
\begin{align*}
\sum_i G_{i, \alpha \beta}^{eq} &= Q_{\alpha \beta}, \\
\sum_i G_{i, \alpha \beta}^{eq} \xi_{i \gamma} &= Q_{\alpha \beta} v_{\gamma}, \\
\sum_i G_{i, \alpha \beta}^{eq} \xi_{i \gamma} \xi_{i \delta} &= Q_{ \alpha \beta} v_{\gamma}v_{\delta},
\end{align*}
while the forcing term $M_{i,\alpha \beta}$ satisfies
\begin{align*}
\sum_i M_{i, \alpha \beta}^{eq} &= \Gamma H_{\alpha \beta} +S_{\alpha \beta},\\
\sum_i M_{i, \alpha \beta}^{eq} \xi_{i \gamma} &= \left( \sum_i M_{i, \alpha \beta}^{eq} \right) v_{\gamma}.
\end{align*}

We finally note that the predictor-corrector scheme has been found to improve numerical stability of the algorithm and to eliminate lattice viscosity effects (usually emerging from the Taylor expansion and appearing in the viscous term, in the algorithms discussed so far) to the second order in $\Delta t$. 
To show this, one can Taylor expand Eq.~\eqref{eqn:evolution_pc_f} to get


\begin{multline}
 (\partial_t + \xi_{i \alpha} \partial_\alpha) f_i(\vec{r}, t) -  \mathcal{C}(\lbrace f_{i} \rbrace) \\= - \dfrac{\Delta t }{2}(\partial_t + \xi_{i \alpha} \partial_\alpha) \left[ (\partial_t + \xi_{i \alpha} \partial_\alpha) f_i -  \mathcal{C}(\lbrace f_{i} \rbrace)  \right]  + \mathcal{O}(\Delta t^2).
\end{multline}
The left-hand side is $\mathcal{O}(\Delta t)$ and coincides with the term in square brackets. One could then write at second order in $\Delta t $
\begin{equation}
 (\partial_t + \xi_{i \alpha} \partial_\alpha) f_i(\vec{r}, t) =  \mathcal{C}(\lbrace f_{i} \rbrace) + \mathcal{O}(\Delta t^2).
\end{equation}
An analogous calculation for $G_{i,\alpha \beta}$ shows that
\begin{equation}
 (\partial_t + \xi_{i \gamma} \partial_\gamma) G_{i,\alpha \beta} (\vec{r}, t) =  \mathcal{C}(\lbrace G_{i,\alpha \beta} \rbrace) + \mathcal{O}(\Delta t^2),
\end{equation}
thus recovering the proper lattice Boltzmann equations.

A hybrid version of the algorithm, widely employed in the study of active matter, solves the Navier-Stokes equation through a predictor-corrector Lattice-Boltzmann approach and the Beris-Edward equation by means of a standard finite-difference method~\cite{henrich2010,Cates2099}.

Further models involving more than just one order parameter have been developed in recent years, such as the theory discussed in Section~\ref{sec:2d}, in which the liquid crystal order parameter (the polarization field) is coupled to the concentration field of a binary fluid mixture. Again a hybrid approach, in which both equations of the concentration and of the polarization have been solved through finite difference methods, has been used~\cite{bone2017,negro2018}.

\subsection{Computational perspectives: stability, efficiency and parallelization}
\label{sec:performance}

\begin{figure}[t]
\center
\resizebox{0.95\columnwidth}{!}{\includegraphics{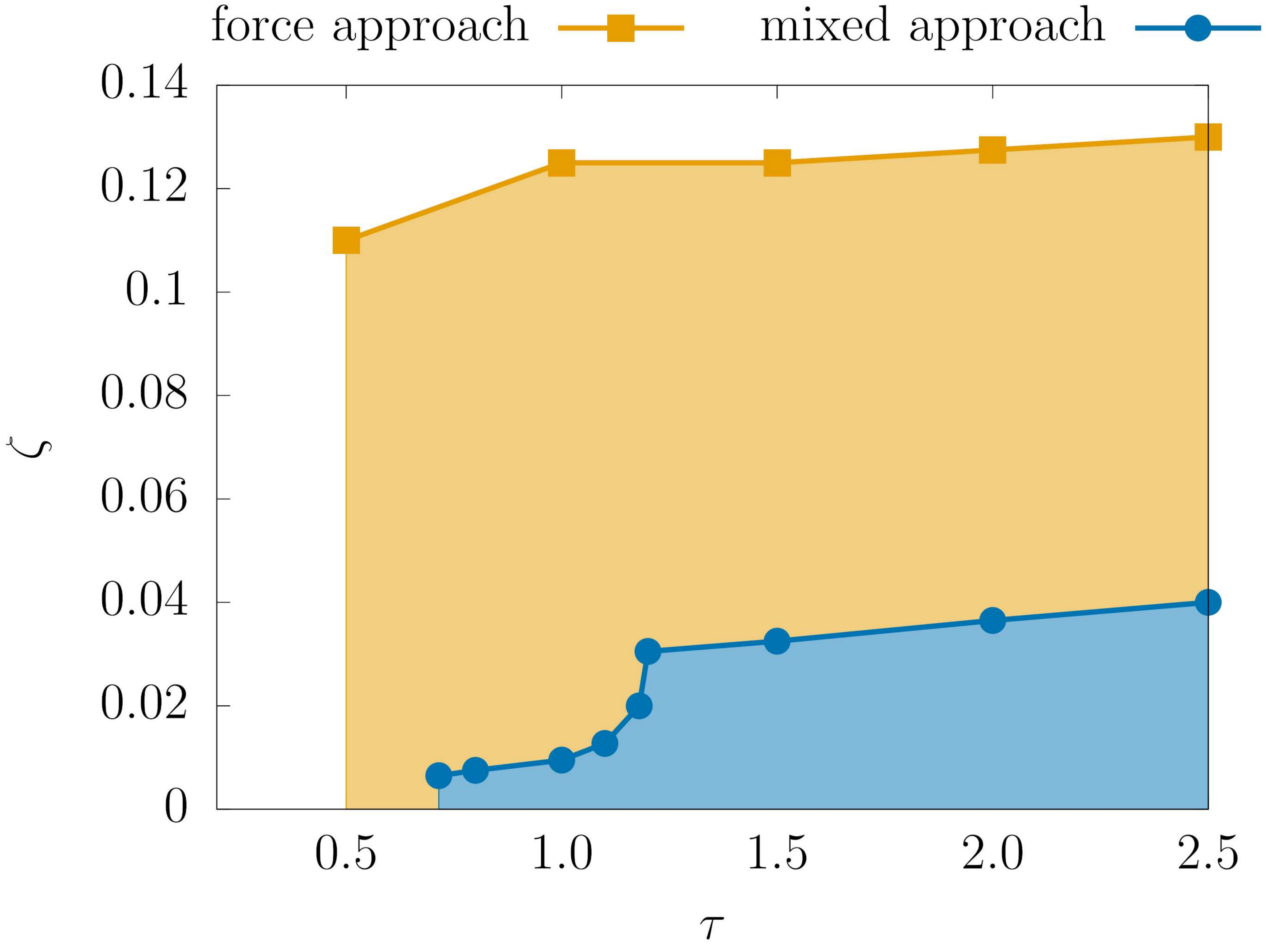}}
\caption{Stability of two hybrid LB codes, for a polar binary mixture, treating the stress tensor by a full-force approach (squared/yellow dots) and a mixed approach (circle/blue dots). The codes are stable for parameters under their corresponding curves.  Simulations were performed on a computational grid of size $64\times 64$, checking stability for $10^5$ LB iterations.}
\label{img:stability_lbm}
\end{figure}

In the previous Sections we presented different LB algorithms for the treatment of the hydrodynamics of complex and active fluids. We will comment here on the stability of two different $d2Q9$ hybrid LB codes solving the equations of an active polar binary mixture (the hydrodynamics is solved by means of LB while the order parameter dynamics is integrated by a finite difference algorithm implementing first order upwind scheme and fourth order derivative accuracy), described by the free energy in Eq.~\eqref{free-energy-mixture}, treating the symmetric part of the stress tensor with two different approaches. The first is a \emph{mixed approach}, presented in the previous Section, where the symmetric part of the stress tensor enters in the definition of the second moment of the distribution functions (see Eq.~\eqref{eqn:general_stress_II_moment}) while the anti-symmetric part is treated by means of the forcing term $p_i$ (see Eqs.~\eqref{eqn:collision_bis} and \eqref{eqn:constraint_forcing_term}). In the second approach the total stress tensor is treated by means of the only forcing term. To compare the stability of the two algorithms we fixed the mesh spacing and the time resolution ($\Delta x=1$, $\Delta t=1$), and we let vary the relaxation time $\tau$ and the intensity of active doping $\zeta$ appearing in the active stress tensor~\eqref{stresspolar}. The results of the stability test in Fig.~\ref{img:stability_lbm} show that the full-force approach is definitely more stable than the mixed one. In this latter case the code is found to be stable for $\tau > 0.715$ in the passive limit ($\zeta=0$) while to simulate active systems ($\zeta>0$), the relaxation time must be accurately chosen to ensure code stability. In the full-force approach the code is found to be stable for $\tau > 0.5$, almost independently of $\zeta$.

The rest of this Section is devoted to a brief discussion of some performance aspects, such as efficiency and parallelization of a LB code. 
LBM is computationally efficient if compared to other numerical schemes.
The reason lies in the twofold discretization of the Boltzmann equation in the physical and velocity space. 
For instance, computational methods such as finite-difference (FD) and pseudo-spctral (PS) methods require high order of precision to ensure stability~\cite{Evans2000} and to correctly compute non-linearities in the NS equation~\eqref{nav}. 
This introduces non-local operations in the computational implementation that reduce the throughput of the algorithm. 
LBM, on the contrary, is intrinsically local, since the interaction between the nodes is usually more confined, according to the particular choice of the lattice,
while non linearities of the NS equation is inherently reproduced at the level of the collision operator.
For instance, while the number of floating point operations needed to integrate the hydrodynamics equations on a $d$-dimensional cubic grid is $\sim L^d$ for LBM, it is instead of order $\sim (\ln L) L^d$ for pseudo-spectral models~\cite{succi1991}. 
Nevertheless LB algorithms are definitely much more memory consuming, since for \emph{each} field to evolve, one needs a number of distribution functions equal to the number of lattice velocities.
From this perspectives, the hybrid version of the code is somewhere in the middle between the two approaches, since it allows one to exploit both computational efficiency and simplicity typical of LB approaches and, at the same time, to keep  the amount of memory to be allocated at runtime lower than that
necessary for a full LB treatment.

LB algorithms are also suitable for parallelization. The reason still lies in the local character of LB, since at the base of the efficiency of any parallelization scheme is the compactness of the data that must be moved among the different devices that take part in the program execution. Parallelization approaches involving both CPUs, \emph{i.e.} MPI or OpenMP, and GPUs, such as CUDA and OpenCL, or even both (CUDA aware MPI) can be used when dealing with LB~\cite{kruger2016}.
Most of them, such as OpenMP or GPU-based approaches, aim at rising the amount of floating operations per unit time, while a different technique consists in splitting the global computational domain in subdomains and assign each of them to a different computational unit (usually threads of one or more processors). This is usually done with MPI. 

\begin{figure}
\center
\resizebox{0.95\columnwidth}{!}{\includegraphics{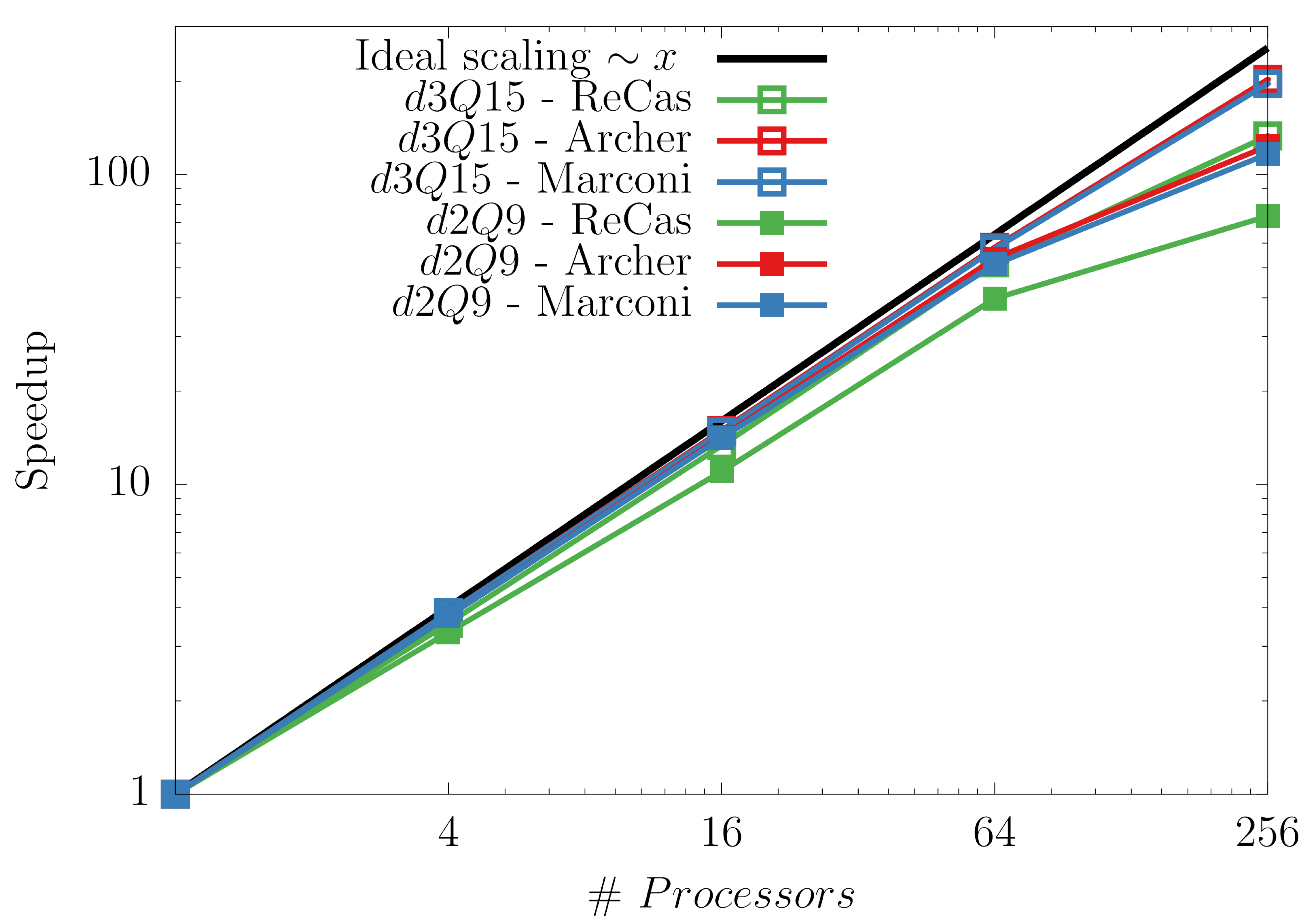}}
\caption{Speedup, as defined in the text, versus number of processors for an MPI parallelized hybrid LBM code, coupled to the dynamics of a concentration scalar field and a polar vector field. Simulations were performed in $2d$ on a square computational grid ($512^2$) and in $3d$ in $128^3$ cubic domain, on different HPC farms: Archer UK National Supercomputing Service (\textit{http://www.archer.ac.uk/}), CINECA Marconi - Skylake partition (\textit{http://www.hpc.cineca.it/}) and ReCas-Bari (\textit{https://www.recas-bari.it/}).}
\label{img:MPI_scaling}
\end{figure}

Fig.~\ref{img:MPI_scaling} shows the results of a strong scaling test performed on a hybrid code integrating the hydrodynamics of a polar binary mixture~\cite{bone2017,negro2018}, implementing the full-force algorithm used for the stability analysis.
This test consists in changing the amount of processors used to perform a certain task, while keeping fixed the size of the computational grid and measuring the speedup, namely the ratio of time spent to perform the operation with only one processor over the time taken when more processors are used. 
Simulations were performed both with $d2Q9$ (hollow dots) and $d3Q15$ (full dots) lattice structures on different computational infrastructures (Archer (red), Marconi (blue) and ReCas (green)). While for a few number of processors the scaling is approximately linear, thus close to the ideal linear behavior (black line), as the number of processors increases, it progressively deviates from the ideal scaling law. This is due to a number of issues that may depend both on the infrastructure characteristics (bandwidth, cache size, latency, \emph{etc.}) and on the program implementation (bottlenecks, asynchrony among processor, \emph{etc.}).
Moreover, code scalability is found to be significantly better in three-dimensional grids than in their bidimensional version. This is because the fraction of time spent by the $3d$ code to perform parallel operations (sending and receiving data, reduction operations, synchronization, \emph{etc.}) is consistently reduced compared with its $2d$ counterpart.

\section{Spontaneous flow}
\label{sec:spont_fl}

\begin{figure}
\centering
\resizebox{.5\textwidth}{!}{
 \includegraphics{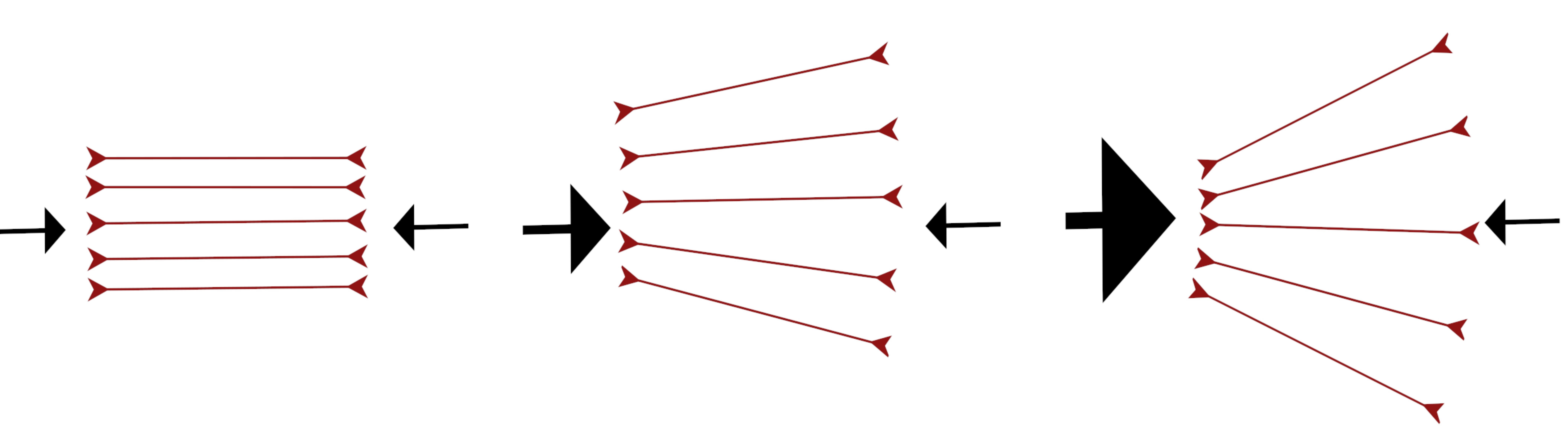}}
 \caption{Sketch of instability and spontaneous symmetry breaking  mechanism for contractile systems. When the system is completely ordered (left panel) force dipoles compensate each other, while if a splay deformation is present (middle panel) the density of contractile forces is greater on the left than on the right. This determines a flow that produces further splay (right panel), resulting in a macroscopic flowing state.}\label{fig:spontaneous_flow}
\end{figure}

\begin{figure*}[t]
\center
\resizebox{1.0\textwidth}{!}{%
  \includegraphics{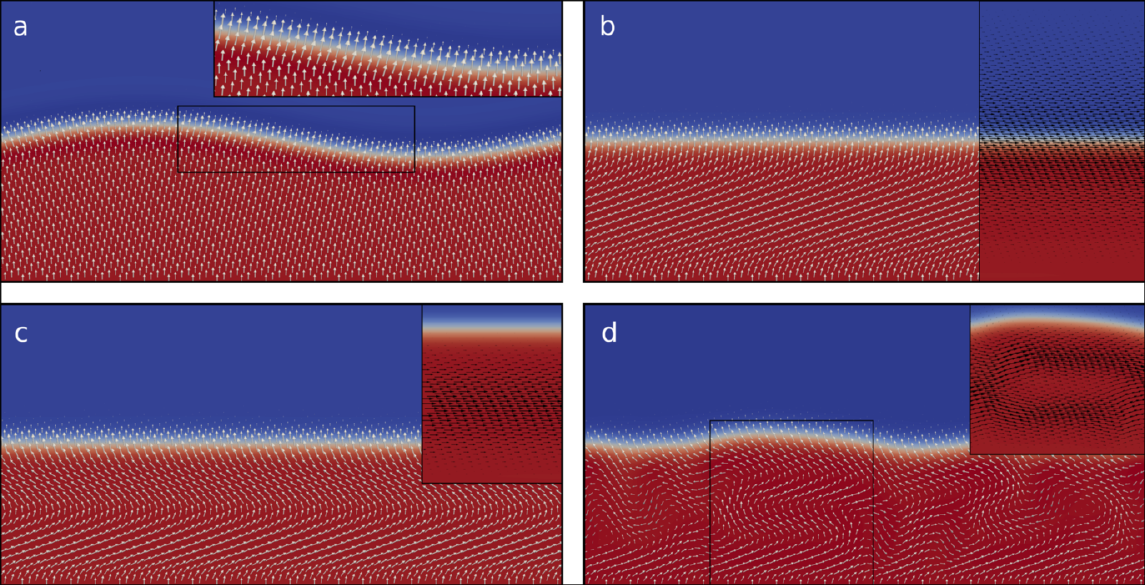}
}
\caption{Instability and spontaneous flow in extensile mixtures. The polarization field is confined in one of the phases of a binary mixture and satisfies homeotropic anchoring both at the lower bound of the channel and at the interface between the two fluid components; moreover the interface is modulated in a sinusoidal fashion, determining a weak splay instability in the polarization field, as shown in panel (a). Inset shows detail of the  polarization at interface. Starting from this configuration and turning on extensile activity, two regimes are found. For weak active doping ($\zeta=5 \times 10^{-4}$), shown in panel (b), the interface relaxes towards a flat profile and the polarization pattern undergoes bending deformations, while the velocity field, shown in the right part of panel (b), is parallel to direction of the channel and confined in proximity of the interface. If activity is raised ($\zeta=10^{-3}$), the bending deformations  are tightened (as clearly visible in panel (c)) and a unidirectional flow field develops in centre of the polar fluid and mostly parallel to the walls (see the corresponding inset). When active doping exceeds a critical threshold ($\zeta=3 \times 10^{-3}$ in panel (d)) the polarization field undergoes instabilities leading to the formation of chaotic non-stationary  patterns. In such condition the interface loses its flat profile, although the velocity field remains roughly parallel to the channel direction. The velocity field plotted in insets of panels (b), (c) and (d) has been rescaled for readability (the  averaged velocity magnitude grows 
 from $|\vec{v}| \simeq 3.5 \times 10^{-3}$ to $\simeq  \times 10^{-2}$ in lattice units
 when $\zeta$ goes from $5 \times 10^{-4}$ to $3\times 10^{-3}$). The free energy 
used in these simulations  is given in Eq.~\eqref{free-energy-mixture} with $a=4\times 10^{-2}$, $k=4\times 10^{-1}$, $\alpha=10^{-3}$, $\kappa=10^{-2}$ and $\beta=10^{-2}$, while mobility $M=10^{-1}$ and aligning parameter $\xi=1.1$.}
\label{fig:stability_extensile}       
\end{figure*}

\begin{figure}[t]
\center
\resizebox{1.0\columnwidth}{!}{%
  \includegraphics{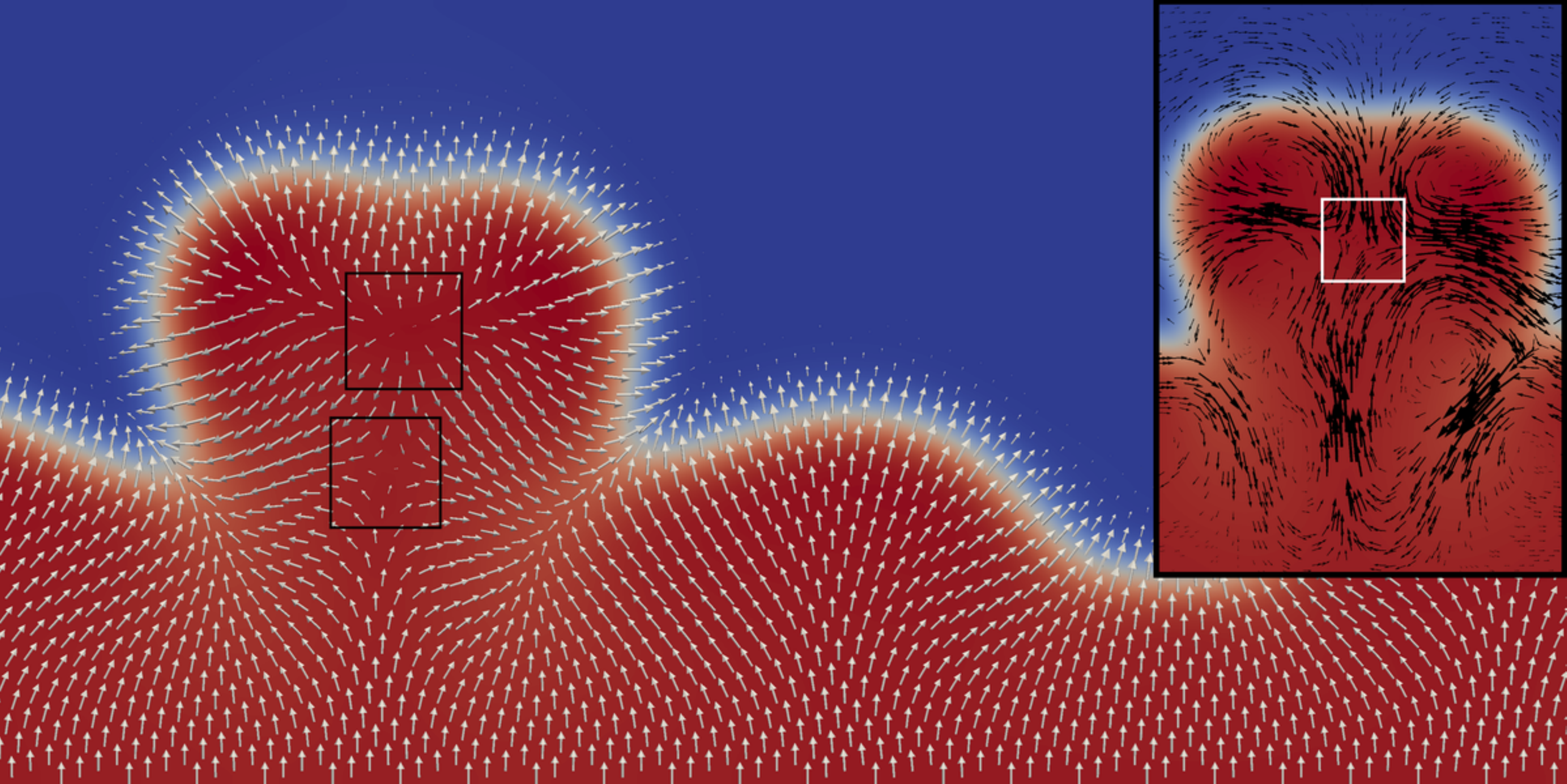}
}
\caption{Splay deformation and defect formation in a contractile mixture  ($\zeta=-3\times 10^{-2}$).  Other free-energy parameters and initial conditions are the same as  in Fig.~\ref{fig:stability_extensile}.
The strong contractile activity leads to a catastrophic dynamics: the polarization splay deformation of the initial condition is tightened until the initial sinusoidal shape in the interface between the two fluids is completely lost and replaced with an undulated profile,
as clearly visible in the center of the system. Notice also the formation of two defects of opposite charge that have been framed with two black squares. Defect formation strongly influences the hydrodynamics of the system, as shown in the inset, where the velocity field develops a quadrupolar flow in their neighborhood.} 
\label{fig:stability_contractile}    
\end{figure}

Many remarkable phenomena in the physics of active fluids are related to the flow behavior induced by
the presence of active forcing in the dynamical description of the system. 
The first effect that was studied, to which this section  is devoted, is the occurrence of spontaneous flow in fluids with sufficiently strong activity. Numerical methods, and LBM in particular, have been essential in the study  of this problem, both for supporting theoretical analysis  of instabilities and for the understanding of non-linear regimes. Here we will put the accent  on numerical studies, while a more complete review of analytical studies of instabilities in active fluids can be found in~\cite{marchetti2013}.

The continuum 
theory  described in Section~\ref{sec:2e} with Eqs.~(\ref{stressnematic}) or (\ref{stresspolar}) was deduced  in~\cite{simha2002} 
where instabilities due to activity   were first analyzed. 
The instability of ordered contractile and extensile gels  to splay and bend perturbations respectively, giving rise to  macroscopic flows~\cite{giomi2008,edwards2009}, can be  qualitatively explained as follows.
Let us consider the  case of contractile dipoles initially perfectly ordered, as in the sketch on the left of 
Fig.~\ref{fig:spontaneous_flow}. In this situation the force dipoles balance each other and the net flow, obtained by the sum of those due to single dipoles as represented  in Fig.~\ref{fig:2}, is null. However, if a small splay deformation is present (middle panel in Fig.~\ref{fig:spontaneous_flow}), the density of contractile forces on the left is larger than that on the right, and a flow sets up. This flow causes further splay which destabilizes the system that starts to flow macroscopically. For extensile activity, under the same splay deformation, the initial flow (directed to the left in this case, see Fig.~\ref{fig:2}) would align the dipoles and no net macroscopic flow would appear. By a similar argument, it can be shown that extensile fluids get unstable to bend deformations.

The spontaneous flow  instability for contractile systems  was analyzed  in~\cite{voituriez2005} for the simple geometry of a bidimensional thin film confined on a one-dimensional substrate, with  planar anchoring on the confining boundaries. 
For small thicknesses or small activity, boundary effects are prevailing and the gel remains in an unperturbed, static, homogeneously polarized  state. Above a critical thickness or a critical activity, a polarization tilt appears and the system flows with a finite shear gradient. The study 
 has been later extended to films where undulations of the free surface are also considered~\cite{sankararaman2009,marchetti2013}. In particular Sankararaman et. al~\cite{sankararaman2009} constructed dynamical equations  for the concentration and the polarization field, and for the height of the film thickness. Activity was found to have two main effects on the evolution of the height field: (i) a splay induced flow that tilts the free surface and (ii) an active contribution to the effective tension. The latter can be understood by noting that active stresses pull (contractile) or push (extensile) the fluid in the direction of the long axis of the particles, giving additional elastic contribution to the stretching along that axis.
By  stability analysis arguments they found that, for contractile stresses, splay destabilizes the surface, while the activity contribution to tension tends to stabilize it. For extensile activity the opposite happens. 

The previous results are illustrated in  Figs.~\ref{fig:stability_extensile} and~\ref{fig:stability_contractile}. 
 Here  the onset of instability and spontaneous flow are shown for the polar active binary mixture  described  in Section~\ref{sec:2d}, in which an active polar gel (red) coexists with a passive component. Homeotropic anchoring is set both at the lower bound of the channel and at the interface between the two fluid components. The system is  numerically studied by means of a hybrid Lattice Boltzmann method which combines a LB treatment for the Navier-Stokes equation with a finite difference algorithm to solve the order parameters dynamics (Section~\ref{sec:LBM4activefluids}). In order to study stability with respect to bending, the interface is initially modulated  by a sinusoidal perturbation. This determines a weak splay pattern in the polarization field, as shown in Fig.~\ref{fig:stability_extensile}(a) and in its inset, in which white arrows 
represent the polarization field.  
As expected, the extensile system is stable under the initial splay deformations, so that,  
by increasing activity,  these  are replaced by stationary  bending  patterns (see Fig.~\ref{fig:stability_extensile}(b)). They are accompanied by   macroscopic flows, as it can be seen  by looking at the velocity field, denoted with black vectors in the zooms of panels~\ref{fig:stability_extensile}(b) and~\ref{fig:stability_extensile}(c), 
with the  magnitude of velocity growing linearly with $\zeta$. 
Then, further increasing activity, the  polarization field becomes unstable (Fig.~\ref{fig:stability_extensile}(d)) and the flow looses 
laminar character. Bends in the pattern give rise to non uniform fluxes, generating complex structures in the velocity field.  
A different behavior results with contractile activity. When it is strong enough, it
tightens splay deformations of the initial condition   of the polar field until, as shown in Fig.~\ref{fig:stability_contractile}, the initial sinusoidal shape of the interface between the two fluids is completely lost and replaced by  an irregular  profile driven by splay polarization deformations.
Fig.~\ref{fig:stability_contractile} also shows the presence  of two defects of opposite charge ($+1$ and $-1$), framed with two black squares. They strongly influence the velocity pattern of the system, as shown in the inset where  a quadrupolar flow can be observed in their neighborhood. 

The first papers where spontaneous flow was systematically  analyzed numerically
are~\cite{marenduzzo2007,marenduzzo2007pre}. A slab of active nematic liquid crystal confined between two fixed parallel plates at a distance $L$ was studied by a hybrid version of  LBM,  with different anchoring conditions. The active nematic model is the same of that described in Section~\ref{sec:2}. 
The dynamics of the order parameter $\underline{\underline{\vec{Q}}}$ is governed by Eq.~\eqref{eq:adv_diff}, with $\mathbf{\Xi}$ replaced by $\underline{\underline{\vec{Q}}}$ and $\mathbf{S}$ given in the last row of Table~\ref{terms}, with an extra active term, besides the active stress term in the Navier-Stokes equations, of the form $\lambda\underline{\underline{\vec{Q}}}$ on the right-hand side of Eq.~\eqref{eq:adv_diff}. This term was suggested on the basis of symmetry considerations  and also obtained by a microscopic derivation in~\cite{ramaswamy2003}. 
Though a linear term in the nematic stress tensor also appears in the molecular field, the extra term here introduced can be regarded as \emph{active}, also because  no counterpart is included in the stress tensor.
Positive (negative) values of $\lambda$ enhance (attenuate) self-aligning features of the nematic network, so that $\lambda > 0$ can be chosen to model 
actomyosin suspensions at high concentration, and $\lambda < 0$ to model dilute emulsions.
\begin{figure}
\centering
\resizebox{.5\textwidth}{!}{
 \includegraphics{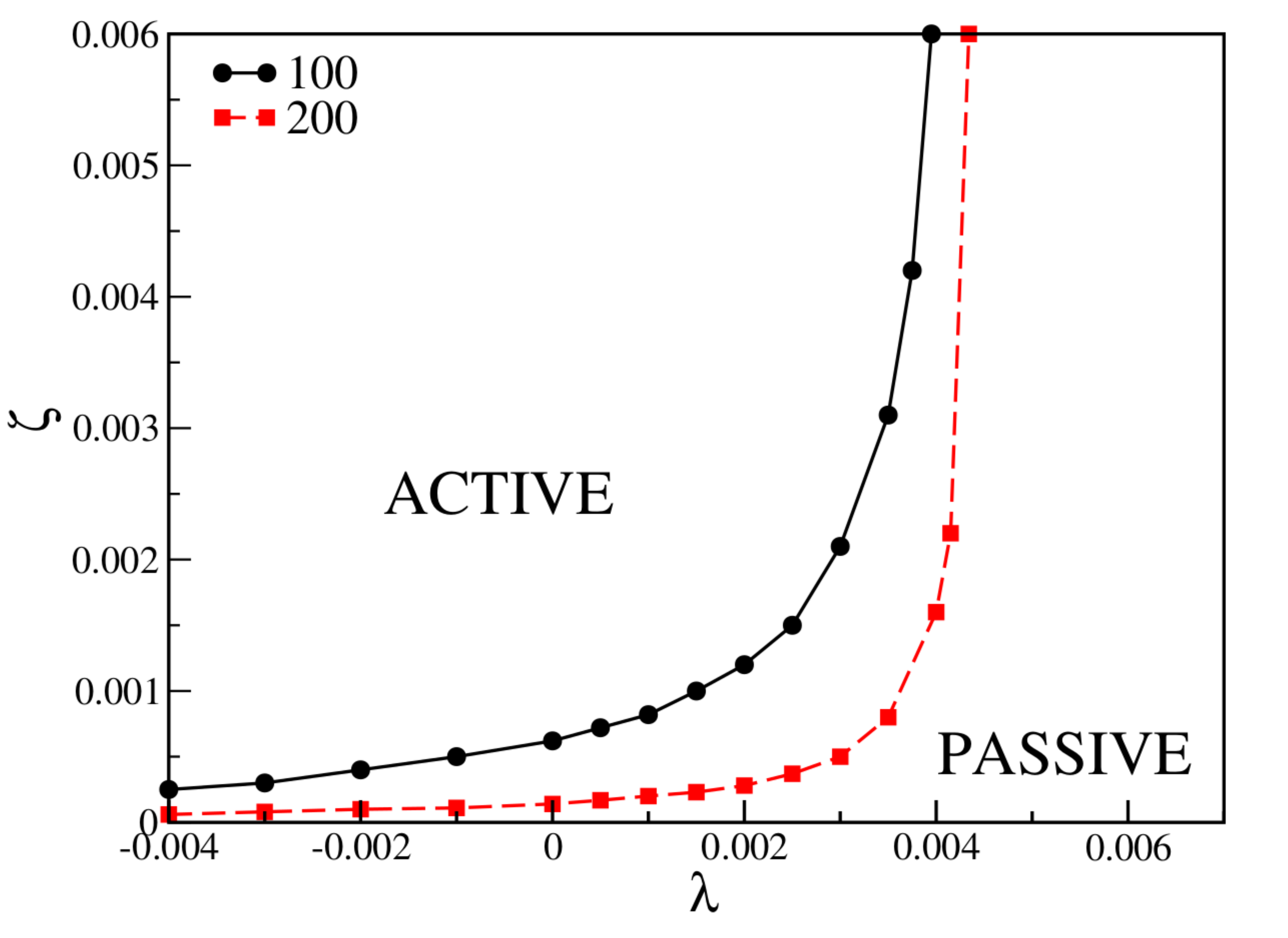}}
 \caption{Phase diagram for spontaneous flow obtained by Marenduzzo \textit{et al.}~\cite{marenduzzo2007pre}, in the two activity parameters plane $(\lambda,\zeta)$ for an active nematic liquid crystal. The lines separate regions of passive immotile state, and active, macroscopic motile state for two different system sizes. A slab of material is considered with homogeneous anchoring at the boundaries and
flow aligning parameter $\xi=0.7$. (Image courtesy of D. Marenduzzo.)}\label{fig:marenduzzo}
\end{figure}

The main results concerning   the occurrence of spontaneous flow are  summarized in Fig.~\ref{fig:marenduzzo} for two different system widths $L=100\textrm{,} 200$, which confirm the presence of a transition between a passive and an active phase as predicted analytically.
Flow properties in the active phase are reported not to depend on the value of $\lambda$.
For small $\zeta$ there is no flow and the polarization field is homogeneous. If $\zeta$ is strong enough, the system sets in the active phase, where a spontaneous flow is observed,  while decreasing $L$ leads to a reduction of the active region in the parameter space.
Alongside the activity parameters $\zeta$ and $\lambda$, the other key parameter is the flow alignment parameter $\xi$ (see Table~\ref{terms}). In fact, the transition is attained for sufficiently extensile suspensions, in the case of flow-aligning ($|\xi|>1$)
liquid crystals, and for sufficiently contractile ones for flow-tumbling materials ($|\xi|<1$).
In the flow-aligning case the velocity profile is characterized by the presence of bands, i.e. areas of constant shear rate, separated by narrow regions where the shear gradient reverses, similar to shear bands in non active materials~\cite{fielding2004} with the number of wavelengths in the channel increasing with $\zeta$. Flow tumbling materials rearrange themselves so that only the two boundary layers flow in steady state. Simulations  with periodic boundary conditions show additional instabilities, with the spontaneous flow appearing as patterns made up of convection rolls.
Boundary conditions for the model in~\cite{marenduzzo2007} are described in detail in~\cite{marenduzzojnnm}, while the numerical  method in~\cite{marenduzzo2011}. The phase diagram was studied, for a quasi-$1d$ system, in~\cite{edwards2009}, extending previous works to the whole $(\xi,\zeta)$ plane, varying also the initial orientation of  the director field.

A detailed numerical study  of the dynamical spontaneous flow transition in polar active films  (not by LBM but directly integrating dynamic equations) is presented in~\cite{giomi2008}. In this work  the effects of varying concentration were explicitly taken into account. The free-energy of the model is similar in spirit to that of Eq.~\eqref{free-energy-mixture} but only one phase for the concentration of the active fluid is considered (the free-energy is at most quadratic in the concentration field 
and no phase separation  can occur).
The transition to spontaneous flow is characterized by a phase diagram in a plane of  two variables, related to activity  $\zeta$ and to a parameter  controlling  self-advection\footnote{{In~\cite{giomi2008}, together with the active stress tensor $
\sigma^{\textrm{active}}_{\alpha \beta}$ of Eq.~\eqref{stresspolar},  the term 
$ \tilde\beta(\partial_{\beta}P_{\alpha}+\partial_{\alpha}P_{\beta})$ is also considered. 
This, primary to polar systems,  arises from
‘‘self-propulsion’’ of the active units, taking into account higher order contributions 
 in gradients in the coarse-graining~onlyonlyonly\cite{simha2002} procedure that leads to Eq.~\eqref{stresspolar}.} This extra contribution complements the modified advected term in the evolution equations of the order parameters (discussed at the end of Section~\ref{sec:2e}) that allows for the description of self advective phenomena.}.
For high values of  activity and self-advection parameters a phase charaterized by spontaneous periodic oscillatory  banded flows is observed.
The latter, accompanied by strong concentration inhomogeneities, can also arise in active nematics, although with a physically distinct origin.



We finally mention another LBM study on spontaneous flow~\cite{bonelli2016}, where the effects of a phenomenological term $\tilde\lambda \vec{v}$ in the dynamic equation for the  polarization field $\vec{P}$ (Eq.~\eqref{eq:adv_diff}) were studied. This term is meant to favor the alignment between the polarization field and the velocity typically observed in experiments. It was shown that the interplay between alignment and activity gives rise to very different behaviors under different boundary conditions and depending on the contractile or extensile character of the fluid. 
With periodic boundary conditions and extensile swimmers, an almost uniform alignment between
$\vec{v}$ and $\vec{P}$ is attained, 
while this is not anymore true for contractile systems. In the former case, when $\zeta$ and $\tilde{\lambda}$ are high enough, domains of full alignment are spaced out by small stripes or, in turn, regions in which bend distortions in the polarization field correspond to an almost zero velocity field. 
The behavior is yet different in bounded contractile systems. These are characterized by mutual alignment of $\vec{P}$ and $\vec{v}$,  except for weak distortions close to walls for small and intermediate values of $\zeta$; moreover, fluctuations are enhanced while increasing $\tilde{\lambda}$.


The occurrence of spontaneous flow 
may also be accompanied by the formation of topological defects. In fact  the dynamics of  order parameter  and velocity fields are interconnected through a feedback loop (see Fig.~\ref{fig:stability_contractile}). The hydrodynamical instabilities give rise to lines of distortions in the order parameter field that are unstable to the formation of defect pairs~\cite{giomi2013}. 
In~\cite{thampi2014_1,thampi2014_2} an extensile active nematic has been considered and the dynamics of defects characterized. Two main stages have been identified: first, ordered regions undergo hydrodynamic instability generating lines of strong bend deformation that relax by forming oppositely charged pairs of defects. Then, annihilation of defect pairs of different charge restores nematic ordered regions which may then undergo further instabilities. 
In passive liquid crystals the coupling between the order parameter and the flow  has  significant effects on the motion of defects, generating a more intense flow around  positively charged defects than for negatively charged ones~\cite{Thot2002}. This phenomenon is still present in active liquid crystals, as suggested by the quadrupolar flow centered around the $+1$ defects in the inset of Fig.~\ref{fig:stability_contractile}. 
The presence of activity gives rise to an even richer phenomenology.
Full defects hydrodynamics in $2d$ polar active fluids was studied by lattice Boltzmann
simulations with a hybrid scheme in~\cite{Elgeti2011}.   
In this paper it was found that extensile activity favors spirals and vortices, like the defect highlighted by a  square in Fig.~\ref{fig:stability_extensile}d, while  contractile activity favors aster-like defects in the polarization field like the ones boxed in Fig.~\ref{fig:stability_contractile}. 
Defect-defect interactions have also been studied.
In a contractile fluid two asters repel each other reaching a  steady state with a fixed distance, that increases at larger activity. In the extensile case two asters turn into two rotating spirals, leading to a final state where the rotation continues at approximately constant rotational velocity. For low activity the angular velocity increases with $\zeta$, while above a critical value an oscillatory behavior is observed, where half clockwise rotation is followed by half anticlockwise one, resembling  the previous cited  oscillatory and then chaotic behaviors appearing in spontaneous flow transition for high activity\footnote{The possibility of having rotating clusters of  swimmers has  been also considered in the context of models of  active brownian particles~\cite{petrelli2018}. It has been shown that aggregates of active dumbbells, at sufficiently high activity, rotate with angular velocity inversely proportional to the average radius of the cluster. }~\cite{marenduzzo2007}.


Confinement of polar/nematic pattern triggers the formation of defects, thus their dynamics is  found to be particularly rich in drops of active fluids. This is not surprising given the complexity of  defect topology known for  passive liquid crystals~\cite{rassegnadifettiliquidcrystalsgocce}. 
 So far, numerical studies  available are mostly concerned with two-dimensional systems, and we will give a review in the next Section. 

\section{Self-propelled droplets and active emulsions}
\label{sec:self_prop}

Geometrical constraints and spontaneous flow are known to significantly alter the hydrodynamics of active fluids. LB simulations show, for example, that, if a sample of active gel is sandwitched between two parallel plates, the onset of a spontaneuos flow crucially depends on the anchoring of the director field at the walls~\cite{marenduzzo2007}. If the active fluid is confined in a spherical geometry, such as that of a droplet, the physics is even richer. Joanny and Ramaswamy~\cite{joanny2012}, for instance, have theoretically demonstrated that the active stress can generate flows driving the spreading process of a droplet, whose shape is significantly affected by the nature of topological defects.


Recent experiments have shown that the presence of an active fluid can favour droplet self-propulsion through several mechanisms, based on chemical reactions~\cite{zhang2015,agladze1984}, spontaneous symmetry breaking and Marangoni effects~\cite{sanchez2012,guillamat2018,herminghaus2014,fadda2017_2}. Such experiments motivated numerous theoretical studies, with the aim of developing minimal  models capable of capturing features of particular relevance in biology (as active droplets can mimick the spontaneous motion of cells~\cite{bray2000,giomi2014,tjhung2012,blow2014}) or in the design of bio-inspired materials~\cite{sanchez2012,herminghaus2014}. In this section we review the theoretical studies in which the physics of an active gel under spherical confinement is described in terms of the continuum Eqs.~\eqref{nav}-\eqref{eq:adv_diff}-\eqref{conc_eq}, numerically solved by means of LB simulations.

Giomi et al.~\cite{giomi2014} have demonstrated that those equations can describe the spontaneous division and motility of an active nematic droplet surrounded by an isotropic Newtonian fluid, a situation closely resembling the functioning of living cells. Motility results from the combination of the initial elongation of the droplet (whose stability is guaranteed by a balance between viscous and pressure forces exherted by the flow on the droplet interface and the resistance due to interfacial tension) and the instability of the active fluid to splay deformations. When the active stress is sufficiently high, splay instability dominates and triggers the formation of a spontaneous flow promoting self-propulsion. Low values of the surface tension disrupt the force balance, and lead to overstretching and eventually division of the droplet.

\begin{figure}
\centering
\resizebox{.48\textwidth}{!}{
 \includegraphics{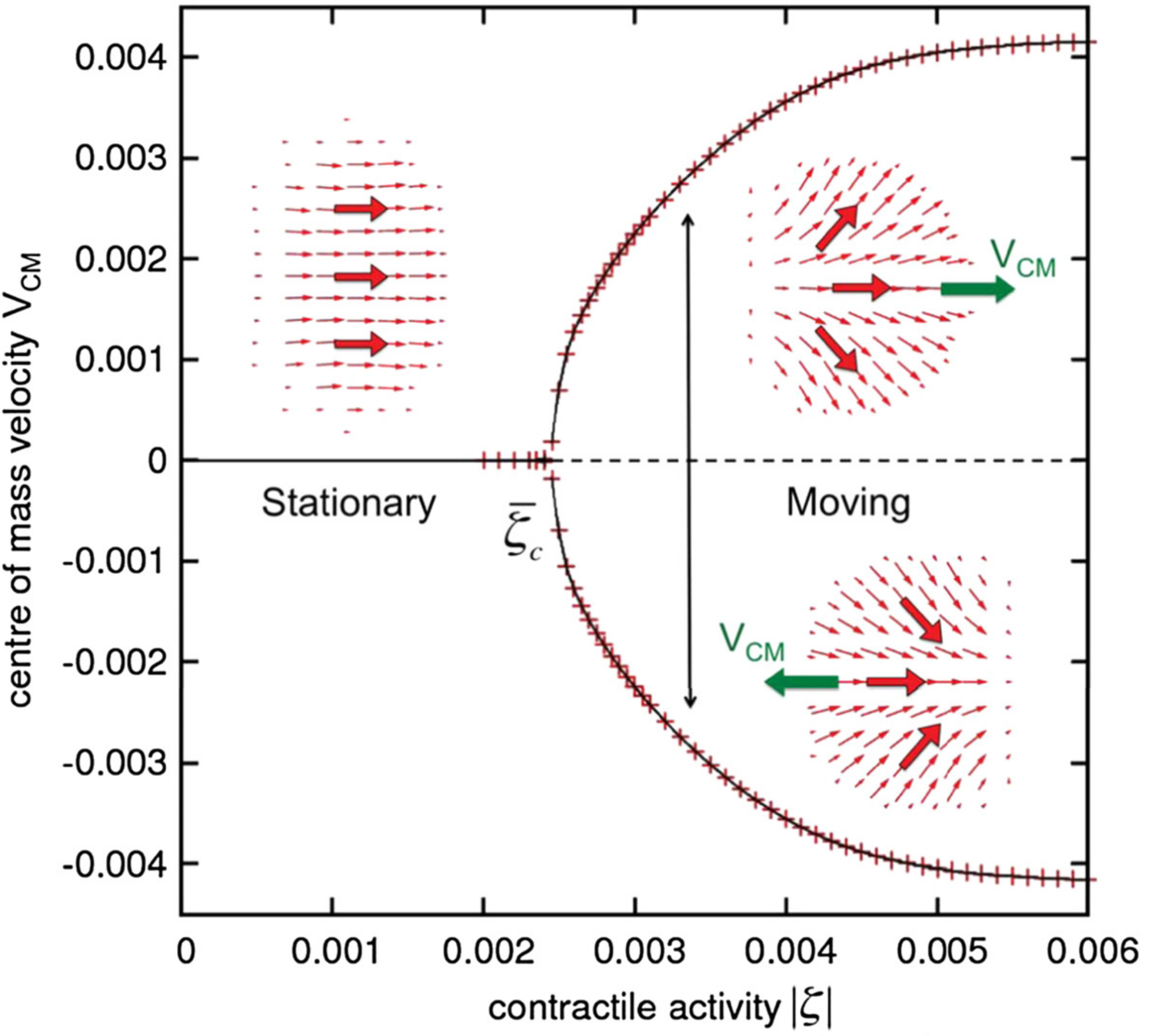}}
\caption{Stationary centre of mass velocity vs activity strength, as reported by Tjhung \textit{et al.}~\cite{tjhung2012}, for an active contractile polar droplet. If $|\zeta|<\bar{\zeta}_C$ the droplet is stationary (no motion), while for $|\zeta|>\bar{\zeta}_c$ the droplet acquires motion along the direction imposed by large splay deformations. In the figure  red arrows indicate the direction of the polar field while green arrows the direction of motion of the droplet. (Image courtesy of D. Marenduzzo.)}\label{fig:cates}
\end{figure}

The generic mechanism of motility of active droplets has been also investigated analytically in~\cite{Whitfield2016} and numerically in~\cite{tjhung2012} by using the polar theory. Here it is shown that the self-propulsion of a contractile droplet stems from the spontaneous symmetry breaking of polarity inversion symmetry, which, in turn, triggers the formation of intense splay distortions that act as a source of kinetic energy leading to motion. The symmetry breaking can be viewed as a continuous nonequilibrium phase-transition from a non-motile to a motile state observed for a sufficiently high activity (Fig.~\ref{fig:cates}). This model offers a simplified example of a cell, as a droplet containing an actomyosin solution, and suggests that motility can arise solely because of myosin contractility, rather than from its combination with actin polymerization, as often occurs~\cite{bray2000}. 
The same hydrodynamic theory has been used to model the physics of an active polar droplet confined on a solid substrate, a situation resembling that of a crawling cell~\cite{nature2015}. Here the droplet is made of an active polar fluid with contractility throughout, but actin polymerization confined in a layer close to the substrate. Such minimal description has been proven capable of capturing shapes (such as the lamellipodium~\cite{bray2000,tjhung2015}) andan exhibiting self-motile properties. More specifically, planar anchoring induces rotational motion in contractile droplets, while normal anchoring has the same effect in extensile ones. In both cases rotation stems from an active torque, due to a pair of bulk elastic distortions whose forma motility regimes (such as oscillatory modulations of shape~\cite{bray2000,tjhung2015}) by only considering a few key ingredients, i.e. actin polymerization, myosin contractility and interface anchoring.

More recently Fialho \emph{et al.}~\cite{fialho2017} investigated the effects of the interface anchoring of the polar field on $2d$ active polar droplets, by means of LB simulations. They found that, for large enough activity, droplets subject to strong anchoring start to spontaneously rotate, rather than exhibiting self-motile properties. More specifically, planar anchoring induces rotational motion in contractile droplets, while normal anchoring has the same effect in extensile ones. In both cases rotation stems from an active torque, due to a pair of bulk elastic distortions whose formation is controlled by a careful balance between activity and interface anchoring conditions.

As already mentioned in Section~\ref{sec:2c}, the active gel theory has been further extended to include chirality, as this is supposed to play a relevant role in many biological systems~\cite{Zhou2014,Naganathan2014}. Actin filaments, for example, are generally twisted in a right-ended direction~\cite{Depue1965} so that, while pulling, myosin motors will tend to rotate them,  creating a torque dipole. The effects of microscopic chirality on the motility of $3d$ contractile  active polar  droplets have been considered in~\cite{Tjhung2017}. Here it has been shown that the combined effect of an (achiral) active force dipole with a (chiral) torque dipole gives rise to a rich dynamical behavior, including helical swimming, run-and-tumble motion and oscillatory swimming, the latter also observed in~\cite{nature2015}. In general, while linear swimming occurs when achiral contractility is dominant, helical trajectories are observed when chiral activity becomes relevant. Such dynamic features may resemble those observed in single-celled protozoa, such as {\it Toxoplasma gondii}\cite{leung2014}.

So far we have analyzed the  behavior of a single active droplet and the mechanisms leading to its self-motility properties. Active emulsions are another challenging class of systems with many potential novel  applications. 
 Examples might range from a two-fluid emulsion encapsulating active matter used for drug delivery to a biomimetic material, such as a soft tissue made up of highly-packed active droplets capable to resist to intense deformations.
Active emulsions may even play a fundamental role in overcoming current major challenges affecting the design of active devices, such as controlling fuel arrival and waste removal~\cite{sanchez2012,herminghaus2014}; this is indeed an essential feature  to create sustainable active materials, namely systems keeping their active properties over long periods of time. 
An active emulsion has been recently realized in the experiments by 
Sanchez \emph{et al.}~\cite{sanchez2012}, 
where  a bundle of active nematic liquid crystal network (more specifically microtubule bundles activated through kinesin motor proteins) is squeezed at the interface of an aqueous droplet emulsified in an oil background. This paves the way towards microscopic confinement of active matter.

LB simulations would be the ideal tool to numerically investigate the hydrodynamics
of a system of many active droplets and active emulsions.
One may use, for instance, a multiphase approach in which
each droplet is described by a different field.
A different possibility is to consider negative values of $k$ in the model presented in Eq.~\eqref{free-energy-mixture}.
This models the presence of a surfactant in a binary mixture, as discussed in Section~\ref{sec:2d}, 
thus favouring 
emulsification
of the two phases. By confining polarization in one of the phases, the active 
behavior is then restricted to a limited portion of the system.
Following this approach, the effects on morphology and hydrodynamics have been studied in a symmetric emulsion, 
made of an equal amount of active and passive fluid, 
by Bonelli \emph{et al.}~\cite{bone2017}, 
while Negro \emph{et al.} \cite{negro2018} focused on highly asymmetric active emulsions, in which the active
phase represents only the $10\%$ of the overall composition. 
Passive lamellar and hexatic orders, respectively for the symmetric and asymmetric cases, are enhanced by small contractile activity, while at more intense active dopings, an emulsion
of passive droplets in an active background appears for the symmetric case.
In the extensile case, a morphological transition from  a stationary state  to a state with  chaotic 
patterns is mediated by the appearence of rotating droplets at intermediate
active doping, regardless of the amount of active material with respect to passive phase.

Spontaneous motility in emulsions where passive droplets are immersed in an active background has also been studied in~\cite{softmatter2014}, via LB simulations. The crucial mechanism leading to generation of active flows is driven by anchoring of polarization on the droplet surface. When this is homeotropic, the formation of a nearby hedgehog defect (due to the conflict anchoring with the orientation of the polar field outside the droplet) is able to drive an active flow which propels the passive droplet forward. If the gel is extensile, motility can even occur with tangential anchoring, favoured by bending deformations in the surrounding active fluid.


\begin{figure}
\centering
\resizebox{.48\textwidth}{!}{
 \includegraphics{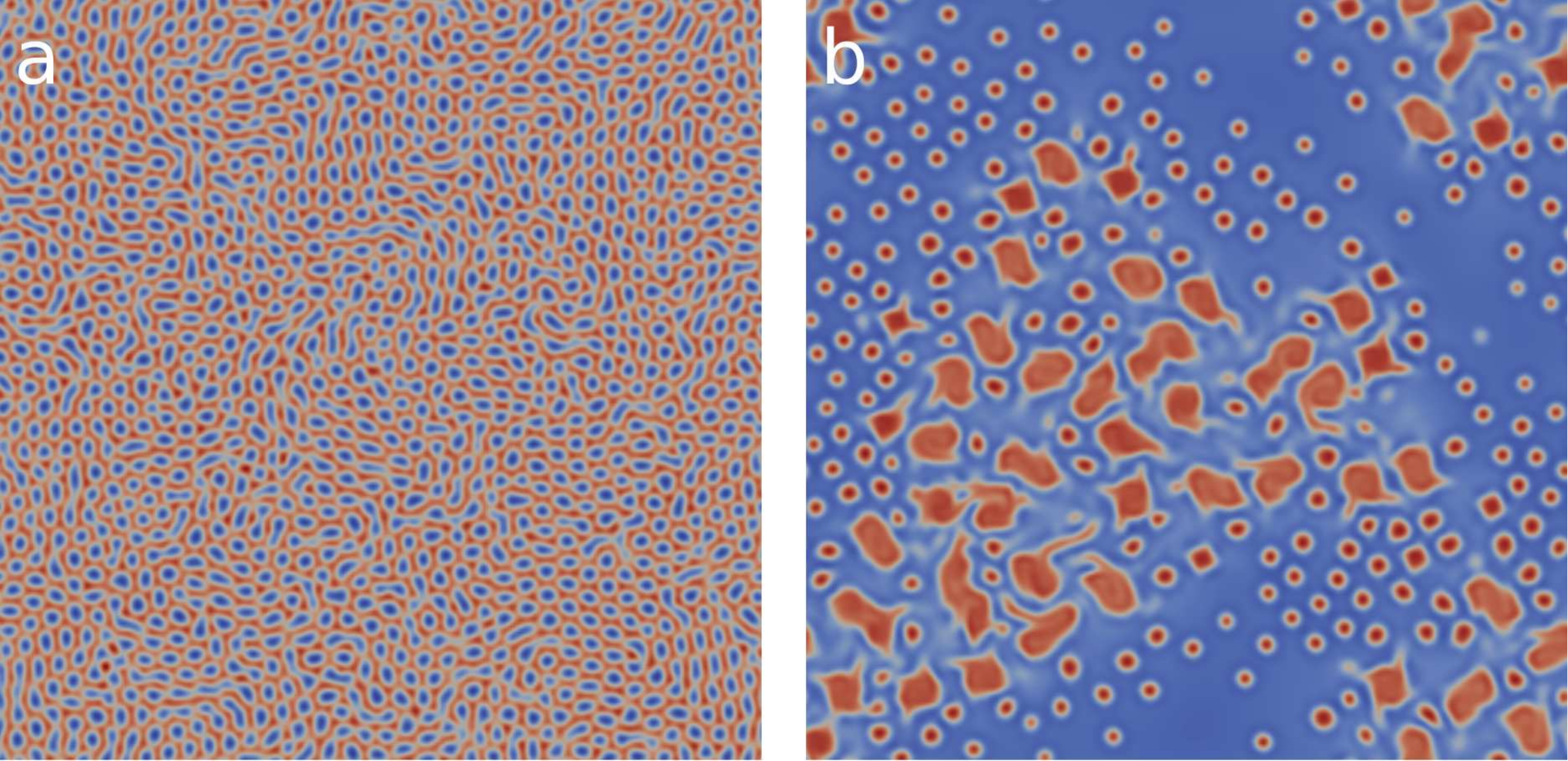}}
\caption{Contour plots of concentration of active material in polar binary mixture for contractile (a) and extensile (b) systems respectively obtained by Bonelli \emph{et al.}~\cite{bone2017} and Negro \emph{et al.}~\cite{negro2018}. 
Activity strongly influences the equilibrium morphology, depending also on the relative amount of active (red) and passive (blue) regions of the mixture.
In a symmetric mixture, where half of the overall system is active, at strong contractile activity ($\zeta=-0.02$), an emulsion of passive droplets in an active background emerges, while moderate extensile doping ($\zeta=0.008$) in an asymmetric mixture ($10 \%$ of active material) leads to the formation of rotating asters.}
\label{fig:binary_lamellar}
\end{figure}

An important field, still in its infancy, is the study of chiral systems. Despite the motility effect of active torque has already been analyzed, as we previously discussed, intrinsically chiral active systems have gain interest only recently~\cite{metselaar2018,carenza2019}.
This route deserves a much deeper investigation: firstly because most biological extracts (such as acto-myosin systems, microtubule bundles, DNA, \emph{etc.}) exhibit indeed an intrinisc cholesteric arrangement, often related to motility properties in living systems, then because the striking variety of competing metastable topologies usually observed at small pitches~\cite{sec2012,fadda2017} is potentially useful for the design of energy-saving active soft composites built from isolated motile cholesteric droplets.

Active droplets also have the potential to exhibit a rich rheological response under shear (from a glassy to a superfluid behavior~\cite{foffanoPRL,PhysRevE.83.041910,cates2008,PhysRevLett.97.268101}), making them suitable candidates for designing materials with new mechanical properties. A crucial question is the following: How the elastic and plastic response of an active emulsion is affected by active stress and external forcing? Decisive dynamic parameters are viscosity, shear rate, elasticity and droplet concentration. 
The rheological behavior of active fluids will be discussed in the next Section.

\section{Rheology}
\label{sec:rheology}
The flux generated by the presence of active constituents in a suspension can deeply interact with external flows modifying the rheological response.
Several experimental studies have shown, in some regime,  a decrease in viscosity for pusher (\textit{Bacillus subtilis} and \textit{Escherichia coli})  and an increase for puller (\textit{Chlamydomonas reinhardtii}) suspensions. However, a comprehensive overview of these studies is found to be far richer and more complex~\cite{saintillanreview2018}.  

Sokolov \textit{et al.}~\cite{sokolov2009} confined \textit{Bacillus subtilis} bacteria in a quasi-$2d$ liquid film. Their experiments gave viscosity estimates below that of the solvent at low concentrations. At higher densities the viscosity was found to increase and exceed that of the solvent. 
Precise measurements using \textit{Escherichia coli} bacteria were obtained by Gachelin \textit{et al.}~\cite{gachelin2013} and 
Lopez \textit{et al.}~\cite{PhysRevLett.115.028301}. In the first case  bacteria are indeed found to decrease the viscosity below that of the solvent for low shear rates. However, the relative viscosity, given by  the ratio of viscosity of the solvent in presence and in absence of suspended bacteria, increases with the shear rate, reaching a maximum above unity before shear thinning again. 
Lopez \textit{et al.} performed their experiments in a conventional Taylor-Couette rheometer specifically build to handle low torques and viscosity. 
The relative viscosity 
displayed a trend with a low shear-rate plateau with relative viscosity less than one, 
followed by shear thickening. Increasing the density of bacteria was found to decrease the value of the plateau towards zero,  meaning an almost vanishing value for the relative viscosity. 
The measured viscosity, within experimental uncertainty, actually vanishes in oxygenated suspensions for volume fractions  
 greater than $1\%$.
 This surprising superfluid-like behavior should not be seen as a violation of thermodynamic principles, as the bacteria consume chemical energy.  It instead suggests that the viscous dissipation in the flowing suspension is macroscopically balanced by the input of energy coming
from swimming, thus allowing for a sustained flow without any applied torque. 
Lopez \textit{et al.} 
 also analyzed the transient stress response upon start-up and cessation of shear flow. 
Turning on shear, they found a sharp increase in viscosity, followed by an exponential decrease towards  steady values. 
When shear  is switched off, the shear stress undergoes a drastic reduction leading to negative values for the apparent viscosity before relaxation to  a null-stress steady state.

The case of puller swimmers has been addressed by Rafai \textit{et al.}~\cite{PhysRevLett.104.098102}, who measured the viscosity of a suspension of \textit{Chlamydomonas reinhardtii} in a Taylor-Couette flow. The suspension viscosity was always found to exceed that of the medium and to increase together with concentration. The effect of activity was spotted comparing the results obtained with suspension of dead and living swimmers; in the latter case the suspension is always found to be significantly more viscous than the one with the same volume fraction of dead cells. Weak shear thinning was also reported for high shear rates. A direct comparison of three different types of swimmers was recently performed by McDonnell \textit{et al.}~\cite{C4SM02742F}, considering \textit{Dunaliella tertiolecta}, \textit{Escherichia coli}, and mouse \textit{spermatozoa}. In each case they compared alive and dead cells.
The viscosity measured were always above those of suspending medium, with living cells algae suspensions  more viscous than dead ones, while both \textit{Escherichia coli} and spermatozoa appeared less viscous when alive than dead.

\begin{figure}
\centering
\resizebox{.48\textwidth}{!}{
 \includegraphics{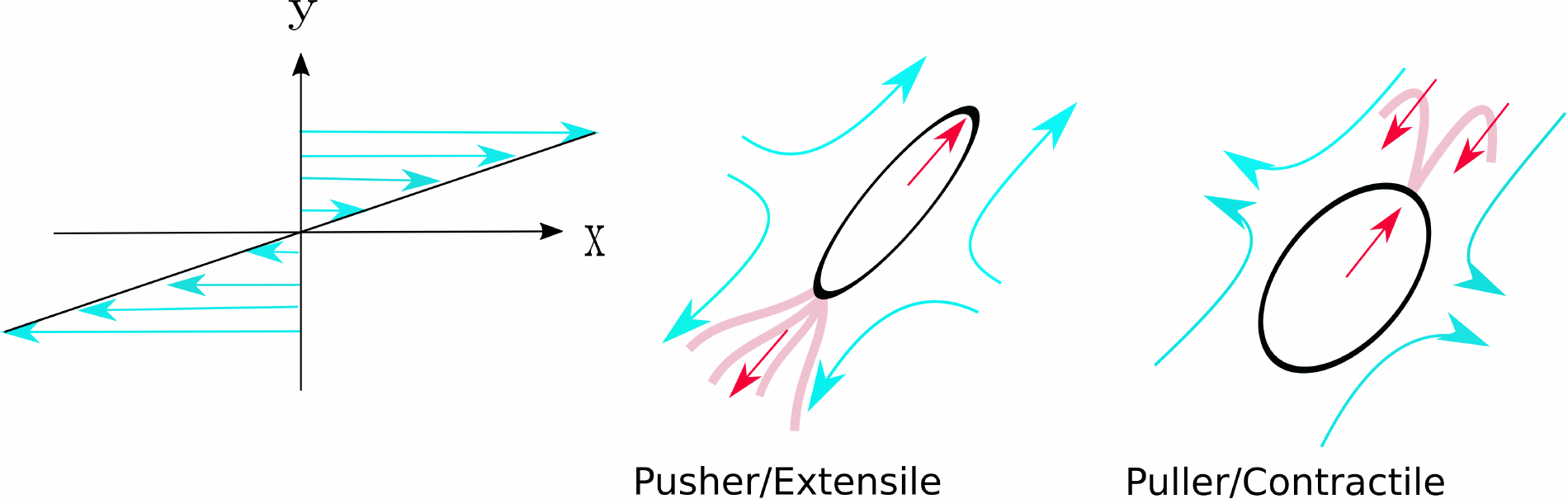}}
 \caption{  Sketch illustrating  the basic mechanism of the rheological response of flow aligning swimmers ($|\xi|>1$) 
 under a uniform  shear flow  (cyan arrows in the left panel). Active forces exerted by the swimmers (red arrows) generate a disturbance flow (cyan arrows) that enhances the applied flow in extensile systems (center panel). The opposite happens for contractile ones 
(right panel).}\label{fig:shear}
\end{figure}

The basic mechanism for viscosity modification in  a suspension of microswimmers was first explained 
by Hatwalne \textit{et al.}~\cite{hatwalne2004}.
It is sketched in Fig.~\ref{fig:shear}. Under an applied shear profile (left panel) flow-aligning extensile systems (middle panel) enforce the applied flow, decreasing the viscosity of the suspension, while the flow generated by contractile systems opposes to the external flow thus resulting in an increase of the apparent viscosity.
They derived the coarse-grained equations governing the rheological behavior 
with several predictions later validated by simulations and experiments.
Applying spatially uniform oscillatory shear 
 with  frequency $\omega$ in the $x-y$ plane 
a linear analysis~\cite{hatwalne2004}  gives for  the $x-y$ component of the stress tensor the expression 
\begin{equation}
\sigma_{xy}=\left[\eta+\frac{\left(L_1-f\right)\xi}{-i\omega+\tau_Q^{-1}}\right]v_{xy}\label{eq:shear_marchetti}
\ ,
\end{equation}
with $v_{xy}=\frac{1}{2}(\partial_x v_y+\partial_y v_x)$,  $L_1$ the nematic elastic parameter in the single
 constant approximation (see Table~\ref{tab:1}),  $\tau_{Q}\sim 1/L_1$  the relaxation time of the order parameter, $\eta$ the viscosity of the suspending medium, and $f$ the active force strength.
We see that viscosity, which is obtained taking the limit $\omega\rightarrow 0$, 
depends on the sign of $f$ and also on the geometry of the microscopic constituents, 
represented by the parameter $\xi$ 
($\xi>0$, $\xi<0$ and $\xi=0$ for rod-like, disk-like and spherical particles, respectively). 
Viscosity is lowered in a suspension of rod-shaped particles by extensile activity~($f>0$), and is increased by contractile activity~($f<0$). For disc-like particles the opposite holds.
Liverpool \textit{et al.}~\cite{PhysRevLett.97.268101} did a similar study for polar active gels, 
obtaining  constitutive equations for the stress tensor in the isotropic phase and in phases with liquid crystalline order. The stress relaxation behavior in the various phases was discussed.
Polar suspensions under uniform shear flows were further studied  by Giomi \emph{et al.} in~\cite{PhysRevE.81.051908},
 analyzing the  rheological behavior  in relation to the shape of the swimmers and their activity.
The analytical treatment of the  stress-strain  linear regime demonstrates that activity
lowers the linear viscosity of both extensile, rod-shaped particle and contractile, disk-shaped particle suspensions, while
it increases the viscosity of contractile, rod-shaped
particle and extensile, disc-like particle suspensions.
For large  activity the rheological response is not expected to be  well described by  linear analysis.
Thus, Giomi \emph{et al.} also integrated equations by means of Euler methods assuming uniformity in the flow direction. 
Depending on the value of activity and strain rate, they suggested the possibility of more exotic scenarios including  hysteresis, yield-stress and non-monotonic stress-strain behaviors, and a superfluid phase with vanishing viscosity. 
The first two regimes have not been confirmed so far by other  experiments or  simulations.

In~\cite{Guo201722505} a minimal phenomenological model to explain the occurrence of vanishing viscosity  was proposed. This is motivated by the observation of unusual symmetric shear banding profiles in bacterial suspensions, and  takes into account  detailed stress balance between  local viscous shear stress and active stress, and  ergodic sampling of asymmetric profiles. 
Shear gradients in conventional complex fluids, such as worm-like micelle solutions and colloidal suspensions~\cite{olmsted2008},  are always positive, while in active fluids, due to the local energy injection, one can also  have regions of the system with  velocity profiles characterized by negative slopes. According to the proposed model, a sheared active fluid samples all allowed shear-banding configurations, leading to a symmetric yet nonlinear shear profile.

The possibility to have states with  negative viscosity, and the formation of macroscale unidirected flows is of  gargantuan importance especially for applications, aiming at the design of active fluid  powered motors~\cite{vizsnyiczai2017}.
These features have been recently explored in two papers.
Slomka \textit{et al.}~\cite{Slomkashear2017} numerically investigated the flow pattern formation and viscosity reduction mechanisms by considering the effective  model presented in Section~\ref{sec:2} (Eq.~\eqref{eqn:swifthonenbergh_st}).
They found, in addition to almost-vanishing viscosity phases,  evidence of spontaneous formation of persistent unidirectional flow.
In another paper,  Loisy \textit{et al.}~\cite{liverpool2018} predicted negative viscosity regimes studying
a model where  the polarization dynamics is coupled to  the flow field obeying the Stokes equation. Hydrodynamic 
equations are solved by  finite difference  methods and parameters  of the model are matched with  experimental data, 
 allowing  a quantitative agreement with experimental behavior.



Lattice Boltzmann methods have been also largely used to analyze numerically  the behavior  of active suspensions under shear.
Actually, the first use of the term \textit{superfluidic} for describing  the behavior of an active fluid is found 
in the paper by Cates et al.~\cite{PhysRevLett.101.068102} based on hybrid LBM simulations. 
Here the rheology of a  slab of active gels is studied close to the isotropic-nematic (I/N) transition.
For contractile gels and \textit{free boundary conditions} (case in which the director is free to rotate at the boundary planes) a divergence of the apparent viscosity was found at the I/N spinodal. 
Close enough to the spinodal, extensile gels  enter a zero-viscosity phase of N/N shear bands, with two regions of well defined nematic order and distinct shear gradients~\cite{marenduzzo2007}.
In the nonlinear regime, both extensile and contractile materials show non-monotonic effective flow curves, with details strongly dependent on initial and boundary conditions. 
Passive nematic liquid crystals ($\zeta=0$) exhibit I and N bands of high and low viscosity~\cite{fielding2004}. Activity affects this behavior, widening (extensile) or narrowing (contractile) the range of shear rates for which bands are stable. 

Other LBM studies on rheology of active fluids have been done by the group in Edinburgh. 
In~\cite{PhysRevE.83.041910}, a $2d$ extensile nematic gel was studied in more detail, comparing results  between planar anchoring 
(\textit{fixed boundary conditions}) and \textit{free boundary conditions}. 
In absence of  applied shear, convection roll patterns would appear due to spontaneous flow, as discussed in Section~\ref{sec:spont_fl}. These structures, in the case of free boundary conditions  and low activity, are destroyed even at low shear rates.  
The flow pattern becomes unsteady and  spatially nonuniform. In this case shear stress takes both positive and negative values, and is characterized by large fluctuations, thus making difficult  a quantitative evaluation.
Increasing  activity at low shear rates, the system displayed   viscosity values close to zero, with less stress fluctuations.
At high shear rates the laminar flow is reestablished. 
For very high activity the chaotic patterns observed without applied shear are only slightly perturbed at small strain rates, with the  appearance of a noticeable upward curvature in the  curve of the stress as a function of the  shear rate.  
In the case of fixed boundary conditions, at low activity and shear rates the flow curves display a linear regime, while for high shear rates or high activity the results are in line with previous case.
 An attempt to link micro- and macro-rheology of active nematics was done in the work 
by Foffano et al.~\cite{foffano2012}.
  To do so, they compared macroscopic shear numerical experiments with the results of  
simulations where a spherical probe particle was dragged through the active fluid.
As already discussed, in sheared nematics the effective viscosity measured by bulk rheology depends on anchoring conditions and also on the system size $L$. 
The effective viscosity increases with $L$ in contractile  fluids and decreases with $L$ in extensile ones. 
Normal anchoring  enhances the apparent viscosity with respect to that measured at zero activity, both for contractile and extensile systems. 
Micro-rheology results agree with the macro-rheological characterization, and give additional informations on the role of local fluid structures on viscosity measurements.
As a first important result, when a spherical particle is dragged through the active fluid, the drag force does not depend linearly on the probe radius, violating the Stokes law.
When the director field is anchored tangentially to the particle surface contractility increases the drag, while extensile activity reduces it  in line with the macro-rheology results. 
In the case of normal anchoring at the probe surface, pulling orthogonally to the far-field director leads to results similar to the tangential anchoring, while dragging the particle along the director leads the particle to move faster in  contractile  than  in extensile nematics.  

Rheology of active suspensions has been found so far to be rich of striking properties with promising perspectives towards future innovative applications.
However, despite the basic mechanisms behind negative viscosity states and superfluidic regime
are understood, 
this subject still deserves further analysis. 
Before such systems can be actually implemented for the design of technological applications, it is thus necessary to clarify some fundamental points such as stability of superfluidic and/or negative viscosity regimes as well as which conditions are necessary to keep the system under control. Numerical simulations, and especially  lattice Boltzmann models, may play a fundamental role in further developments to this extent.


\section{Active turbulence}
\label{sec:turb}
Spontaneous flow in active suspensions  evolves, for sufficient strength of the active component, 
into complex patterns for the velocity field that, at qualitative level,
looks chaotic, resembling that of simple fluids in the passage from laminar to turbulent behavior, as suggested by streamlines of the velocity field in Fig.~\ref{fig:defects_simulation}.  
This observation, as mentioned in the Introduction, first came from experiments conducted on highly concentrated bacterial suspensions, where the velocity field was found to develop coherent vortical structures as well as transient jets at high velocities~\cite{kessler1997,mendelson1999,dombrowski2004}.
Jet-like fluid motion was found in~\cite{dombrowski2004}, with 
the speed of the jets $ > 100 \mu m/s$, significantly  larger than the speed of an individual
bacterium 
($ \sim 10 \mu m/s$). 
The fluid pattern observed on length scales of $100-200 \mu m$  is reminiscent
of a von Karman vortex street, which arises in simple fluids
when the Reynolds number is much larger ($\sim 50$) compared with that of a single bacterial cell ($\sim 10^{-4})$.
This justifies the terms \emph{bacterial turbulence} and \emph{active turbulence},
 used to denote this kind of flow at low Reynolds numbers~\cite{dombrowski2004}.
Similar behaviors have also been reported in experiments on suspensions of both puller and pusher swimmers~\cite{dombrowski2004,sokolov2007, lauga2012}, as well as microtubules bundles~\cite{sanchez2012} or acto-myosin systems and microalgae~\cite{drescher2010}. 

A theoretical description of this complex behavior was first presented by a phenomenological model based on the Stokes equations, adapted to take into account the swimming mechanism of bacteria, each acting as a dipole stress on the fluid~\cite{wolgemuth2008}. This model, solving the equations in two dimensions by using realistic parameters, empirically reproduces the observed velocity field. 

\begin{figure}
\centering
\resizebox{.35\textwidth}{!}{
 \includegraphics{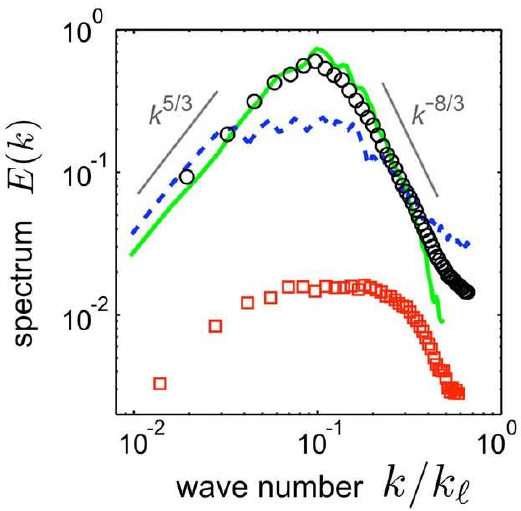}}
 \caption{Experimental and numerical turbulent energy spectra (Wensink \emph{et al.}~\cite{wensink2012}). $2d$ SPR simulations (blue dashed line), $2d$ continuum model (green line) and quasi-$2d$ experimental data of dense suspensions of \textit{Bacillus subtilis} (black circles) agree on power-law scaling at small and large wave numbers. Energy spectra from full $3d$ bacterial systems (red squares) qualitatively agree with the bidimensional case, but exhibit a plateau region at intermadiate wavenumbers.}\label{fig:wensink}
\end{figure}

The first  attempt of a quantitative analysis of turbulent-like behaviors was done by Wensink \emph{et al.} in~\cite{wensink2012}, with a combination of experiments, simulations of self-propelled rods (SPR) and continuum theory aimed at identifying the statistical properties of self-sustained meso-scale turbulence in dense suspensions of \textit{Bacillus subtilis}.
Experimental and numerical data coming from SPR simulations for the energy spectrum, exhibit power-law scaling regimes better observed in two dimensions for both small ($k^{5/3}$) and large ($k^{-8/3}$) wave numbers (see Fig.~\ref{fig:wensink}); however, the power-law in the inertial energy  range differs from the characteristic $k^{-5/3}$ decay of $2d$ high Reynolds number turbulence, as it will be discussed later. 
Similar results  were  also found making use of the continuum
theory of Toner and Tu~\cite{toner1998}, supplemented with the Swift-Hohenberg stress tensor presented in Eq.~\eqref{eqn:swifthonenbergh_st}, containing higher order derivatives of the velocity gradient.
Experimental data in $3d$ look qualitatively similar, but show  
an intermediate plateau region absent in two-dimensional systems, to indicate the spreading of kinetic
energy over a wider range of scales.
The same model was used in a subsequent paper~\cite{dunkel2013}
to  consistently  reproduce  velocity statistics
and correlations  in a highly concentrated $3d$ suspension of \textit{Bacillus subtilis}, while in~\cite{dunkel2013_2} the  linear stability analysis was presented.

Beside bacteria, eukaryotic cells with self-motile properties also show  self-sustained flows. 
In~\cite{creppy2015} a suspension of  spermatozoa was found to develop a directed energy cascade characterized by a power-law scaling $k^{-3}$ at high wave numbers, a behavior that is also found  in $2d$ turbulent flows and is due to non linear transfer of enstrophy, rather than energy~\cite{Boffetta2012}.
Recently, active turbulence features were also found in a system 
of self-assembled ferromagnetic Nickel microparticles dispersed at water-air interface, while subjected to an external oscillating 
magnetic field~\cite{kokot2017}. 
Self-assembled spinners locally inject energy in the solvent via
generation of local vortex flows, thus leading to the subsequent
energy cascade towards larger scales.
The hydrodynamic state is characterized by a power-law decay $k^{-5/3}$ of the energy spectrum at low packing fractions, but when this increases, steric and magnetic interactions become important so that even the exponent starts to deviate, while the system undergoes a transition toward a new phase where spinners are replaced by non-rotating clusters.
The experimental observations
 were qualitatively reinforced through numerical simulations
of disks suspended in $2d$ geometries whose dynamics was solved using  
the multiparticle collision approach~\cite{gompper2009}.


The model of Wensink \emph{et al.}~\cite{wensink2012}, previously discussed, has the advantage of simplicity with respect to the active gel theory of Section~\ref{sec:2e}.
In that model the scale of variation for the velocity field is set phenomenologically, by varying the ratio between coefficients in Eq.~\eqref{eqn:swifthonenbergh_st}, thus the explicit dynamics of the active components is  neglected. 
Beside being pioneristic in the characterization of  bacterial turbulence,  whithin Wensink's approach it is not possible to derive any information on the relations between  the velocity, the active component fluctuations,  and  the order parameter patterns with its inherent defect dynamics. 
A more complete analysis of these features  can be performed  by considering the nematic gel theory of Section~\ref{sec:2e} as first done  in~\cite{giomi2012}.
Here, decay spectra for kinetic energy and enstrophy were calculated for a relatively small $2d$ system with periodic boundary conditions. Making use of finite difference	 methods to solve the Beris-Edwards and the Navier-Stokes equations of Section~\ref{sec:2e}, Giomi carefully analyzed the statistics of vortices in~\cite{giomi2015}, finding that their areas are exponentially distributed and giving  insights into the relations between the topological properties of the nematic director and the geometry of the flow. 
In particular he found that the chaotic regime is due to a feedback mechanism between the advection of nematic $\pm 1/2$ disclinations and vortical flows, whose structure is in turn closely related to presence of disclinations in their neighborhood, as demonstrated in Refs.~\cite{Thot2002,giomi2014_2}.  The number of both $+1/2$ disclinations and vortices is found to be linearly proportional to the strength of the active injection, while  creation and annihilation rates of oppositely charged defects exhibit a quadratic dependence.  
Defects are advected
along the edge of vortices at an angular velocity that is 
linearly dependent on the activity. 
As a consequence, the annihilation lifetime of disclinations is inversely proportional to the angular velocity hence, to the activity. 
The energy and enstrophy spectra were respectively observed to decay as $k^{-4}$ and $k^{-2}$, in the limit of high wave numbers,  with no significant dependence neither on the value nor on the kind of activity (extensile or contractile).

The importance of relation between defects and vorticity has also been highlighted in Ref.~\cite{putzig2016}, where a phenomenological continuum theory for overdamped active nematic liquid crystals is introduced. Three different non-equilibrium steady states are found according to the mutual effects of active torques ($ \sim \nabla \times \nabla \underline{\underline{\mathbf{Q}}}$) and active convection: when the first is dominant over the second, an array of nematic defects is developed, where the comet tails of $+1/2$ disclinations are aligned throughout the system; when, instead, active convection dominates over torques, an undulated nematic free-of-defect state is formed. Nevertheless, if these contributions are equally important, a turbulent nematic state emerges, being characterized by a sharp increase both in vorticity and in the number of defects.
 
The growing interest in active turbulence over the last decade is motivated by some peculiar aspects: firstly, fluid active systems mostly evolve in the low Reynolds number regime, so that the development of a turbulent state due to hydrodynamic energy transfer is highly surprising; secondly, the search for an universality class for this phenomenum is extremely challenging. Moreover, even the physical mechanism that drives active systems toward the turbulent state is remarkable: energy injection takes place on  small scales $l$ of the same order of the length of microswimmers, setting an upper bound for the spectral range of energy injection $k_l \sim 2\pi/ l$, while the typical wavenumbers $k$ over which the turbulent flows take place, namely those regions where energy spectra exhibit power-law decay, are such that $k/k_l <1$. Hence, energy, that has been injected by the active components on microscales, is transported on much larger scales, through complex mechanisms. This is somehow reminescent of two-dimensional  classical turbulence in which energy pumped into the system by an external force at a certain scale $l_f$ is not dissipated by viscosity, but transferred to larger scales giving rise to an \emph{inverse energy cascade}~\cite{Boffetta2012}. By requiring a scale independent energy-flux, one finds the Kolmogorov solution for the energy spectra $\sim k^{-5/3}$, in the inertial range $k \ll k_f=2 \pi /l_f$. Yet another inertial region is expected in the direct-cascade range $k \gg k_f$, where the requirement of constant enstrophy flux gives a solution for the energy spectrum $\sim k^{-3}$.

In spite of the efforts taken up to now, a complete characterization for the variety of behaviors observed in the scope of bacterial turbulence is still lacking. Indeed, beside the qualitative similarity between bacterial and classical turbulent flows, it is evident, from the wide phenomenology presented before, that arguments valid for the classical high-$Re$ turbulence do not hold for active matter systems, because of the plurality of behaviors observed, nor any alternative theory has already been able to fully provide a consistent explanation of such behavior.

To fully characterize turbulent flows it is fundamental to clarify phenomena at the base of exchange of energy between different scales. For instance, as said above, in Kolmogorov theory of turbulence, energy is transported by means of the advective term leading to direct (inverse) energy cascade in $3d$ ($2d$) systems.
An insight into the mechanism due to energy injection and dissipation in active fluids has been outlined by Bratanov \emph{et al.} in~\cite{Bratanov2015}. In this work the spectral properties of the same model as in~\cite{wensink2012} are analyzed. A power-law energy spectrum was found at large scales (commonly addressed as \emph{energy-containing range}), even in absence of an inertial range, namely a region of constant energy flux. The presence of further non-linearities introduced in the model, with respect to the only advective term in the Navier-Stokes Eq.~\eqref{nav}, provides additional freedom to the energy exchange between different scales, due to forcing and energy consuming terms. Despite the model manages to replicate some phenomenological features of active turbulent systems, advective non-linear transfer plays a significantly role, that one would expect to be substantially negligible in low-Reynolds number environments. %
Such non-trivial hydrodynamic behavior has as  consequence that exponents of the energy power-law scaling  do not exhibit any universal character, since they depend on the rate and on the mean length-scale of energy injection.

\begin{figure}
\centering
\resizebox{.5\textwidth}{!}{
 \includegraphics{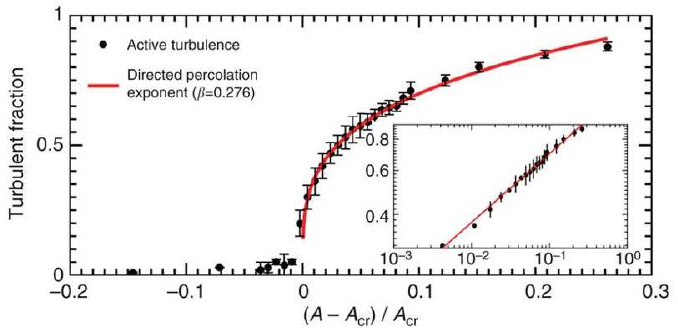}}
 \caption{Turbulent fraction versus active number $A$, presented by Doostmohammadi \emph{et al.} in~\cite{Doostmohammadi2017}. When a critical threshold $A_{cr}$ is exceeded, the turbulent fraction (black dots) grows with power-law $\sim (A-A_{cr})^{0.275\pm 0.043}$. Red line shows the critical universal behaviour for the  directed percolation. (Image courtesy of J.M. Yeomans.)}\label{fig:yeomans}
\end{figure}

Recently an insight into bacterial turbulence has been given
by Doostmohammadi \emph{et al.} in~\cite{Doostmohammadi2017} and by Shendruk \emph{et al.}~\cite{Shendruk2017}, where continuum equations for the active nematic theory of Section~\ref{sec:2b} were solved through a hybrid lattice Boltzmann method. Shendruk found that, by confining active nematics in a microchannel, the system exhibits various morphological behaviors depending on the intensity of active doping. At small activities they found an unidirectional spontaneous flow regime, followed by an oscillating laminar flowing state, as active doping is increased, that becomes turbulent for enough intense values of the activity parameter. One of the results presented in this work is the appearance of an intermediate state between the laminar regime and the turbulent one, named \emph{Ceilidh dance}. This represents the motion of paired disclinations in the nematic pattern moving along the channel, that, advected by the vortical flow, exchange partners, producing a dynamical ordered state that is reminescent of Ceilidh dancing.
The transition to the turbulent state is fully analyzed in the work of Doostmohammadi \emph{et al.} where vorticity puffs are used to characterize the transition. Unlike the inertial puffs that drive high-$Re$ systems toward a full turbulent state and are initiated by external forcing, in this case  puffs  are driven by the internal injection of energy on small scales.
Far from the turbulent transition, the turbulent fraction, namely the area fraction occupied by the active puffs,  is almost null, since they often vanish before splitting, but when the active doping becomes more intense while approaching the critical threshold, their lifetime grows and splitting becomes more likely.
As shown in Fig.~\ref{fig:yeomans}, if the critical threshold is exceeded the turbulent fraction grows with power-law $\sim (A-A_{cr})^{0.275\pm 0.043}$, where the adimensional active number $A=\sqrt{\zeta h^2/K}$, being $h$ the channel width and $A_{cr}$ the turbulent threshold. The exponent characterizing the transition closely matches the universal critical exponent of the directed percolation ($\beta=0.276$). Moreover, this is in turn very close to the exponent measured in Couette flows for inertial turbulence~\cite{Lemoult2016} (since 	in that case $\beta = 0.28 \pm 0.03$).
A similar characterization of $3d$ turbulence in a confined active nematics has been performed in~\cite{shendruk2018}, where the crossover from quasi-$2d$ to $3d$ turbulence is controlled and driven by the deformation and twisting of disclination lines.

We finally observe that many numerical simulations performed so far~\cite{Bratanov2015,wensink2012,linkmann2018} show energy transfer between different lengthscales due to hydrodynamic non-linearities, but they are not able to correctly reproduce the low-Reynolds number regime observed in experiments.
Moreover active turbulence is mediated by energy injection due to self-organization of active constituents, so that one could expect that active terms and elastic interactions may play some role in energy transfer, as it happens in elastic turbulence~\cite{Larson2000,Burghelea2006,morozov2007}, where the turbulent state results from elastic instabilities in viscoelastic (passive) media. No studies have been able to clarify this point yet.
The works of Shendruk and Doostmohammadi and Yeomans had shed light on the route to follow to fully charachterize the turbulent-like behavior, but still a number of questions has to be answered. 
Is active turbulence actually \emph{turbulence}? Which mechanism drives the system towards the chaotic state? How energy is exchanged between different lengthscales? Is it possible  to confine and tune the turbulent-like behavior?
These 
are matters for  further research.

\section{Particles and fluids}
\label{sec:coll}
In this last section we briefly review simulation studies investigating the modeling of particle suspensions (such as colloids) dispersed in a fluid, in the context of active matter and performed via Lattice Boltzmann methods.

Inspired by natural examples such as bacteria and sperm cells~\cite{marchetti2013}, a growing interest has been dedicated on producing synthetic particles capable of a similar autonomous motion~\cite{ebbens2016,patterson2016,elgeti2015,bechinger2016}. Systems studied in experiments range from bi-metallic nanorods, self-propelled via electrophoresis (by catalytically decomposing hydrogen peroxide~\cite{paxton2006,wang2006}), to spherical active Janus particles, 
acquiring motion through self-diffusiophoresis, electrokinesis~\cite{ebbens2014} and bubble nucleation~\cite{wang2014,gregory2015}.

The numerical studies  performed so far 
can be divided in two broad classes. The first one, on which large part of the recent research has been addressed, encompasses those made up of active particles dispersed in a passive fluid, while the second one, whose physics remains still largely unexplored, includes those in which passive particles are dispersed in an active medium.
  
Several models, pertaining to the first group, have been proposed to simulate microscopic swimmers interacting with a fluid at low Reynolds number. Ramachandran {\it et al.}~\cite{ramachandran2006}, for instance, described a swimmer by extending the method introduced by Ladd~\cite{ladd1994_1,ladd1994_2,ladd2001},  in which, for the first time, a solid particle was computationally modeled within a LB mesh. In this approach a solid particle can be created by initially defining a closed surface, represented by a set of boundary links joining neighbouring fluid nodes, located outisde and inside the surface. The former are fluid nodes, whereas the latter are solid nodes. At the surface,
stick boundary conditions, bouncing back the density of the fluid moving along a boundary link, are implemented. For computational convenience the interior of the particle remains fluid (i.e. lattice nodes inside and outside the particle are treated in the same way), a condition overall acceptable for a single-phase fluid as inertial effects are negligible\footnote{Extensions of the model, in which the internal fluid is excluded, have been also implemented~\cite{nguyen2002}. }. The active contribution is included by adding a force, exerted by the particle to the fluid, distributed on each boundary node located between the fluid and the solid particle. The sum of these forces  must be zero for the Newton's third law, a condition fulfilled solely by a force dipole. Self-propulsion is then achieved by modifying the symmetry of these forces. An asymmetric force distribution generates a net force on the particle (due to the difference of the fluid velocity in its back and in its front) and promotes motion, whereas a symmetric distribution does not yield a force unbalance, although the particle still generates a non-zero quadropolar velocity field in the fluid. The former particles are called movers and the latters are called shakers.




A complementary mechanism promoting particle self-propulsion was proposed in Ref.~\cite{Llopis2006}, and consisted in adding a constant amount of momentum with fixed magnitude to the particle. To restore momentum conservation, the same amount of momentum is subtracted to the fluid nodes connected to the solid ones. This approach has been used to study the collective dynamics of self-propelled particles dispersed in a fluid solvent, and was found to reproduce, for instance, the formation of transient aggregates of particles, as well as the transition of the particle mean square displacement from ballistic towards diffusive motion at low concentration~\cite{Llopis2006}. A similar mechanism has been also used to investigate the hydrodynamics of active rotors~\cite{Llopis2008}, systems capturing, for example, the rotating motion of molecular motors on the cell membrane~\cite{noji1997}.

An alternative approach has been afterwards used to simulate suspensions of swimming particles (again built using the Ladd's method) starting from the squirmer model introduced by Lighthill~\cite{lighthill1952,blake1971}, in which particle's motion is now triggered by a predefined axis-symmetric tangential velocity distribution, imposed on the surface of the particle~\cite{alarcon2013}. By using such velocity distribution as boundary condition of the Stokes and the continuity equations (as the inertia of the fluid is neglected), the mean fluid flow induced by a minimal squirmer can be computed as 
\begin{equation}\label{axi_vel}
 u(\theta)=B_1\sin(\theta) + \frac{B_2}{2}\sin(2\theta),
\end{equation}
where the terms on the right hand side derive from an expansion with Legendre polynomials truncated to the second mode. Here $\theta$ is a polar angle and $\theta=0$ defines the swimming direction of the squirmers. The dimensionless parameter $\epsilon=B_2/B_1$ defines the type of squirmer: $\epsilon>0$ describes a puller (or contractile squirmer) while $\epsilon <0$ describes a pusher (or extensile squirmer). The case with $\epsilon=0$ corresponds to an apolar squirmer (or shaker).

Lattice Boltzmann has been also applied to investigate the hydrodynamics of active three-bead linear swimmers~\cite{earl2007}, a system for which an analytic solution is available~\cite{najafi2004}. Here the swimmer is constituted of three spheres of predefined radius linked with sufficiently thin rods to neglect hydrodynamic interactions. Lengths and angles between the rods can be modified in a periodic and time irreversible manner by the action of internal forces and torques, which favor change in the swimmer shape and potentially motion when coupled to the fluid. In this model there is no net interface separating the fluid from the bead, whose interior, as in the model proposed by Ramachandran {\it et al.}~\cite{ramachandran2006}, is then essentially fluid. 
The swimmer-fluid interaction is incorporated in three steps, the first of which computes the total linear and angular momenta of the swimmer. Afterwards position and shape of the swimmer are updated with respect to the original one (such that linear and angular momentum are conserved) and finally the motion of the swimmer is coupled to the fluid to obtain updated fluid velocities used to calculate the equilibrium distribution functions of the LBM~\cite{earl2007,pooley2008}. Such approach was found to reproduce analytical results with sufficient accuracy, in particular when swimmers with a high number of beads (even more than three) and close to each other (or near a surface) are considered.
An alternative and perhaps computationally simpler way to model solid object on a grid was suggested by Smith and Denniston~\cite{smith2007}. Here the surface of the swimmer is represented by a large number of point particles, which ensure a non-slip condition by introducing a drag force between the particles and the nodes of the lattice Boltzmann mesh~\cite{pooley2008}. Although this method was found efficient in simulating the swimming velocity of a single swimmer, it suffered of large discretization errors affecting inter-swimmer hydrodynamic interactions.

Due to a limited computational efficiency, the models described so far have been rarely applied to simulate highly anisotropic particles. This drawback has been recently circumvented by de Graaf \emph{et al.}~\cite{degraaf2016,degraaf2016_2}, that modeled active colloids of arbitrary shape as clusters of spheres coupled to the LB fluid through the scheme introduced by Ahlrichs and D\"uenweg~\cite{ahlrichs1999}. Unlike the grid-based Ladd's method, in this approach a particle is described through a set of points coupled to the fluid through a frictional force, which depends on the relative velocity (in analogy to the Stokes formula for a sphere in a viscous fluid) between the fluid and the points. Due to this coupling and to the interpolation of the force 
on the LB mesh at the particle position, a \emph{hydrodynamic shell} forms around the points, which now acquire an effective hydrodynamic radius. A solid object is then obtained  when a high enough number of points is considered~\cite{degraaf2015,degraaf2015_2}.
 Self-propulsion is achieved by means of a persistence force applied along a direction vector assigned to the particle, while an equal and opposite counter force is applied to the fluid~\cite{nash2008,hernandez2005,saintillan2007}. Its location sets the nature of the swimmer (whether contractile or extensile) and the structure of the fluid flow in its surrounding. Besides being a facile approach, this method has the advantage to incorporate hydrodynamic interactions with higher order multipole moments, (in addition to the usual dipole ones), and has been found to well reproduce far-field theoretical results in system with periodic boundary conditions and in a spherical cavity with no-slip walls~\cite{deGraaf2017}. A likewise computationally efficient method, in which swimmers are described using the point-force implementation developed by Nash {\it et al.}~\cite{nash2008}, has been used to perform unprecedented large-scale LB simulations of $\sim 10^6$ hydrodynamically interacting swimmers in a cubic box with periodic boundary conditions~\cite{stenhammar2017}. In this work the authors show that swimmers move in a correlated fashion well below the transition to bacterial turbulence, and elaborated a novel kinetic approach capturing such behavior with results in quantitative agreement with LB simulations.

More recently, lattice Boltzmann methods have been extended to simulate more complicated physical systems in which active particles play, once more, a relevant role. A remarkable example is the process of cross-streamline migration  (often called margination~\cite{schmid1980}) of stiff active particles (such as synthetic nanoparticles used for drug-delivery) migrating towards the vessel walls when moving in blood flows. Such phenomenon, studied in Ref.~\cite{gekle2016}, is attributed to hydrodynamic interactions of red blood cells with stiff particles, and disappears when red blood cells are absent. Active particles are modeled by triangulated spheres whose internal grid nodes are massless points flowing with the local fluid velocity and connected each other through stiff harmonic springs and bending potential to preserve the grid arrangement. This approach is often referred as immersed boundary method~\cite{vahid2014,vahid2015}. A force is then applied to the center of mass of the particles (which become active) and along the opposite direction with respect to the fluid flow. 

A much less investigated system is that in which a passive particle is embedded in active fluid. This has been done in Refs.~\cite{foffanoPRL,foffano2012}, where LB simulations show a violation of the Stokes' law  when a particle (modeled by using the Ladd's method) is dragged in an active fluid. Even more strikingly, a negative viscous drag in the steady state of a contractile fluid is found for large enough particles. Such simulated microrheological experiment, in principle realizable in the laboratory, highlights the fact that, although at its infancy, the study of these systems may unveil novel and intriguing physics, potentially useful for the design of novel active soft materials.

Clearly many efforts have been addressed to model systems which (i) correctly describe the hydrodynamics of active particles and (ii) capture new dynamical and mechanical properties of great importance for future practical applications. Several directions of research can be envisaged starting from the results achieved so far. A largely unexplored field is that of the active rheology, in which suspensions of particles (e.g. passive) are subject to an external perturbation (such as a shear flow) in an active liquid crystal. Here the nature of the active system (either contractile or extensile) as well as the particle volume fraction and the shear stress may dramatically affect the rheological response. In addition, the presence of topological defects, due to the conflicting anchoring of the liquid crystal with that on particle surface, may foster the formation of turbulent-like velocity patterns, whose physics is still under investigation. LB simulations performed so far usually neglect thermal fluctuations, which may however be crucial especially when the particle size decreases up to nanometers. Even more intriguing seems the possibility of a multiscale coupling of the LB with other simultation methods, such as dissipative particle dynamics or immersed boundary method, to capture, for instance, the near-contact interactions of active colloids at the interface of a binary fluid.

\section{Conclusions}

In this review we have offered
an introduction to theoretical approaches in continuum modeling of  active fluids together with
the description of the most appropriate lattice Boltzmann techniques for numerical simulations of this new fascinating category of soft matter. 
This has been done in Sections~\ref{sec:2} and~\ref{sec:LBM}, respectively.
One of the most immediate consequences of activity is the occurrence of spontaneous flow. We have illustrated in Section~\ref{sec:spont_fl} that this emerges as a result of bending instability of the polarization field in extensile systems and of splay deformations in contractile
fluids.  The role played by  nematic or polarization defects    has been also discussed.
The presence of geometrical constraints and/or external forcing makes the behavior of active fluids or mixtures even richer, with many interesting and somehow unexpected features. 
Active fluids confined in droplets can spontaneously flow or spread
faster, if compared with similar passive cases, and this opens the way to many  futuristic applications coming from the design of new bio-inspired materials. 
Self-propelled droplets can be even model systems  for living cells and this also justifies the enormous recent research interest towards this matter,
reviewed in Section~\ref{sec:self_prop}. In Section~\ref{sec:rheology} the rheological behavior of active fluids is discussed. Unidirected motion and  superfluidic behavior are examples of results in this field whose theoretical and applicative consequences are matter for future research.  We have discussed  these issues  with  the more recent perspectives of research in these fields.
Large self-propulsive forces  induce large flows with chaotic  velocity patterns. This behavior  has been termed active turbulence even if the possibility of a quantitative  description similar  to that of standard turbulence at high Reynolds number is still a matter to
be properly understood. The status  of research in this field has been reviewed in Section~\ref{sec:turb}. Finally, in Section~\ref{sec:coll}, it is shown how mixed LBM-particle algorithms  can be  used to simulate the effects of the active bath on particles in suspension.
In all these subjects  numerical simulations have demonstrated to be an essential tool
for the characterization and  for the general understanding of the phenomena considered. 
Between different numerical methods  the role of LBM has been prominent
as shown by the considerable amount of results 
in literature  coming from LBM studies.
The reason of this is mainly in  the easy adaptation  of
lattice Boltzmann algorithms to the study  of complex fluids and active fluids.
We hope that this review  can be useful to other researchers in this new field of soft matter.

\section*{Acknowledgments}
We warmly thank L. Biferale, P. Digregorio, D. Marenduzzo, E. Orlandini, I. Petrelli for many fruitful discussions. Simulations supporting the theories presented in this review were run at Bari ReCaS (\textit{https://www.recas-bari.it/}) e-Infrastructure funded by MIUR through PON Research and Competitiveness 2007-2013 Call 254 Action I, at ARCHER UK National Supercomputing Service (\textit{http://www.archer.ac.uk}) through the program HPC-Europa3 and at Marconi (CINECA - \textit{http://www.hpc.cineca.it/}) under CINECA-INFN agreement (Project No. INF19-fldturb).

\section*{Contribution Statement}
All the authors have equally contributed to conceive and write this review. L.N.C. and G.N. have performed the simulations needed for Figs. 3,6,7,9,10.

\bibliographystyle{unsrt}
\bibliography{review_LBM_active.bbl}

\appendix

\section{Biaxial nematics}
\label{sec:biaxial}
In this Appendix we briefly discuss how to extend the uniaxial order parameter to biaxial nematics and how biaxiality provides information about the localization of topological defects in nematic liquid crystals.

A biaxial nematic is a nematic liquid crystal with three distinct optical axis and, unlike an uniaxial liquid crystal, it does not have any axis of complete rotational symmetry. Hence one can define three perpendicular axes ${\bf n}$, ${\bf m}$ and ${\bf l}$ (two are sufficient, since the third one would be perpendicular to the others), or director fields, for which there is a reflection symmetry. The order parameter in three dimensions is then
\begin{equation}
  Q_{\alpha\beta}=S(n_{\alpha}n_{\beta}-\frac{1}{3}\delta_{\alpha\beta})+\varepsilon(m_{\alpha}m_{\beta}-\frac{1}{3}\delta_{\alpha\beta}),
\end{equation}
where $S$ and $\varepsilon$ are called scalar order parameters. This representation of $\underline{\underline{\vec Q}}$ is a traceless symmetric second order rank tensor with five independent components. If the smaller of the two scalar order parameters is very small (like in many systems), one recovers the $\underline{\underline{\vec Q}}$ tensor in the uniaxial approximation. 

For a biaxial nematic the three eigenvalues are in general different, and the diagonal representation of the $\underline{\underline{\vec Q}}$ tensor is 
\begin{equation}
\vec{\underline{\underline{Q}}}= \Diag \left(\dfrac{2}{3}S \ , \ -\dfrac{1}{3}S+ \varepsilon \ , \ -\dfrac{1}{3}S - \varepsilon \right)
\end{equation}
where $\varepsilon \leqslant S$, with $\varepsilon$ gauging the degree of biaxiality.

The investigation of nematic defects in the context of a Landau-de Gennes theory has shown that their core presents a heavy degree of biaxiality~\cite{schopohl1987}. By following the approach of Ref.~\cite{callan2006}, this can be measured by computing three scalars, $c_l = \tilde\lambda_1-\tilde\lambda_2$,  $c_p = 2(\tilde\lambda_2-\tilde\lambda_3)$ and $c_s = 3\tilde\lambda_3$, where $\tilde\lambda_1$, $\tilde\lambda_2$ and $\tilde\lambda_3$ (with $\tilde\lambda_1\ge\tilde\lambda_2\ge\tilde\lambda_3$) are three eigenvalues of the diagonalised matrix $G_{\alpha\beta}=Q_{\alpha\beta}+\delta_{\alpha\beta}/3$. These parameters fulfill the following properties: $0\le c_l, c_p, c_s\le 1$ and  $c_l+c_p+c_s=1$. An ordered unixial nematic will give $c_l \simeq 1$, while the isotropic state, where both $S$ and $\eta$ are approximatively null, corresponds to $c_s \simeq 1$. Finally the biaxial case implies $c_p \simeq 1$.

\section{Chapman-Enskog expansion}
\label{sec:appB}
In Section~\ref{sec:LBM_sf} we presented a LB algorithm to solve the hydrodynamics of a simple fluid. We show here that Eqs.\eqref{conc_eq} and~\eqref{nav} can be recovered in the continuum limit, starting from the evolution equation~\eqref{eqn:evolution_equation_LBM} for the distribution functions $f_i$.
Two approaches can be followed. The first one starts from a Taylor expansion of the left-hand side of Eq.~\eqref{eqn:evolution_equation_LBM}~\cite{Nourgaliev2003}, whereas the second one, discussed below, uses a Chapman-Enskog method, that is an expansion of the distribution functions about equilibrium, which assumes that successive derivatives are of increasingly high order 
in the Knudsen number $\epsilon=\lambda/L$. This is a dimensionless number defined as the ratio between the molecular mean free path $\lambda$ and a characteristic length $L$ of the system. This number determines whether a macroscopic continuum mechanics or a microscopic statistical mechanics formulation of fluid dynamics should be used. For small values of $\epsilon$ ($\epsilon \ll 1$) the mean free path is much smaller than $L$ and a continuum theory is a good approximation. 
To take into account both ballistic and diffusive scales, spatial density fluctuations of order $\mathcal{O}(\epsilon^{-1})$ are assumed to relax over time scales of order $\mathcal{O}(\epsilon^{-2})$. A suitable expansion for temporal and spatial derivatives as well as for distribution functions is
\begin{eqnarray}
f_i = f_i^{(0)} + \epsilon f_i^{(1)}+ \epsilon^2 f_i^{(2)},  \label{eqn:CE_1}\\
\partial_t = \epsilon \partial_{t_1} + \epsilon^2 \partial_{t_2}, \label{eqn:CE_2}\\
\partial_\alpha = \epsilon \partial_{\alpha_1}, \label{eqn:CE_3}
\end{eqnarray}
built assuming that there is a diffusion time scale $t_2$ slower than the convection one $t_1$.

We start by expanding the left-hand side of equation~\eqref{eqn:evolution_equation_LBM}
to the second order in $\Delta t$:
\begin{equation}
\begin{split}
\Delta t ( \partial_t + \xi_{i\alpha} \partial_{\alpha}) f_i  + \dfrac{(\Delta t)^2}{2} \left(\partial_t^2 + 2 \xi_{i \alpha} \partial_\alpha \partial_t + \xi_{i\alpha} \xi_{i\beta} \partial_\alpha \partial_\beta \right) f_i \\ =
- \dfrac{1}{\tau} (f_i - f_i^{eq}),
\end{split}
\label{eqn:eq1_ce}
\end{equation}
where Eq.~\eqref{eqn:BGK_simple_fluid} has been used to express the collision operator.
By substituting Eq.~\eqref{eqn:CE_1},~\eqref{eqn:CE_2} and~\eqref{eqn:CE_3} into
Eq.~\eqref{eqn:eq1_ce} one obtains
\begin{eqnarray}
&&\Delta t \left[ (\epsilon \partial_{t_1} + \epsilon^2 \partial_{t_2})+ \epsilon \xi_{i \alpha} \partial_{\alpha_1} \right] (f_i^{(0)} + \epsilon f_i^{(1)}+ \epsilon^2 f_i^{(2)}) \nonumber\\
&&+  (\Delta t)^2  \left[ \dfrac{1}{2} (\epsilon \partial_{t_1} + \epsilon^2 \partial_{t_2})^2 + \epsilon \xi_{i \alpha} \partial_\alpha (\epsilon \partial_{t_1} + \epsilon^2 \partial_{t_2}) \right. \nonumber\\ &&\left. + \dfrac{1}{2}\epsilon^2 \xi_{i\alpha} \xi_{i\beta} \partial_{\alpha_1} \partial_{\beta_1} \right] (f_i^{(0)} + \epsilon f_i^{(1)}+ \epsilon^2 f_i^{(2)}) \nonumber\\
&&=- \dfrac{1}{\tau} (f_i^{(0)} + \epsilon f_i^{(1)}+ \epsilon^2 f_i^{(2)} - f_i^{eq}).
\label{eqn:eq2_ce}
\end{eqnarray}
By retaining at most terms of second order in $\epsilon$, the previous equation reads
\begin{eqnarray}
&&\epsilon \Delta t  \left(\partial_{t_1} f_i^{(0)} +  \xi_{i \alpha} \partial_{\alpha_1} f_i^{(0)} \right) \nonumber\\
&+& \epsilon^2\left[ \Delta t \left(\partial_{t_1} f_i^{(1)} +  \xi_{i \alpha} \partial_{\alpha_1} f_i^{(1)}  +\partial_{t_2} f_i^{(0)} \right)  \right.\nonumber\\
&+& \left. \Delta t ^2 \left( \dfrac{1}{2} \partial_{t_1}^2 + \dfrac{1}{2} \xi_{i \alpha} \xi_{i \beta}  \partial_{\alpha_1} \partial_{\beta_1} + \xi_{i \alpha} \partial_{\alpha_1} \partial_{t_1} \right) f_i^{(0)} \right]\nonumber \\
&=&- \dfrac{f_i^{(0)}- f_i^{eq}}{\tau} - \dfrac{ \epsilon f_i^{(1)}+ \epsilon^2 f_i^{(2)}}{\tau}.
\label{eqn:eq3_ce}
\end{eqnarray}
Finally, grouping terms of same order in $\epsilon$, we get
\begin{eqnarray}
&&f_i^{(0)} = f_i^{eq} + \mathcal{O}(\epsilon) , \label{eqn:eps0} \\
&&\partial_{t_1} f_i^{(0)} +  \xi_{i \alpha} \partial_{\alpha_1} f_i^{(0)} = - \dfrac{1}{ \tau \Delta t}f_i^{(1)} + \mathcal{O}(\epsilon), \label{eqn:eps1}\\
&&\partial_{t_1} f_i^{(1)} +  \xi_{i \alpha} \partial_{\alpha_1} f_i^{(1)} +\partial_{t_2} f_i^{(0)} \nonumber\\
&&+  \dfrac{\Delta t}{2} \left( \partial_{t_1}^2 + 2 \xi_{i \alpha} \partial_{\alpha_1} \partial_{t_1}
+ \xi_{i \alpha} \xi_{i \beta}  \partial_{\alpha_1} \partial_{\beta_1} \right) f_i^{(0)} \nonumber\\&& = - \dfrac{1}{\tau \Delta t}f_i^{(2)} + \mathcal{O}(\epsilon).\label{eqn:eps2}
\end{eqnarray}
In the following paragraphs we will use these relations to recover continuum equations up to second order in the Knudsen number.

\paragraph{Recover Continuity Equation}
To recover the continuity equation one can start by summing Eq.~\eqref{eqn:eps0} over lattice velocities with the constraints given in Eqs.~\eqref{eqn:mass_constraint_eq} and~\eqref{eqn:momentum_constraint_eq}. One then gets
\begin{equation}
\sum_i f_i^{(0)} = \rho, \qquad \sum_i f_i^{(1)}=\sum_i f_i^{(2)}=0,
\label{eqn:knudsen_mass_contraint}
\end{equation}
and, by using again Eq.~\eqref{eqn:momentum_constraint_eq}
\begin{equation}
\sum_i f_i^{(0)} \xi_{i \alpha} = \rho v_\alpha, \qquad \sum_i f_i^{(1)}\xi_{i \alpha}=\sum_i f_i^{(2)}\xi_{i \alpha}=0.
\label{eqn:knudsen_momentum_contraint}
\end{equation}
By performing a summation over lattice velocities in Eq.~\eqref{eqn:eps1}, one gets
\begin{equation}
\partial_{t_1} \rho + \partial_{\alpha_1} (\rho v_{\alpha} ) = 0 + \mathcal{O}(\epsilon),
\label{eqn:CE_continuum_eq_eps1}
\end{equation}
which is the continuity equation at first order in the Knudsen number. To recover the complete time derivative according to Eq.~\eqref{eqn:CE_2}, we need to explicitly compute the term $\partial_{t_2} \rho$. By applying the  differential operators $\epsilon \partial_{t_1}$ and $ \epsilon \xi_{i \beta} \partial_{\beta_1}$ to Eq.~\eqref{eqn:eps1} we obtain
\begin{equation}
  \partial^2_{t_1} f_i^{(0)} +  \xi_{i \alpha} \partial_{\alpha_1} \partial_{t_1} f_i^{(0)} = - \dfrac{1}{\tau \Delta t} \partial_{t_1} f_i^{(1)} + \mathcal{O}(\epsilon^2), \label{eqn:eps1'}
\end{equation}
\begin{equation}
\xi_{i \beta} \partial_{\beta_1} \partial_{t_1} f_i^{(0)} +  \xi_{i \alpha} \xi_{i \beta} \partial_{\alpha_1} \partial_{\beta_1} f_i^{(0)}  = - \dfrac{1}{\tau \Delta t} \xi_{i \beta} \partial_{\beta_1} f_i^{(1)} + \mathcal{O}(\epsilon^2)\label{eqn:eps1''},
 \end{equation}
and, summing both equations:
\begin{equation}
\begin{split}
(\partial^2_{t_1}+ 2\xi_{i\alpha} \partial_{\alpha_1} \partial_{t_1} + \xi_{i \alpha} \xi_{i \beta} \partial_{\alpha_1} \partial_{\beta_1})f_i^{(0)}\\ = - \dfrac{1}{\tau \Delta t} ( \partial_{t_1} + \xi_{i \beta} \partial_{\beta_1}) f_i^{(1)}+\mathcal{O}(\epsilon^2).
 \end{split}
\label{eqn:eps1'''}
\end{equation}
Note that the left-hand side of this equation is exactly the term in round brackets of Eq.~\eqref{eqn:eps2}, that now becomes
\begin{equation}
\begin{split}
\partial_{t_2} f_i^{(0)} +\left(1-\dfrac{1}{2\tau} \right) (\partial_{t_1}+\xi_{i \alpha} \partial_{\alpha_1}) f_i^{(1)} \\ = - \dfrac{1}{\tau \Delta t} f_i^{(2)} + \mathcal{O}(\epsilon).
\end{split}
\label{eqn:eps2'}
\end{equation}
By summing over lattice directions and using  Eqs.~\eqref{eqn:knudsen_mass_contraint} and~\eqref{eqn:knudsen_momentum_contraint}, we get
\begin{equation}
\partial_{t_2} \rho = 0 + \mathcal{O}(\epsilon^2),
\end{equation}
that, together with Eq.~\eqref{eqn:CE_continuum_eq_eps1}, reads
\begin{equation}
(\epsilon\partial_{t_1} + \epsilon^2 \partial_{t_2}) \rho + \epsilon \partial_{\alpha_1} (\rho v_{\alpha} ) = 0 + \mathcal{O}(\epsilon^2).
\end{equation}
Finally, after restoring the canonical differential operators (through Eqs.~\eqref{eqn:CE_2} and~\eqref{eqn:CE_3}), we get the continuity equation
\begin{equation}
\partial_t\rho + \partial_{\alpha} (\rho v_{\alpha} ) = 0+ \mathcal{O}(\epsilon^2),
\end{equation}
at second order in the Knudsen number.

\paragraph{Recover Navier-Stokes Equations}
The procedure to recover the Navier-Stokes equation is analogous, albeit less straightforward, than that used for the continuity equation. We will proceed by calculating the first-order moment of Eq.~\eqref{eqn:eps1} and Eq.~\eqref{eqn:eps2}. First multiply by $\xi_{i\beta}$ both members of Eq.~\eqref{eqn:eps1} and sum over index $i$, to get
\begin{equation}
\partial_{t_1}(\rho v_\alpha) +  \partial_{\beta_1} \left( \dfrac{c^2}{3} \rho \delta_{\alpha \beta} + \rho v_{\alpha} v_{\beta} \right) = 0 + \mathcal{O}(\epsilon).
\label{eqn:NSE_I}
\end{equation}
To get the Navier-Stokes equation to second order in the Knudsen number we need to calculate the first-order moment of equation~\eqref{eqn:eps2}.
We can then multiply Eq.~\eqref{eqn:eps2'} by $\xi_{i \gamma}$ to obtain
\begin{equation}
\begin{split}
\partial_{t_2} \xi_{i \gamma} f_i^{(0)} +\left(1-\dfrac{1}{2\tau} \right) (\partial_{t_1}- \xi_{i \alpha} \partial_{\alpha_1}) \xi_{i \gamma} f_i^{(1)} \\ = - \dfrac{1}{\tau \Delta t} \xi_{i \gamma} f_i^{(2)} + \mathcal{O}(\epsilon^2),
\end{split}
\end{equation}
and, by summing over lattice velocities, we are left with
\begin{equation}
\partial_{t_2} (\rho v_\alpha) - \left( 1-\dfrac{1}{2\tau}\right)  \partial_{\beta_1} \left[\sum_i f_i^{(1)} \xi_{i \alpha} \xi_{i \beta} \right]=0.
\label{eqn:eq50}
\end{equation}
Now we must determine an expression for the summation in square brackets. From Eqs.~\eqref{eqn:eps0} and~\eqref{eqn:eps1} we note that
\begin{equation}
\begin{split}
\sum_i f_i^{(1)} \xi_{i \alpha} \xi_{i \beta}&= - \tau \Delta t (\partial_{t_1}+ \xi_{i \gamma} \partial_{\gamma_1} ) \left(\sum_i f_i^{eq} \xi_{i \alpha} \xi_{i \beta} \right) \\
&=  - \tau \Delta t \left[ \partial_{t_1}\left(\dfrac{c^2}{3} \rho \delta_{\alpha \beta}+ \rho v_\alpha v_\beta\right) \right. \\  & \qquad +  \left. \partial_{\gamma_1} \left(\sum_i f_i^{eq} \xi_{i \alpha} \xi_{i \beta} \xi_{i \gamma} \right) \right],
\end{split}
\label{eqn:eq51}
\end{equation}
where we have used Eq.~\eqref{eqn:II_moment_condition} in the second equality.
The second term of the second line of Eq.~\eqref{eqn:eq51} can be written in terms of the equilibrium distribution functions given in Eq.~\eqref{eqn:expansion_eq_df_simple_fuid} and of the related coefficients in Eq.~\eqref{eqn:Icoeff}
\begin{equation}
\begin{split}
\partial_{\gamma_1} & \left( \sum_i f_i^{eq} \xi_{i \alpha} \xi_{i \beta}  \xi_{i \gamma} \right) = \\ & \qquad \qquad \dfrac{c^2}{3} \partial_{\gamma_1} \left[ \rho (\delta_{\alpha \beta} v_\gamma + \delta_{\alpha \gamma} v_\beta +\delta_{\beta \gamma} v_\alpha) \right],
\end{split}
\label{eqn:eq52}
\end{equation}
while the first round bracket in the second line of Eq.~\eqref{eqn:eq51} can be written by means of Eq.~\eqref{eqn:CE_continuum_eq_eps1} and Eq.~\eqref{eqn:NSE_I} as
\begin{equation}
\begin{split}
\partial_{t_1}\left(\dfrac{c^2}{3} \rho \delta_{\alpha \beta}+ \rho v_\alpha v_\beta\right)
= -\dfrac{c^2}{3} \partial_{\gamma_1} (\rho v_\gamma)  \delta_{\alpha \beta} \\+ v_\beta \partial_{t_1} (\rho v_\alpha) + v_\alpha \partial_{t_1} (\rho v_\beta) - v_\alpha v_\beta \partial_{t_1} \rho \\
\simeq -\dfrac{c^2}{3} \left[ \partial_{\gamma_1} (\rho v_\gamma)  \delta_{\alpha \beta} + (v_\alpha\partial_{\beta_1}  \rho+   v_\beta\partial_{\alpha_1} \rho) \right].
\end{split}
\label{eqn:eq53}
\end{equation}
In the last line terms of order $v^3$ were neglected, an approximation valid 
as far as the Mach number is kept small. Now substituting Eqs. \eqref{eqn:eq53} and~\eqref{eqn:eq52} into Eq.~\eqref{eqn:eq51} we find, after some algebra, that
\begin{equation}
\sum_i f_i^{(1)} \xi_{i \alpha} \xi_{i \beta}= - \tau \Delta t \dfrac{c^2}{3}  \rho \left[ \partial_{\beta_1} v_\alpha  +   \partial_{\alpha_1} v_\beta \right].
\label{eqn:eq54}
\end{equation}
This term, in turn, enters Eq.~\eqref{eqn:eq50}, which now reads  
\begin{equation}
\partial_{t_2} (\rho v_\alpha) - \left( \tau-\dfrac{1}{2}\right) \Delta t \dfrac{c^2}{3} \partial_{\beta_1} \left[ \rho \left( \partial_{\beta_1} v_\alpha  +   \partial_{\alpha_1} v_\beta \right) \right]=0.
\end{equation}
Finally, summing this equation with Eq.~\eqref{eqn:NSE_I} and using the canonical differential operators (i.e. Eqs.~\eqref{eqn:CE_2} and~\eqref{eqn:CE_3}), we obtain the Navier-Stokes equation
\begin{eqnarray}
&&\partial_t (\rho v_\alpha) + \partial_{\beta} \left( \rho v_{\alpha} v_{\beta} \right) =\\&&  -\partial_{\alpha} p  + \Delta t \left( \tau-\dfrac{1}{2}\right)  \dfrac{c^2}{3} \partial_\beta \left[ \rho \left( \partial_{\beta} v_\alpha  +   \partial_{\alpha} v_\beta \right) \right],
\end{eqnarray}
where $p=(c^2/3) \rho$ is the isotropic pressure
and the shear viscosity is given by
\begin{equation}
\eta = \dfrac{\rho c_s^2 \Delta t}{3} \left(\tau -\dfrac{1}{2} \right) .
\end{equation}



\section{Recovering continuum equations for the algorithm described in Section~\ref{sec:external_force} }
\label{sec:appC}
In this Appendix we show the calculations to recover the continuum equations obtained by means of the forcing method algorithm discussed in Section~\ref{sec:external_force}.   

We start with a Chapman-Enskog expansion of the distribution functions and of the derivatives according to relations~\eqref{eqn:CE_1}-\eqref{eqn:CE_3}, and for the forcing term we assume that
\begin{equation}
\mathcal{F}_i = \epsilon \mathcal{F}_{i 1}.
\label{eqn:CE_4}
\end{equation}.
By Taylor expanding the evolution Eq.~\eqref{eqn:evolution_equation_LBM}, we get
\begin{multline}\label{tayl_exp}
\Delta t ( \partial_t + \xi_{i \alpha} \partial_\alpha)f_i + \dfrac{\Delta t^2}{2} (\partial_t^2 + 2 \xi_{i \alpha} \partial_\alpha  \partial_t +  \xi_{i \alpha} \xi_{i \beta} \partial_\alpha \partial_\beta ) f_i  \\= \Delta t \mathcal{F}_i -  \dfrac{f_i - f_i^{eq}}{\tau}.
\end{multline}
We now substitute Eqs.~\eqref{eqn:CE_1}-\eqref{eqn:CE_3} and~\eqref{eqn:CE_4} in Eq.~\eqref{tayl_exp} and, after grouping terms of the same order in $\epsilon$, we get
\begin{eqnarray}
f_i^{(0)}=f_i^{eq} + \mathcal{O}(\epsilon) , \label{eqn:eps0_app} \\
 ( \partial_{t_1} + \xi_{i \alpha} \partial_{\alpha_1}) f_i^{(0)} = - \dfrac{1}{\tau \Delta t }f_i^{(1)} + F_{i1} + \mathcal{O}(\epsilon), \label{eqn:eps1_app}\\
\begin{aligned}
\partial_{t_1} f_i^{(1)} +  \xi_{i \alpha} \partial_{\alpha_1} f_i^{(1)} +\partial_{t_2} f_i^{(0)} \\
+  \dfrac{\Delta t}{2} \left( \partial_{t_1}^2 + 2 \xi_{i \alpha} \partial_{\alpha_1} \partial_{t_1}
+ \xi_{i \alpha} \xi_{i \beta}  \partial_{\alpha_1} \partial_{\beta_1} \right) f_i^{(0)} \\ = - \dfrac{1}{\tau \Delta t}f_i^{(2)} + \mathcal{O}(\epsilon).
\end{aligned} \label{eqn:eps2_app}
\end{eqnarray}
From these equations one gets the zeroth-order moments of the distrubution functions
\begin{equation}
\sum_i f_i^{(0)} = \sum_i f_i^{eq} = \rho \qquad \sum_i f_i^{(1)} = \sum_i f_i^{(2)} = 0,
\label{eqn:eq117}
\end{equation}
the first-order ones
\begin{align}
\sum_i f_i^{(0)} \xi_{i_\alpha} &= \sum_i f_i^{eq} \xi_{i_\alpha}= \rho v_\alpha, \\
\sum_i f_i^{(1)}\xi_{i_\alpha} &= - \dfrac{\Delta t}{2} F_{1 \alpha}, \\
\sum_i f_i^{(2)}\xi_{i_\alpha}& = 0.
\end{align}
and the  zeroth, first and second moments of the forcing term 
\begin{eqnarray}
&&\sum_i \mathcal{F}_i =0 , \label{eqn:CE_F0}\\
&&\sum_i \mathcal{F}_i \xi_{i \alpha} = \left(1 - \dfrac{\Delta t}{2 \tau} \right) F_\alpha, \label{eqn:CE_F1}\\
&&\sum_i \mathcal{F}_i \xi_{i \alpha} \xi_{i \beta} = \left( 1 - \dfrac{\Delta t}{2 \tau} \right)  (v_\alpha F_\beta + v_\beta F_\alpha).\label{eqn:CE_F2}
\end{eqnarray}
These relations can be explicitly calculated by using Eq.~\eqref{eqn:force_exp}.
Now combining the zeroth-order moment of Eq.~\eqref{eqn:eps1_app} with Eqs.~\eqref{eqn:eq117} and~\eqref{eqn:CE_F0}, one gets
\begin{equation}
\partial_{t_1} \rho + \partial_{\beta_1} (\rho v_\beta)= 0,
\label{eqn:cont_1order}
\end{equation}
the continuity equation at first order in the Knudsen number.
To recover it at second order we apply the differential operators $\partial_{t_1}$ and $\xi_{i \alpha} \partial_{\alpha_1}$ to Eq.~\eqref{eqn:eps1_app} and then, 
by perfoming the difference between the two resulting equations, one has
\begin{multline}
\partial_{t_1}^2 f_i^{(0)} =  \xi_{i \alpha} \xi_{i \beta}  \partial_{\alpha_1} \partial_{\beta_1} f_i^{eq} - \dfrac{1}{\tau \Delta t } (\partial_{t_1} - \xi_{i \gamma} \partial_{\gamma_1}) f_i^{(1)}  \\+  (\partial_{t_1} - \xi_{i \gamma} \partial_{\gamma_1})\mathcal{F}_{i 1}.
\label{eqn:eq125}
\end{multline}
Now we substitute the latter into Eq.~\eqref{eqn:eps2_app} and, by using Eq.~\eqref{eqn:eps1_app}, we obtain
\begin{multline}
(\partial_{t_1} +  \xi_{i \alpha} \partial_{\alpha_1}) f_i^{(1)} +\partial_{t_2} f_i^{(0)}
- \dfrac{1}{2 \tau} \left( \partial_{t_1} +	 \xi_{i \gamma} \partial_{\gamma_1} \right) f_i^{(1)}
 \\ = - \dfrac{1}{\tau \Delta t }f_i^{(2)} - \dfrac{\Delta t}{2} (\partial_{t_1} + \xi_{i \gamma} \partial_{\gamma_1}) \mathcal{F}_{i 1}.
 \label{eqn:eq126}
\end{multline}
Finally summing this one over index $i$, by means of Eqs.~\eqref{eqn:eps0_app}-\eqref{eqn:CE_F1}, we get
\begin{equation}
\partial_{t_2} \rho = 0.
\end{equation}
By summing this equation with Eq.~\eqref{eqn:cont_1order} we obtain 
the continuity equation at second order in the Knudsen number.

To reproduce the Navier-Stokes equation, we start by computing the first moment of Eq.~\eqref{eqn:eps1_app} that, after using Eqs.~\eqref{eqn:eq117}-\eqref{eqn:CE_F2}, reads
\begin{equation}
\partial_{t_1}( \rho v_\alpha) + \partial_{\beta_1} \left( \dfrac{c^2}{3} \rho \delta_{\alpha \beta} + \rho v_\alpha v_\beta \right) = F_{\alpha}.
\end{equation}
Following the same procedure for Eq.~\eqref{eqn:eq126} we get
\begin{equation}
\partial_{t_2} (\rho v_\alpha) = \dfrac{c^2}{3} \Delta t \left( \tau -\dfrac{1}{2} \right) \partial_{\beta_1} \left[ \rho (\partial_{\beta_1} v_\alpha + \partial_{\alpha_1} v_{\beta}) \right].
\end{equation}
Finally, summing these two equations, one can restore the Navier-Stokes equation
\begin{multline}
\partial_{t}( \rho v_\alpha) + \partial_{\beta} \left( \rho v_\alpha v_\beta \right) = \partial_\beta \sigma_{\alpha \beta} \\ + \dfrac{c^2}{3} \Delta t \left( \tau -\dfrac{1}{2} \right) \partial_{\beta_1} \left[ \rho (\partial_{\beta_1} v_\alpha + \partial_{\alpha_1} v_{\beta}) \right],
\end{multline}
where we require that
\begin{equation}
F_\alpha = \partial_\beta \left[ \sigma_{\alpha \beta} + \dfrac{c^2}{3} \rho \delta_{\alpha \beta}    \right],
\end{equation}
and the kinematic viscosity is
\begin{equation}
\nu = \dfrac{c^2}{3} \Delta t \left(\tau - \dfrac{1}{2} \right).
\end{equation}

\section{Boundary conditions}
\label{sec:appD}
In many practical situations, such as in a system under shear flow, one may be interested in studying the physics of the system within a confined geometry. Here we describe the implementation of boundary conditions of a sheared bidimensional fluid defined on a lattice of size $L_x\times L_y$ and confined between two parallel flat walls located at $y=0$ and $y=L_y$.
Two key requirements are necessary for a correct description of the physics:
\begin{itemize}
\item no flux accross the walls,
\item fixed velocity $v_x^*$ along the walls,
\end{itemize}
which correspond to the following relations on the wall sites:
\begin{equation}
\sum_i f_i \xi_{i x} = \rho v_x^*, \qquad \sum_i f_i \xi_{i y} = 0.
\label{eqn:conditions_bb}
\end{equation}
Assuming a $d2Q9$ lattice geometry (see Fig.~\ref{img:lattices}) with the walls located along the lattice links (i.e. along the lattice vectors $\vec{\xi}_1$,$\vec{\xi}_3$), one can explicitly write the previous relations at $y=0$ (the bottom wall):
\begin{align}
f_2 + f_5 + f_6 - f_4 - f_7 - f_8 &=0, \label{eqn:expl_conditions_bb_1}\\
f_1 + f_5 + f_8 - f_3 - f_6- f_7 &=\rho v_x^*.
\label{eqn:expl_conditions_bb_2}
\end{align}
Note that after the propagation step, functions $f_0$, $f_1$, $f_4$, $f_7$ and $f_8$ are known, so that one can use relations~\eqref{eqn:expl_conditions_bb_1} and~\eqref{eqn:expl_conditions_bb_2} to determine 
the three unknown distribution functions $f_2,f_5,f_6$. 
This system of equations can be closed by adding the bounce-back rule:
\begin{equation}
f_2 = f_4.
\label{eqn:further_condition_bounce_back}
\end{equation}
The two remaining distribution functions $f_5$ and $f_6$ are then given by~\cite{zou1997}
\begin{align}
f_5 = \dfrac{1}{2} ( 2 f_7 + f_3 - f_1 + \rho v_x^*) \label{eqn:f5_1}\\
f_6 = \dfrac{1}{2} ( 2 f_8 + f_1 - f_3 - \rho v_x^*) \label{eqn:f6_1}.
\end{align}
With this choice for inward-pointing distributions, the desired momentum at the boundary is achieved. Unfortunately this scheme does not allow for the local conservation of mass since, after the collision step, inward-pointing distributions are not streamed. In~\cite{lamura2001} an improvement of this scheme was proposed to overcome such a problem. In the following we will use notation $f_{pi}$ to identify the outgoing distribution function in a wall lattice site at time $t-\Delta t$, while $f_i$ denotes those streamed from neighboring sites at time $t$. 
Besides conditions in~\eqref{eqn:conditions_bb} it is required that the fraction of mass moving towards the wall or eventualy still on a wall site at time $t-\Delta t$ is the same that moves from the wall or stay still on the walls at time $t$. This is expressed for a bottom-wall site by the following relation:
\begin{equation}
f_{p0}+f_{p7}+f_{p4}+f_{p8} = f_0 + f_5 + f_2 + f_6,
\label{eqn:lamura2000}
\end{equation}
where $f_0$ must be determined by solving the system of Eqs.~\eqref{eqn:conditions_bb} and~\eqref{eqn:lamura2000} together with the bounce-back condition~\eqref{eqn:further_condition_bounce_back}. This leaves unchanged the solutions for the unknown $f_5$ and $f_6$ in Eqs.~\eqref{eqn:f5_1} and~\eqref{eqn:f6_1}, but provides a new expression for $f_0$ that is thus given by:
\begin{equation}
f_0 = \rho - (f_1+f_3) - 2( f_4+f_7+f_8).
\end{equation}

Such scheme can be easily adjusted to the case of the pure forcing method presented in Section~\ref{sec:external_force}. The only difference lies in the momentum conservation relations~\cite{kahler2015} that in such case read as follows,
\begin{equation}
\sum_i f_i \xi_{i x} + \dfrac{\Delta t}{2} F_x = \rho v_x^*,  \qquad \sum_i f_i \xi_{i y} + \dfrac{\Delta t}{2} F_y = 0.
\label{eqn:kahler2015}
\end{equation}
The system of Eqs.~\eqref{eqn:kahler2015} together with Eq.~\eqref{eqn:lamura2000} admits the following solutions:
\begin{align}
f_5 &= \dfrac{1}{2} \left( 2 f_7 + f_3 - f_1 + \rho v_x^* - \dfrac{\Delta t}{2} (F_x + F_y) \right), \label{eqn:f5_2}\\
f_6 &= \dfrac{1}{2} \left( 2 f_8 + f_1 - f_3 - \rho v_x^* + \dfrac{\Delta t}{2} (F_x - F_y) \right), \label{eqn:f6_2} \\
f_0 &= \rho - (f_1+f_3) - 2( f_4+f_7+f_8) + \dfrac{\Delta t }{2} F_y,
\end{align}
where the outward-pointing distribution $f_2$ was fixed by the bounce back condition~\eqref{eqn:conditions_bb}.

\end{document}